\documentclass[fleqn,usenatbib]{mnras}

\usepackage{newtxtext,newtxmath}
\usepackage{physics}
\usepackage{hyperref}
\usepackage{multirow, tabularx}
\usepackage{array}
\usepackage[table, svgnames, dvipsnames]{xcolor}
\usepackage{makecell, cellspace, caption}

\usepackage[T1]{fontenc}
\usepackage{mathtools}
\usepackage{orcidlink}

\DeclareRobustCommand{\VAN}[3]{#2}
\let\VANthebibliography\thebibliography
\def\thebibliography{\DeclareRobustCommand{\VAN}[3]{##3}\VANthebibliography}
\def\beq{\begin{equation}}
\def\eeq{\end{equation}}

\def\softwarenamestyle[#1]{\textsc{#1}}

\defcitealias{moTwophaseModelGalaxy2024}{Paper-I}
\defcitealias{chenTwophaseModelGalaxy2024a}{Paper-II}
\defcitealias{luEmpiricalModelStar2014}{L14}

\usepackage{graphicx}	% Including figure files
\usepackage{amsmath}	% Advanced maths commands
\usepackage{mathtools}
\usepackage[dvipsnames]{xcolor}
\usepackage{enumitem}
\usepackage{url}
\usepackage{braket}
\usepackage{threeparttable}
\usepackage{hyperref}

\def\dex{\,{\rm dex}}
\def\Kelvin{\,{\rm K}}
\def\Mpc{\,{\rm Mpc}}
\def\pc{\,{\rm pc}}
\def\mpc{\, h^{-1}{\rm {Mpc}}}

\def\Kpc{\, {\rm {kpc}}}
\def\kpc{\, h^{-1}{\rm {kpc}}}

\def\Myr{\,{\rm Myr}}
\def\Gyr{\,{\rm Gyr}}

\def\kms{\,{\rm {km\, s^{-1}}}}
\def\Msun{{\rm M_\odot}}
\def\msun{\, h^{-1}{\rm M_\odot}}
\def\msunperyr{\, {\rm M_\odot}/{\rm yr}}

\def\perccm{\, {\rm cm}^{-3}}
\def\Msunperpcsq{\,{\rm M_\odot}{\rm pc}^{-2}}
\def\breakpara{\bigskip\noindent}
\def\figcapem{\bf}
\def\Zsun{\,{\rm Z}_{\odot}}
\title[Formation of Globular Clusters]{A two-phase model of galaxy formation: III. The formation of globular clusters}

\author[Yangyao Chen et al.]{
Yangyao Chen\orcidlink{0000-0002-4597-5798},$^{1,2}$\thanks{E-mail: yangyaochen.astro@foxmail.com}
Houjun Mo\orcidlink{0000-0001-5356-2419}$^{3}$
and
Huiyuan Wang\orcidlink{0000-0002-4911-6990}$^{1,2}$
\\
$^{1}$School of Astronomy and Space Science, University of Science and Technology of China, Hefei, Anhui 230026, China\\
$^{2}$Department of Astronomy, University of Science and Technology of China, Hefei, Anhui 230026, China\\
$^{3}$Department of Astronomy, University of Massachusetts, Amherst, MA 01003, USA
}

\date{Accepted XXX. Received YYY; in original form ZZZ}

\pubyear{2024}

\begin{document}
\label{firstpage}
\pagerange{\pageref{firstpage}--\pageref{lastpage}}
\maketitle

\begin{abstract}
We develop a model of globular cluster (GC) formation within the cosmological 
hierarchy of structure formation. The model is rooted in the `two-phase' scenario 
of galaxy formation developed in Paper-I, where the fast accretion 
of dark matter halos at high redshift leads to the formation 
of self-gravitating, turbulent gas clouds that subsequently fragment into 
dynamically hot systems of dense sub-clouds with masses $\sim 10^6$--$10^7 \Msun$. 
Here we elaborate on the formation, evolution, and fate of these sub-clouds, 
and show that some of the sub-clouds can be compactified via two distinctive 
channels into a `supernova-free' regime to form two distinct populations of GCs. 
The model is simple, characterized by a small number of free parameters 
underpinned by physical considerations, and can be efficiently implemented into 
cosmological N-body simulations to generate a coherent sample of halos, 
galaxies, and GCs. 
Calibrated with observations, 
our model can reproduce a range of observational statistics,
including those for GC masses, sizes, metallicities, spatial distributions,
and the relation of GC systems with host galaxies/halos.
Significant discrepancies between model results and existing observations are 
discussed in connection to processes implemented in the model.
Predictions for GCs are made for both the local Universe 
and for redshift up to $z \approx 10$, and can be tested by upcoming 
observations.

\end{abstract}


%%%%%%%%%%%%%%%%%%%%%%%%%%%%%%%%%%%%%%%%%%%%%%%%%%

%%%%%%%%%%%%%%%%% BODY OF PAPER %%%%%%%%%%%%%%%%%%

%%%%%%%%%%%%%%%%%%%%%%%%%%%%%%%%%%%%%%%%%%%%%%%%%%%%%%%%%%%%%%%%%%%%%%%%%%%%%%%%

\section{Introduction}
\label{sec:intro}

% Introduce GCs as an important population of cosmic objects. 
% They are old objects, may contain information about formation at early time, 
% They are found in different environments, and may study environmental effects, and how formation and environments are linked. They are also extreme, some of them are metal poor, and 
% very dense, testing of star formation in extreme conditions.

Globular clusters (hereafter GCs) are massive and dense stellar systems observed in 
nearly every galaxy \citep{harrisGlobularClustersGalaxies1979,
krumholzStarClustersCosmic2019}. Many GCs have old age, 
low metallicity and high density compared to their host galaxies, implying
that they are remnants of star formation in the early 
Universe \citep{peeblesOriginGlobularStar1968,fallTheoryOriginGlobular1985} and thus provide 
fossils for cosmological archaeology. At a distinct and important rank  
in the cosmic structure hierarchy, GCs are crucial to understanding galaxy formation, 
extending its boundaries to extreme conditions on small scales.

Current observations have already provided critical information about the formation 
of GCs. They are observed in galaxies of all masses \citep{
blakesleeDependenceGlobularCluster1997,blakesleeGlobularClustersDense1999,
harrisGlobularClustersGalaxies1979,harrisPhotometricSurveyGlobular2023} 
and morphology \citep{harrisDarkMatterHalos2015} in diverse environments 
\citep{forbesGlobularClustersComa2020,janssensGlobularClustersStar2022,
saifollahiImplicationsGalaxyFormation2022,
jonesGasrichFieldUltradiffuse2023,
forbesUltraDiffuseGalaxies2024}
in the local Universe. The discovery of young massive star clusters 
\citep[YMSCs;][]{whitmoreLuminosityFunctionYoung1999,
figerMassiveStarsQuintuplet1999,
schweizerYoungGlobularClusters1999,
zhangMassFunctionYoung1999,
figerYoungMassiveClusters2004,
bastianYoungStarCluster2006} 
suggests that the conditions for GC formation are still 
present in the local Universe, particularly in starburst and interacting/merging 
galaxies \citep[see][for a review]{krumholzStarClustersCosmic2019}.

The ubiquitous presence of GCs implies that their formation may be driven 
by some general formation mechanisms rather than rare conditions associated with 
special galaxies or with rare stochastic processes. On the other hand, observations also show 
that the population of GCs within a galaxy, commonly referred to as the globular cluster 
system (GCS) of the galaxy, shows strong correlations with the properties of the host galaxy 
and host halo \citep{harrisGlobularClustersGalaxies1979}, suggesting that 
the diversity and complexity of the GC population may be related to those 
of the galaxy population. The most intriguing correlation is the linear relationship 
between the total mass of GCS and the host halo mass observed in the local Universe
\citep[e.g.][]{blakesleeDependenceGlobularCluster1997,blakesleeGlobularClustersDense1999, 
spitlerNewMethodEstimating2009,harrisCatalogGlobularCluster2013,hudsonDarkMatterHalos2014,harrisDarkMatterHalos2015,
harrisGalacticDarkMatter2017,dornanInvestigatingGCSMRelation2023,
harrisPhotometricSurveyGlobular2023}. Exceptions seem to occur in dwarf galaxies 
with halo masses $ \sim 10^{10} \Msun$, where the number of GCs is very small 
and has large scatter \citep{saifollahiImplicationsGalaxyFormation2022,jonesGasrichFieldUltradiffuse2023}, 
and in the brightest cluster galaxies (BCGs), where contributions from satellite
galaxies and the intracluster population can introduce significant ambiguity 
\citep{spitlerNewMethodEstimating2009,hudsonDarkMatterHalos2014,
dornanInvestigatingGCSMRelation2023}.

This linear relation was inferred to be a result of the central limit theorem 
applied to massive galaxies, where the total number of GCs brought in by merger
events naturally normalize the GCS mass with the halo mass
\citep{boylan-kolchinGlobularClusterdarkMatter2017,valenzuelaGlobularClusterNumbers2021}. 
However, the establishment of such a relation in dwarf galaxies and whether
or not this relation also holds at high redshift (hereafter high $z$) 
remain poorly understood \citep[e.g.][]{bastianGlobularClusterSystem2020}. 
Alternative explanations have also been proposed, such as higher
gas fractions and enhanced GC formation rates in halos of lower masses
\citep[e.g.][]{choksiFormationGlobularCluster2018,
choksiOriginsScalingRelations2019,
el-badryFormationHierarchicalAssembly2019,
chenModelingKinematicsGlobular2022,
chenFormationGlobularClusters2023,chenCatalogueModelStar2024}.
Within individual galaxies, GCs are found to have a bimodal distribution in color 
\citep{fahrionFornax3DProject2020a,
harrisPhotometricSurveyGlobular2023,hartmanComparingGlobularCluster2023}. 
This bimodality is interpreted as a result of the bimodal distribution of
metallicity, as metal-poor and metal-rich stars follow different evolutionary 
paths in the color-magnitude diagram, 
which is also supported by observations with spectroscopically calibrated
color-metallicity conversions \citep{fahrionFornax3DProject2020,hartmanEffectAgeStellar2024}.
The observed bimodal distribution signifies mechanisms that correlate
non-linearly with metallicity, such as galaxy-galaxy merger
\citep{liModelingFormationGlobular2014,choksiFormationGlobularCluster2018,valenzuelaGlobularClusterNumbers2021} 
and high-density/pressure 
condition \citep{kruijssenFractionStarFormation2012,maSelfconsistentProtoglobularCluster2020},
or a bimodal pair of mechanisms that drives the formation and 
evolution of GCs. There is thus a strong motivation to investigate how the formation
and evolution of GCs are correlated with the metallicity of the interstellar medium (ISM).

Previous theoretical modeling and numerical simulations of GC formation have 
reached a consensus that GCs in general emerge from turbulent, highly pressurized 
density peaks in the ISM 
\citep[see e.g.][for a review]{kruijssenGlobularClusterFormation2014}. 
Cosmological zoom-in simulations, such as those conducted by 
\citet{maSelfconsistentProtoglobularCluster2020}, showed that such conditions 
are naturally produced at high $z$. During the structure formation 
at high $z$, fast mass accretion by a dark matter halo can drive 
large amounts of gas inflow into the galaxy that has formed at the halo center.  
Such a collapsing process, coupled with the high ISM density and supernova 
feedback, can generate supersonic turbulence that leads to the formation of dense 
gas clumps and GCs. At low $z$, when halo accretion becomes slow and 
the overall ISM density becomes too low to form a large number of dense gas clouds,  
GC formation can still be triggered by occasional galaxy mergers and interactions
\citep{kruijssenGlobularClusterFormation2014}.
Simulations of individual giant molecular clouds (GMCs) provide further clues 
about the conditions needed for GC formation. As shown
in \citet{fernandezSlowCoolingLowmetallicity2018}, dense gas clumps can 
be produced in a metal-poor ISM under the influence of a radiation background,  
even in the absence of compression by strong initial turbulence.
In this case, the GMC is unable to fragment and form stars due to 
inefficient cooling; instead, it contracts and collapses further to 
form a dense gas core, providing conditions needed for the formation of a GC. 
These simulations provide support to observational inferences 
that GCs have formed predominantly in metal-poor ISM in the early Universe. 
They also demonstrate that the formation of GCs should and can be investigated 
in the hierarchy of cosmological structure formation.

The main challenge in modeling GC formation in a cosmological context is 
the large gap between the properties of GCs and their host galaxies. GCs are 
relatively small, with a typical size of $\sim 10\pc$ 
\citep[e.g.][]{krumholzStarClustersCosmic2019} that is about two orders 
of magnitude smaller than a typical dwarf galaxy 
\citep[e.g.][]{wangEvolutionaryContinuumNucleated2023}. Consequently, 
certain approximations are necessary to simulate GC formation as a part of 
galaxy formation in the cosmic density field. 
Along this line, two semi-analytic approaches have been developed, 
each incorporating different physical inputs derived from numerical simulations.
The first approach involves post-processing cosmological hydrodynamic 
simulations, or their zoom-in runs, with subgrid physics specific to GC 
formation. Depending on the resolution of the simulation used, the input conditions 
for GC formation can range from global properties of individual GMCs 
\citep[e.g.][]{grudicGreatBallsFIRE2023} to the smoothed field of the gas 
properties \citep[e.g.][]{kruijssenModellingFormationEvolution2011,
pfefferEMOSAICSProjectSimulating2018,
kruijssenEMOSAICSProjectTracing2019}. 
The second approach more closely resembles traditional semi-analytic models of 
galaxies based on dark matter halos and their merger trees. The global properties 
of individual galaxies are obtained from a semi-analytic model  
\citep{el-badryFormationHierarchicalAssembly2019,deluciaOriginGlobularClusters2024} 
or an empirical model of galaxy formation 
\citep{liModelingFormationGlobular2014,choksiFormationGlobularCluster2018,valenzuelaGlobularClusterNumbers2021}, 
while the criteria for GC formation are based on some properties of the assembly 
histories of galaxies/halos, such as mergers of galaxies  
\citep{el-badryFormationHierarchicalAssembly2019} and rates of halo mass assembly 
\citep{liModelingFormationGlobular2014,choksiFormationGlobularCluster2018,
valenzuelaGlobularClusterNumbers2021,
chenModelingKinematicsGlobular2022}. Both approaches have been successful in 
reproducing some observed properties of GCs.
Despite the progress, significant knowledge gaps still exist in our understanding 
of the GC population. Indeed, our understanding of galaxy formation, in particular 
of star formation and feedback processes, is still incomplete, and so the large-scale conditions 
for the formation of GCs are still uncertain. In addition, small-scale processes driving 
the formation and disruption of GCs are also not well understood.   

Recent observations by the James Webb Space Telescope (JWST) have opened 
a new avenue to study galaxy formation in the early Universe. 
These observations indicated that early galaxies may differ significantly from 
their counterparts in the local Universe, suggesting that the processes driving 
galaxy formation at high $z$ may be different from those in the low-$z$ Universe. 
In particular, the efficiency of gas consumption by star formation seems to be 
enhanced significantly at high $z$ \citep{bouwensEvolutionUVLF2023,
finkelsteinCEERSKeyPaper2023,xiaoAcceleratedFormationUltramassive2024,weibelGalaxyBuildupFirst2024,
wangMAssiveGalaxiesCOSmic2024};
supermassive black holes (SMBHs) appear to be over-massive for their host galaxy mass
\citep{pacucciJWSTCEERSJADES2023,maiolinoJADESDiversePopulation2024,
mattheeLittleRedDots2024,
pacucciRedshiftEvolutionRelation2024};
there seems to be an inversion of metallicity gradients in both the ISM and 
circumgalactic medium \citep[CGM;][]{wangEarlyResultsGLASSJWST2022,
linMetalenrichedNeutralGas2023}; 
and the total mass of bound star clusters in galaxies
seems to be excessively large \citep{vanzellaEarlyResultsGLASSJWST2022,
vanzellaJWSTNIRCamProbes2023,
welchRELICSSmallscaleStar2023,
linMetalenrichedNeutralGas2023,
claeyssensStarFormationSmallest2023,
adamoBoundStarClusters2024,
messaPropertiesBrightestYoung2024,
fujimotoPrimordialRotatingDisk2024,
mowlaFireflySparkleEarliest2024,
whitakerDiscoveryAncientGlobular2025}. Given that many GCs are old, these discoveries at high $z$
have important implications for understanding GC formation. 

Motivated partly by these discoveries, \citet[][hereafter Paper-I]{moTwophaseModelGalaxy2024}
recently proposed a `two-phase' model of galaxy formation that highlights the 
transition from an early phase of fast halo accretion to a later phase of slow accretion. 
A critical part of this model is the formation of self-gravitating and turbulent gas 
clouds (SGCs) during the fast phase. Such an SGC is expected to fragment subsequently 
and form a dynamically hot system of dense sub-clouds (SCs), each comparable to a GC in terms 
of both size and mass. This model is capable of reproducing numerous properties of galaxies 
observed over a large range of redshift. In particular, its predictions for the 
sub-cloud formation are directly related to the formation of GCs. 
As we will show in this paper, the model naturally provides conditions for
two channels for the formation of GCs, and the GCs produced in the two channels have 
distinctive properties. Our model thus provides a unified cosmological framework to 
understand the formation of both galaxies and their GCs, allowing the model 
to be tested on spatial scales ranging from star clusters to large-scale structures, 
and on timescales from $z \approx 0$ to $z \approx 10$.

This paper is organized as follows. In \S\ref{sec:galaxy-model}, we describe 
the context of galaxy formation within which GCs emerge. In \S\ref{sec:gc-model}, 
we elaborate on our model of GC formation. In \S\ref{sec:results}, we describe 
numerical implementations of our model, present its predictions and 
compare them with observations and with other models. In \S\ref{sec:summary}, 
we summarize our results and discuss their implications.
Throughout this paper, we adopt a flat $\Lambda$CDM cosmology with parameters 
derived from the Planck2015 results \citep{planckcollaborationPlanck2015Results2016}:
the density parameters $\Omega_{\rm M,0}=0.3089$, 
$\Omega_{\rm B,0}=0.0486$, and $\Omega_{\rm \Lambda,0}=0.6911$; 
the Hubble constant $H_0 = 100\, h\,{\rm km\,s^{-1}\,Mpc^{-1}}$, with $h=0.6774$; 
the Gaussian initial density field with a power spectrum $P(k)\propto k^n$ with 
$n=0.9667$; and a perturbation amplitude specified by $\sigma_8=0.8159$. 
To avoid confusion, we use `log' to denote the base-10 logarithm, and `ln' to 
denote the natural logarithm.
We use the terms 1, 2, and 3-$\sigma$ ranges to denote those 
centralized at the median value
and covering $68\%$, $95\%$, and $99.7\%$, respectively, of the probability mass 
for any probability distribution. 

%%%%%%%%%%%%%%%%%%%%%%%%%%%%%%%%%%%%%%%%%%%%%%%%%%%%%%%%%%%%%%%%%%%%%%%%%%%%%%%%

\section{The context of galaxy formation} \label{sec:galaxy-model}

% The contents:
% The two-phase model: the basic picture; halo merger trees; SGCs;
% SF and SMBH growth; metallicity evolution; SCs. 
% For SCs we just mention the basic formation scenario expected in the context 
% of structure formation, and mention that we leave the details related to 
% GCs to the next section.

\begin{figure*} \centering
    \includegraphics[width=0.825\textwidth]{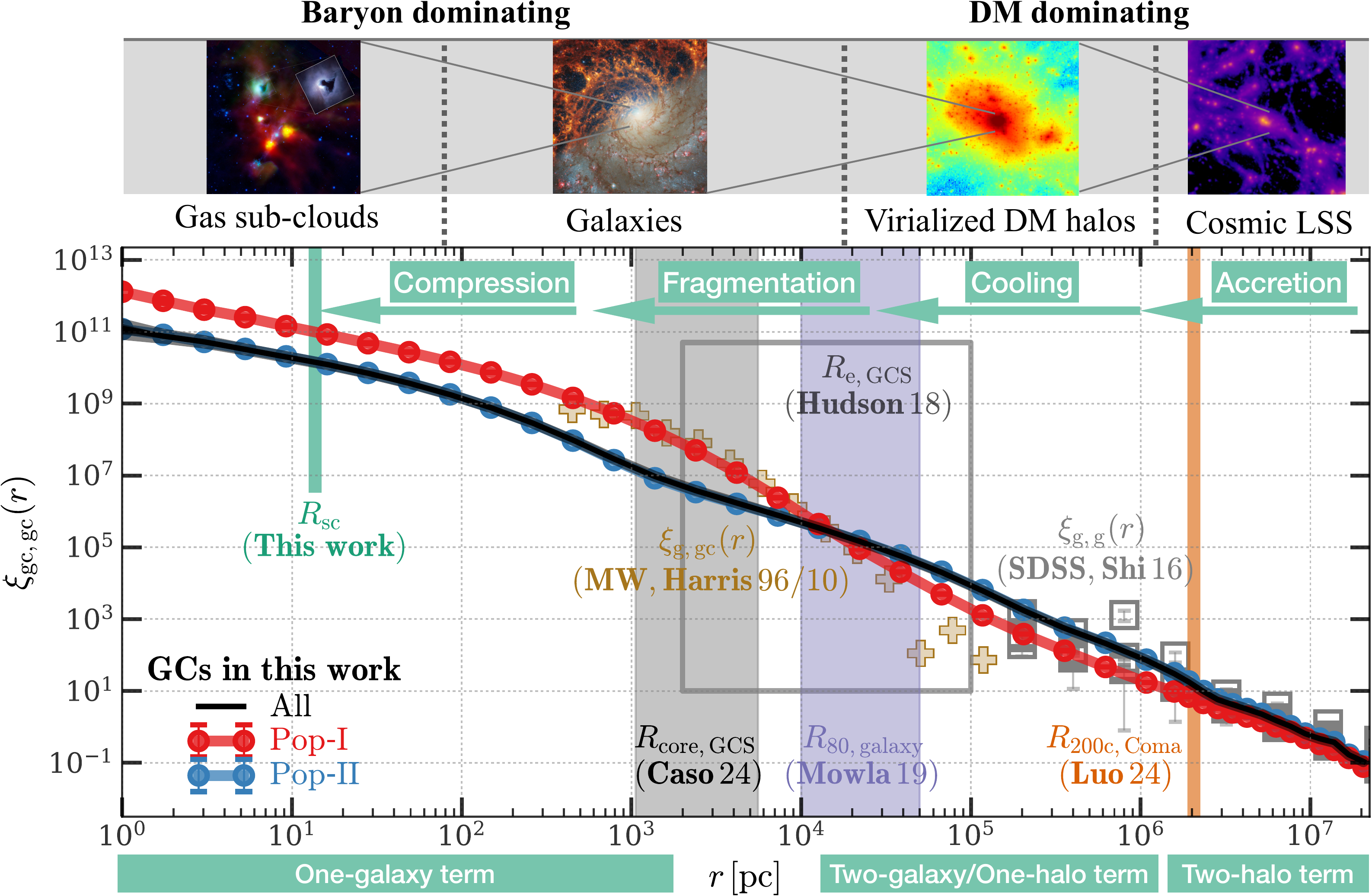}
    \caption{
        Two-point auto-correlation functions of 
        GCs at $z=0$ predicted by the model in this paper.
        {\bf Black}, {\bf red}, and {\bf blue} curves show the results for all, Pop-I 
        and Pop-II GCs, respectively, with errorbars or shaded
        areas showing 
        the standard deviation computed using 20 bootstrap samples.
        {\bf Vertical lines/shaded areas} mark the typical scales of structures.
        {\bf Green line} indicates $R_{\rm sc}$, 
        the typical size of GC-forming sub-clouds derived 
        in this paper (\S\S\ref{ssec:sf-in-scs} and \ref{ssec:size-mass}). 
        {\bf Grey shaded} area indicates the core radius, 
        $R_{\rm core, GCS}$, below which GC profile presents 
        flattening, obtained by 
        \citet{casoScalingRelationsGlobular2019},
        \citet{debortoliScalingRelationsGlobular2022} 
        and \citet{casoScalingRelationsGlobular2024} 
        for a sample of early-type galaxies with $M_* \gtrsim 10^9 \Msun$
        in the local volume. Here we show the 
        $5^{\rm th}$-$95^{\rm th}$ percentile range of their sample.
        {\bf Grey rectangle} indicates
        the effectively radius, $R_{\rm e, GCS}$, of GC systems 
        for galaxies with $M_* > 10^{10.5} \Msun$, compiled 
        by \citet[][see their figure 12]{hudsonCorrelationSizesGlobular2018}.
        {\bf Purple shaded area} indicates the range of 
        $R_{\rm 80,galaxies}$, the 80\%-light radius in optical band,
        for galaxies with $M_* > 10^{10.5} \Msun$, 
        obtained by \citet{mowlaMassdependentSlopeGalaxy2019}.
        {\bf Orange line} indicates the radius of the Coma cluster simulated
        by \citet{luoELUCIDVIIISimulating2024} based on the constrained density 
        field. Other markers show the observations at different scales. 
        {\bf Brown markers} show the galaxy-GC cross-correlation
        function obtained by the Milky Way GCs compiled by
        \citet{rodriguezGreatBallsFIRE2023} based on the catalog of
        \citet{harrisCatalogParametersGlobular1996,harrisMassiveStarClusters2010}.
        {\bf Grey markers} show the real space 
        galaxy-galaxy auto-correlation functions obtained by 
        \citet[see their figure 10 and table 3]{shiMappingRealSpace2016} 
        for a sample of flux-limited SDSS galaxies 
        in a number of bins of $M_{\rm r}$, 
        based on their correction method for the redshift-space distortion.
        This figure demonstrates the overall context of our model
        for the formation of halos, galaxies and 
        star clusters. See \S\ref{sec:galaxy-model} and 
        \S\ref{sec:gc-model} for the details of the model, 
        and \S\ref{ssec:spatial-dist} for the description of this figure.
    }
    \label{fig:tpcf}
\end{figure*}

\begin{figure*} \centering
    \includegraphics[width=0.975\textwidth]{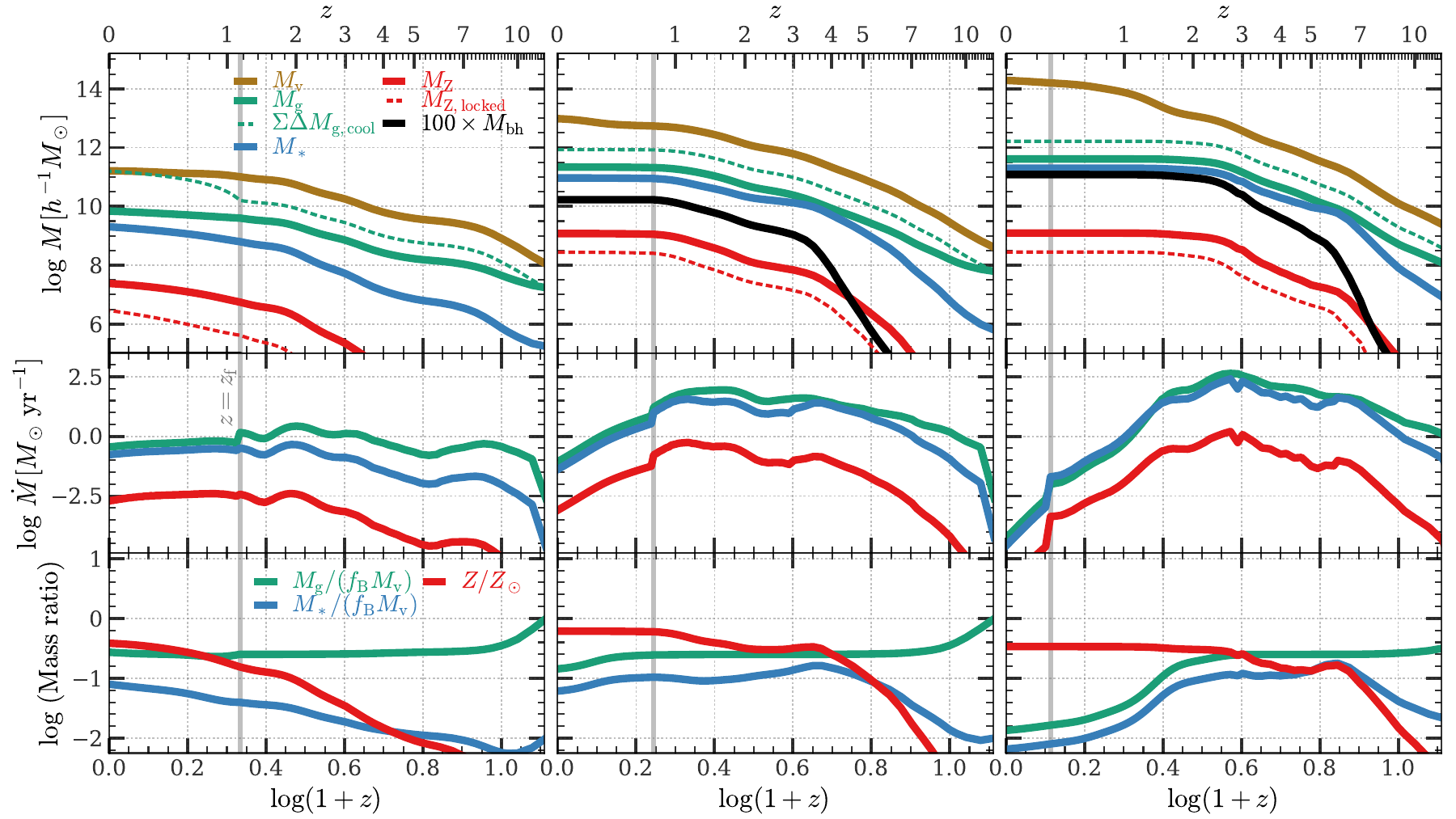}
    \caption{
        History of different mass components predicted by the galaxy model. 
        {\figcapem Each column} shows the results for an example halo and 
        its central galaxy. 
        Three {\bf rows} show the cumulative mass, the increase rate of 
        mass, and the ratio of two masses, respectively, as 
        indicated in the legends.
        {\bf Brown}, {\bf green}, {\bf blue}, {\bf red} and {\bf black} curves show the results for 
        halo, gas, stars, gas-phase metal and central SMBH, respectively.
        For gas, the {\bf solid} and {\bf dashed} curves show the total gas 
        mass within the galaxy, and the cumulative mass of cooled gas.
        For metal, the {\bf solid} and {\bf dashed} curves show the total 
        oxygen mass within the galaxy, and the cumulative mass of oxygen
        locked into stars and SMBHs.
        In each panel, the {\bf vertical grey} line marks the transition 
        redshift, $z_{\rm f}$, at which halo assembly transits from the `fast 
        phase' to the `slow phase'.
        See \S\ref{sec:galaxy-model} for the detailed definition of masses,
        their evolution equations, and the description of this figure. 
    }
    \label{fig:history_of_masses}
\end{figure*}

Our goal is to develop a paradigm to understand the formation of GCs
in the current cosmological framework of galaxy formation.  
Due to the large dynamic range covered by our model, we structure it in a hierarchical way, 
from large scales of low density to small scales of high density.
% An engineering approach - think it this way.
The basic idea behind the hierarchy is to introduce different physical and numerical
models to describe the `jumps' in characteristic scales, and to join them with 
recipes describing their interfaces. Thus, we start with modeling the structure formation 
at large scales, and use the results as boundary conditions, 
i.e. `environments', for the next step on smaller scales. The first scale, describing the 
large-scale structures of the cosmic density field and its interface with the second scale, 
which describes dark matter halos, is modeled using results from cosmological N-body 
simulations. Models of galaxy formation and the growth of SMBH are built 
on halos and their merger trees, using the two-phase paradigm described in 
\citetalias{moTwophaseModelGalaxy2024}, which bridges the interface between halos and galaxies. 
Gas fragmentation and star formation are modeled by a new 
recipe developed in this paper, which bridges the gap to the fourth scale, sub-clouds
in galaxies, where globular clusters form. Fig.~\ref{fig:tpcf} shows the overall context of our 
model, where the two-point auto-correlation function of globular clusters,  
predicted by our model, is shown to represent the scales and densities of the structures 
in question. The spatial and density scales of relevant gas processes 
(e.g. mass accretion, gas cooling, fragmentation and compression) and the corresponding 
structures (large-scale structure, halos, galaxies, and sub-clouds) are marked for reference.
This plot demonstrates schematically how the formation of GCs fits in the general 
cosmological context, and how the related physical processes define the characteristic 
scales involved in our problem. In what follows, we describe these processes in more detail. 

\subsection{The assembly of dark matter halos}
\label{ssec:halo-assembly}

Our modeling of the halo population is based on a sample of halos and 
merger trees constructed from numerical simulations of the current 
$\Lambda$CDM cosmology, as described in \S\ref{sec:results}.  
% Subhalos, which are bound structures within   
For each subhalo merger tree rooted in a `root subhalo' 
(i.e. the subhalo without a descendant), we decompose it into a set of disjoint 
branches using the method described in 
\citet[see their \S3.2.1]{chenConditionalAbundanceMatching2023}.
This is equivalent to traversing over the tree in a depth-first order
starting from the root subhalo, and extracting each subsequent segment of the 
tree bracketed by a subhalo and its main leaf. For each branch, we 
find the last snapshot when it is a central subhalo, and define this snapshot
as the `infall snapshot'. The assembly history at and before this snapshot 
is defined as the `central stage' of the branch, while that after it is defined 
as the `satellite stage'. 

In the central stage, galaxy formation (and SMBH growth) are modeled according
to the two-phase model described in \citetalias{moTwophaseModelGalaxy2024},
In this paper, we extend the previous model to include the recipes for the
satellite stage. A model for the formation of globular clusters is then 
built upon this extended galaxy model, as to be described in \S\ref{sec:gc-model}.

For each branch in the merger tree, we smooth its virial mass history, $M_{\rm v}(z)$, by a running 
Gaussian kernel with a width equal to the dynamical 
timescale,
$t_{\rm dyn} = R_{\rm v}(z) / V_{\rm v}(z) = 1/[10 H(z)]$,
where $R_{\rm v}(z)$ and $V_{\rm v}(z)$ are, respectively, 
the virial radius and virial velocity of the halo,
and $H(z)$ is the Hubble parameter, all evaluated at the redshift $z$ 
in question. 
Smoothed versions of other virial quantities
are defined using the smoothed $M_{\rm v}(z)$, while histories of other relevant 
halo quantities are smoothed using the same window as that used for $M_{\rm v}(z)$. 
The smoothing is required to ensure the numerical stability of our model implementation.
Note that we do not aim to model the short-time-scale variations in
the properties of halos, galaxies and star clusters, but intend to model the 
long-time-scale averages. The results of our model should thus be understood as 
an `effective' description over a timescale of $\gtrsim t_{\rm dyn}$.

As discussed in \citetalias{moTwophaseModelGalaxy2024}, the assembly history of a halo
can in general be separated into a fast phase, where $V_{\rm v}$ increases rapidly with 
time, followed by a slow phase where $V_{\rm v}$ remains roughly constant 
\citep[e.g.][]{zhaoGrowthStructureDark2003}, and galaxy formation in these two 
phases is expected to have very different characteristics.  We thus model 
the two phases separately using different recipes.
To this end, we identify, for the central stage of each branch, a transition 
redshift, $z_{\rm f}$, at which the halo assembly transits from the `fast phase'
to the `slow phase'. The details can be found in \citetalias{moTwophaseModelGalaxy2024}.
Briefly, we first fit the halo mass assembly history to the following functional form,
\begin{equation} \label{eq:mah-fitting}
    \ln\,M_{\rm v}(z) = c_0 + c_1 {z\over 1+z} + c_2 \ln(1+z) + c_3 z\,,
\end{equation}
where $c_0$, $c_1$, $c_2$ and $c_3$ are fitting parameters.  
We then obtain the specific halo growth rate, defined as  
$\gamma(z) \equiv \dot{V}_{\rm v}(z) / [ H(z)V_{\rm v}(z) ]$, from the 
fitting function. Finally, we determine the redshift at which $\gamma(z)$ falls 
below a threshold, $\gamma_{\rm f}$, and define this redshift as $z_{\rm f}$.
As our default choice, we set $\gamma_{\rm f}=3/16$, the central value between 
the Low-$z_{\rm f}$ and High-$z_{\rm f}$ variants used in \citetalias{moTwophaseModelGalaxy2024}.

The brown curves in the top row of Fig.~\ref{fig:history_of_masses} show the mass assembly
histories of three individual halos of different masses. The transition redshift, $z_{\rm f}$,
defined above is indicated by the vertical grey line in each panel. 
For each halo, the mass assembly rate is higher at higher $z$ and becomes roughly 
flat at $z < z_{\rm f}$. More examples can be found in \citetalias{moTwophaseModelGalaxy2024},
where the transition can be seen more clearly in the history of the halo virial velocity, 
$V_{\rm v} (z)$. To cover the large dynamic range relevant to our model, we use analytical 
approximations in our description to gain an intuitive understanding of the problem. 
The actual implementations are performed numerically (see \S\ref{ssec:impl}). In our modeling, many 
processes are connected to halo properties. We thus list some of the most important 
halo quantities, their typical values and redshift dependence in the following. 
Based on the spherical collapse model, the virial velocity of a halo can be approximated as
\begin{align}\label{eq:v-vir}
    V_{\rm v}  
    & = \left( \frac{\Delta_{\rm v}}{2} \right)^{1/6}\left[G M_{\rm v} H(z)\right]^{1/3}  \\
    & = \begin{dcases}
        \left[ 80.0 \kms \right] M_{\rm v,10}^{1/3} (1+z)_{10}^{1/2}       \,,\\
        \left[ 138.6 \kms \right] M_{\rm v,11.5}^{1/3} (1+z)_{3}^{1/2}     \,,
    \end{dcases}
\end{align}
where 
$1+z = 10 (1+z)_{10} = 3(1+z)_{3}$,
$M_{\rm v} = 10^{10} \Msun M_{\rm v,10} = 10^{11.5}\Msun M_{\rm v,11.5}$
and $\Delta_{\rm v} = 200$ \citep[e.g.][]{moGalaxyFormationEvolution2010}.
% all updated to Planck15
% $h = 0.7\,h_{0.7}$, 
% $\Omega_{\rm m} = 0.3\,\Omega_{\rm m,0.3}$ 
The virial radius is 
\begin{align}\label{eq:r-vir}
    R_{\rm v} &= \left(\frac{2 G M_{\rm v}}{\Delta_{\rm v}}\right)^{1/3} H^{-2/3}(z)  \\
    &= \begin{dcases}
        \left[ 6.7 \Kpc \right] M_{\rm v, 10}^{1/3}(1+z)_{10}^{-1}  \,,\\    
        \left[ 70.8 \Kpc \right] M_{\rm v, 11.5}^{1/3}(1+z)_{3}^{-1} \,.
    \end{dcases}
\end{align}
The accretion rate of a halo, calibrated by simulations \citep{dekelToyModelsGalaxy2013}, can be  approximated as 
\begin{equation}\label{eq:m-vir-dot}
    \dot{M}_{\rm v} =
    \begin{cases}
        \left[ 49.8\msunperyr \right] M_{\rm v,10}^{1.14}(1+z)_{10}^{2.5} \,,\\
        \left[ 125.9\msunperyr \right] M_{\rm v,11.5}^{1.14}(1+z)_{3}^{2.5} \,.
    \end{cases}
\end{equation}

\subsection{The collapse of the gas component}
\label{ssec:collapse-gas}

%I have add additional words for the cold-mode/filamentary accretion, as 
%it is asked by almost all people listening to our talks (e.g. Shude Mao, 
%Fangzhou Jiang). The analytical approximations are updated to the Planck15
%cosmology.
%

As gas gets accreted into a halo, it can cool and contract until it becomes 
self-gravitating or supported by rotation. One important conclusion of the 
two-phase model developed in \citetalias{moTwophaseModelGalaxy2024}
is the association of the formation of dynamically hot stellar systems and SMBHs with 
the self-gravitating, turbulent gas cloud (referred to as the SGC) expected during the fast 
accretion phase of the dark matter halo, while the formation of a stable, angular-momentum 
supported disk is associated only with the slow phase. The basic arguments can be 
summarized as follows.
\begin{itemize}[topsep=0pt,parsep=0pt,itemsep=0pt]
    \item 
    During the early, fast-accreting stage of halo assembly 
    ($z \geqslant z_{\rm f}$) and if $M_{\rm v} (z) \lesssim 10^{12}\Msun$, 
    the cooling timescale of the halo gas is shorter than the free-fall timescale
    ($t_{\rm cool,halo} < t_{\rm ff, halo}$),
    so that cold gas can flow in free-fall to the center of the halo.
    Thus, before stars and SMBHs can drive a large fraction 
    of the gas from the halo, the galaxy that forms in the halo center 
    is expected to be gas-rich, and the baryon fraction of the galaxy     
    roughly follows the universal baryon fraction, $f_{\rm B}\sim 0.15$. With 
    such a high fraction, the contraction of the cooled halo 
    gas will become self-gravitating (thus forming an SGC) before 
    it can be supported by angular momentum. Gas motion is expected to be turbulent in 
    such an SGC because of the fast mass assembly and effective cooling.
    In the presence of effective cooling, the SGC is also expected to fragment 
    into sub-clouds due to gravitational instabilities, and the sub-clouds formed 
    in this way should inherit the turbulent motion generated by the rapid gravitational collapse. 
    Shocks associated with the turbulent motion 
    can compress sub-clouds to higher density, which allows them to move ballistically
    without being affected significantly by ram pressure and cloud-cloud collision. 
    Stars that form in the dense sub-clouds also inherit the random motion,  
    forming a dynamically hot stellar system such as the bulge of a galaxy.
    Sub-clouds with sufficiently low angular momentum can reach the halo center to feed 
    the central SMBH, producing feedback effects that can reduce the amount of cold gas.
    %In other cases where the inflow gas is not spherically symmetric, for example, is in cold filaments as suggested by numerical simulations of structure formation at high redshift, the fragmentation of the SGC, the formation and compression of sub-clouds, and their random motion, are expected to be stronger.
    \item 
    In the intermediate stage where feedback from stars and SMBHs can heat/eject 
    large amounts of the gas from the galaxy, a disk may form. However, 
    if the halo is still in the fast accretion regime, frequent interactions and 
    mergers can thicken or even destroy the disk. If the galaxy remains gas-rich 
    (with the gas fraction much higher than the effective spin of the gas) 
    after transiting to the slow accretion phase, the formed disk structure will be too dense 
    to be stable, which may lead to the transformation of the disk 
    into a bar or a pseudo-bulge. Thus, a stable and thin disk is not expected 
    in this stage, while other dynamically hot stellar components, as well as the 
    central SMBH, may continue to grow.
    \item 
    In the late stage where the halo has transited to the slow accretion phase
    and feedback has reduced the gas fraction to a value that is comparable to or lower than 
    the effective spin of the gas, a long-lasting, angular momentum supported thin disk forms.
\end{itemize}
In any of the above stages, the galaxy may temporarily or permanently stop forming 
stars and become quenched, if the halo has grown massive enough 
($M_{\rm v} \gg 10^{12} \Msun$) so that cooling becomes inefficient,
and/or if feedback effects have ejected most of the gas from the halo.  
In addition, if the galaxy is accreted into another halo of larger mass and becomes 
a satellite, environmental processes may also play a role in quenching the galaxy.

The above `two-phase' scenario leads to a prediction that most of the central 
galaxies have undergone a transition from a dynamically hot phase to 
a dynamically cold phase driven by the transition in the halo assembly history. 
Therefore, many of the equations in the following will be presented separately for the 
two phases: the fast one at $z \geqslant z_{\rm f}$ and the slow one at $z < z_{\rm f}$. 
The formulations and fiducial parameters are designed and calibrated as in 
\citetalias{moTwophaseModelGalaxy2024}. The extensions needed to model the formation of 
globular clusters (GCs) are described in the next section. 

The model starts from the gas accretion associated with the halo assembly.
At each time step, the total amount of halo gas available for subsequent 
processes is determined by the mass assembly history of the halo: 
\begin{equation}\label{eq:delta-m-g-avail}
    \Delta M_{\rm g, avail} = 
    \begin{dcases}
        f_{\rm B} \Delta M_{\rm v}
        \,, \ \ \ \ \ &\text{if } z \geqslant z_{\rm f} \,, \\
        \frac{f_{\rm B} M_{\rm v}}{\tau(z)} \Delta t
        \,, &\text{if } z < z_{\rm f} \,.
    \end{dcases}
\end{equation}
where $f_{\rm B}$ is again the cosmic baryon fraction, 
$\tau(z) = (1+z)^{-3/2} / (10 H_0)$ is an approximation to the dynamical timescale 
of the halo, and $\Delta t$ is the time span of the step.
Note that the equation in the second line expresses the available gas mass 
in terms of the halo mass, $M_{\rm v}$, instead of $\Delta M_{\rm v}$. 
This is motivated by the fact that the halo mass may stop increasing 
significantly in the slow phase, while the gas already in the halo 
is still available to feed star formation and SMBH growth. 

The initial density of the gas accreted into the halo is below the threshold
of star formation. As it cools, the halo gas contracts to a radius of 
about $f_{\rm gas} R_{\rm v}$ where it becomes self-gravitating, forming an SGC
\citepalias[see][]{moTwophaseModelGalaxy2024}. The typical density of SGC is thus 
\begin{align} \label{eq:n-sgc}
    n_{\rm sgc} 
    & = \frac{f_{\rm str} f_{\rm gas} M_{\rm v}}{ (4\pi/3) (f_{\rm gas} R_{\rm v})^3 \mu m_{\rm p} } \nonumber \\
    & = f_{\rm str} \frac{n_{\rm B,0} \Delta_{\rm v}}{\mu f_{\rm gas}^2 f_{\rm B}} (1+z)^3 \nonumber \\ 
    & = f_{\rm str,4} f_{\rm gas, 0.04}^{-2} \times \begin{cases}
        \left[ 663.24 \perccm \right]\, (1+z)_{10}^3  \,,  \\
        \left[17.91 \perccm\right]\, (1+z)_3^3 \,,
    \end{cases}
\end{align}
where the cosmic baryon number density 
$n_{\rm B,0} = 3 H_0^2 f_{\rm B} \Omega_{\rm m} / (8\pi G m_{\rm p}) \approx 2.48 \times 10^{-7} \perccm $,
$\Delta_{\rm v} = 200$, $f_{\rm gas} = 0.04 f_{\rm gas, 0.04}$, $\mu =1.2$.
The factor $f_{\rm str}$ characterizes the `streamness' of the gas inflow. For 
spherically symmetric contraction, $f_{\rm str} = 1$, and the typical value 
of $f_{\rm gas}$, $0.04$, is regulated by feedback \citep[see][hereafter Paper-II]{chenTwophaseModelGalaxy2024a}.
For cold-mode streams, $1 < f_{\rm str} \leq f_{\rm B}/f_{\rm gas} \approx 4 $, 
as gas can penetrate into a galaxy along filaments and a larger fraction of gas 
can pile up around the halo center. \citet{mandelkerColdFilamentaryAccretion2018}
and \citet{dekelEfficientFormationMassive2023} estimated that the effective radius of 
streams is about $ 0.043 R_{\rm v}$, similar to the contraction factor 
based on $f_{\rm gas}$. 
%The later is believed as a common case at $z \gtrsim 1$-$2$ in halos below the 
%cooling threshold \citep{dekelToyModelsGalaxy2013}.
% Note: this is similar to what we guess to happen in the Q2 phase, where
% gas fraction is high, but dissipation is somehow significant. However, the 
% dissipation here takes place, rapidly, in the filaments, and does not require a 
% small $\gamma$. Structure of galaxy in this case may also be a slow rotator, 
% such as pseudo-bulge.
The SGC density given by Eq.~\eqref{eq:n-sgc} is much lower than the 
supernova-free threshold $n_{\rm snf} \approx 10^{3.5} \perccm$ required 
to form GCs (see \S\ref{ssec:gc-channels}) at $z=2$, and marginally touches the 
threshold at $z = 9$. Thus, additional processes within SGCs are required for  
the formation of GCs in galaxies. 

For halos with $M_{\rm v} \gtrsim 10^{12} \Msun$, the cooling time of the halo gas 
exceeds the free-fall time, so that only a fraction of the gas can cool within a 
free-fall timescale. We use a factor, $F_{\rm cool}$, to account for the 
reduction of the cold gas due to inefficient cooling,
\begin{equation} \label{eq:delta-m-g-cool}
    \Delta M_{\rm g,cool} = F_{\rm cool} \Delta M_{\rm g, avail}       \,.
\end{equation}
We follow \citetalias{moTwophaseModelGalaxy2024} to model $F_{\rm cool}$:
\begin{equation} \label{eq:f-cool}
    F_{\rm cool} = \begin{dcases}
        \left[
            1+  \left(\frac{M_{\rm v}}{M_{\rm cool}}\right) ^{\beta_{\rm cool, f}}
        \right]^{-1}
        \,, \ \ \ \ \ &\text{if } z \geqslant z_{\rm f} \,, \\
        \left( 1 + \frac{M_{\rm v}}{M_{\rm c}} \right)^{-\beta_{\rm cool, s}(z)} 
        \,, &\text{if } z < z_{\rm f} \,,
    \end{dcases}
\end{equation}
where we take $M_{\rm cool} = 10^{13} \msun$, $\beta_{\rm cool,f} = 4$,
$\beta_{\rm cool,s} = 3.6 (1+z)^{-0.72}$, and $M_{\rm c} = 10^{11.9} \msun$
as our fiducial values of the model parameters. 
The redshift dependence in the fast phase comes implicitly from the 
redshift dependence of the halo mass, while in the slow phase, it comes explicitly from the 
redshift dependence of the power index \citep[as in Model-II of][hereafter L14]{luEmpiricalModelStar2014}. 
The remaining fraction, $1 - F_{\rm cool}$, of the halo gas is referred to as 
the `hot' gas.

In the top row of Fig.~\ref{fig:history_of_masses}, the dashed green lines 
show the time evolution of $M_{\rm g, cool}$, the cumulative mass of the cooled gas,
in central galaxies of three example halos of different masses. 
For halos with mass smaller than the cooling threshold ($M_{\rm cool}$) and 
in the fast phase ($z \geqslant z_{\rm f}$), the amount of cooled gas tightly follows 
the total available baryons, with the mass roughly proportional to $f_{\rm B} M_{\rm v}(z)$, 
as shown in the left and middle panels. For massive halos, 
cooling becomes inefficient and the accumulation of cooled gas is stalled 
once its mass reaches $M_{\rm cool}$, as shown in the right panel.
In the slow phase and if the halo mass is sufficiently small,  
as is the case shown in the left panel, the gas that is heated and ejected from 
the galaxy may cool down and recycle back to the galaxy to feed further star formation. 
This is why the accumulative inflow mass of cooled gas can exceed $f_{\rm B} M_{\rm v}(z)$,
as seen at $z < 1$ in the left panel.

\subsection{Star formation, SMBH growth and feedback} \label{ssec:star-smbh-formation}

The amount of cooled gas can be affected by the feedback from 
supernova (SN) explosions associated with star formation and active galactic nuclei 
(AGN) associated with the growth of SMBHs. Thus, the amount of star-forming gas, $\Delta M_{\rm g,sf}$, 
is modulated by two additional factors, one for each source of feedback:
\begin{equation}
    \Delta M_{\rm g,sf} = F_{\rm sn} F_{\rm agn} \Delta M_{\rm g,cool}\,.
\end{equation}
Part of the feedback-affected gas is ejected from the galaxy (ejected gas), 
while the rest remains in the galaxy as a hot medium and is prevented from forming stars
(prevented gas). Following \citetalias{moTwophaseModelGalaxy2024}, we model the amounts of 
ejected and prevented gas as
\begin{equation}
    \Delta M_{\rm g,ej} = f_{\rm ej} (1 - F_{\rm sn} F_{\rm agn}) \Delta M_{\rm g,cool} \,,
\end{equation}
and
\begin{equation}
    \Delta M_{\rm g,prev} = (1-f_{\rm ej}) (1 - F_{\rm sn} F_{\rm agn}) \Delta M_{\rm g,cool} \,,
\end{equation}
respectively. 
Here, $F_{\rm sn}$ is given by
\begin{equation} \label{eq:f-sn}
    F_{\rm sn} = 
    \begin{dcases}
        \frac{
            \alpha_{\rm sn,f}  + (V_{\rm g}/V_{\rm w})^{\beta_{\rm sn,f}}
        }{
            1 +(V_{\rm g}/V_{\rm w})^{\beta_{\rm sn,f}}
        }
        \,, \ \ \ \ \ &\text{if } z \geqslant z_{\rm f} \,, \\
        \left(
            \frac{ M_{\rm v} / M_{\rm c} }{R + M_{\rm v} / M_{\rm c}}
        \right)^{\beta_{\rm sn, s}}
        \,, &\text{if } z < z_{\rm f} \,,
    \end{dcases}
\end{equation}
where $\alpha_{\rm sn,f} = 0$, $\beta_{\rm sn,f} = 2.5$, $V_{\rm w}=250 \kms$,
$R = 10^{-0.96}$, $\beta_{\rm sn, s} = 1.9$, and 
$V_{\rm g} = V_{\rm max}$ is the typical velocity dispersion of the galaxy.
$F_{\rm agn}$ is given by
\begin{equation} \label{eq:f-agn}
    F_{\rm agn} = 
    \begin{dcases}
        1- {\alpha_{\rm agn,f} M_{\rm bh} c^2 \over M_{\rm g} V_{\rm g}^2}
        \,, \ \ \ \ \ &\text{if } z \geqslant z_{\rm f} \,, \\
        \left(
            \frac{ R + M_{\rm v} / M_{\rm c} }{ 1 + M_{\rm v} / M_{\rm c} }
        \right)^{\beta_{\rm agn,s}}
        \,, &\text{if } z < z_{\rm f} \,,
    \end{dcases}
\end{equation}
where $\alpha_{\rm agn,f} = 10^{-3}$, $\beta_{\rm agn,s}=1.8$,
and $M_{\rm bh}$ and $M_{\rm g}$ are the masses of central SMBH and the gas 
within the galaxy (equations describing their evolution are given below), respectively.
Modeled in this way, the effects of the SN and AGN feedback are constrained by the 
gravitational potential of the host halo, which is needed to reproduce the observed
$M_*$-$M_{\rm v}$ relation at the low-mass ($\approx 10^{8} \Msun$) end 
\citep{yangEvolutionGalaxyDarkMatter2012} and the scaling relation of 
$M_{\rm bh}$-$M_{\rm *,bulge}$ \citep{grahamAppreciatingMergersUnderstanding2023}.
The fraction of ejected gas is set to be $f_{\rm ej} = 0.75$ for the fast phase. 
This choice was made in \citetalias{moTwophaseModelGalaxy2024} to ensure 
that the gas fraction in a feedback-regulated galaxy is comparable to the 
typical spin, $\lambda \approx 0.04$, so that a switch from bulge formation to disk 
formation can occur after the fast phase. For the slow phase, we set $f_{\rm ej} = 1$.  
Note that in the slow phase, the hot ISM component represented by 
$\Delta M_{\rm g, prev}$ is degenerate with the remaining part of 
$\Delta M_{\rm g, sf}$ after the consumption of star formation, 
so that $f_{\rm ej}$ is degenerate with $F_{\rm sn}$ and $F_{\rm agn}$, and 
the star formation efficiency of the star-forming gas, $\Delta M_{\rm g, sf}$. 
Once the star formation efficiency is fixed, the choice of $f_{\rm ej}$ does affect 
the gas fraction in the galaxy. The roughly constant gas fraction shown by the green 
curve in the bottom left panel of Fig.~\ref{fig:history_of_masses} indicates
a gas fraction of $M_{\rm g}/M_{\rm v} \approx \lambda \approx 0.04$.
Such a gas fraction allows a transition from bulge to disk at the end of the fast phase,
and the formation of a stable disk in the slow phase
\citep[see \S3.2 of][]{moFormationGalacticDiscs1998},
suggesting that the gas fraction predicted by our choice for $f_{\rm ej}$ 
is reasonable.

The star formation rate is assumed to be proportional to the generation rate of 
star-forming gas:
\begin{equation} \label{eq:delta-m-star}
    \Delta M_* = \epsilon_* \Delta M_{\rm g,sf} \,,
\end{equation}
where $\epsilon_* = 0.75$ and $0.32$ for the fast and slow phases, respectively, 
as calibrated in \citetalias{moTwophaseModelGalaxy2024}.
Here, we assume instantaneous evolution, as the timescale of stellar evolution is 
usually much shorter than the dynamical timescale of the halo.
Thus, the stellar mass modeled above should be treated as the remaining mass 
modulated by stellar evolution. The instantaneous star formation rate can be obtained by 
\begin{equation} \label{eq:mass-return}
    {\rm SFR} = \frac{1}{1-R} \frac{\Delta M_*}{\Delta t}\,,
\end{equation}
where $R = 0.4$ is the returned fraction of mass 
assuming a Chabrier IMF \citep{bruzualStellarPopulationSynthesis2003}.
For small-scale processes, such as the formation of globular clusters, the timescale of 
stellar evolution is critical in setting the time window for star formation. 
We will discuss this in detail in \S\ref{sec:gc-model}.

The growth of SMBH is determined by the fraction of low-angular-momentum
gas that can reach the halo center to feed the central SMBH. Following the 
arguments in \citetalias{moTwophaseModelGalaxy2024}, we model it as
\begin{equation} \label{eq:delta-m-bh}
    \Delta M_{\rm bh} = \alpha_{\rm cap} { M_{\rm bh} \over M_{\rm g} } 
        F_{\rm en} \Delta M_{\rm g, sf}    \,,
\end{equation}
where $\alpha_{\rm cap} = 2.5$ in the fast phase, and $\alpha_{\rm cap} = 0$ 
in the slow phase due to the full mixing of gas. 
The enhancement factor, $F_{\rm en}$, which describes the enhancement of 
turbulence by SN feedback, is modeled as   
\begin{equation}
    F_{\rm en} = {\alpha_{\rm en} + (M_{\rm v}/M_{\rm en})^{\beta_{\rm en}}
        \over 1+ (M_{\rm v}/M_{\rm en})^{\beta_{\rm en}}}\,,
\end{equation}
where $\alpha_{\rm en} = 3$, $\beta_{\rm en}=2$, and 
$M_{\rm en} = 10^{11.5}\msun$. The dependence on halo mass comes from the 
evidence that SN feedback on galactic scales is only effective in low-mass halos.
As discussed in \citetalias{moTwophaseModelGalaxy2024}, this enhancement factor 
is needed to boost the formation of the seeds for growing SMBHs.

Finally, the change in the amount of gas within a galaxy is determined by the 
combination of cooling, star formation, SMBH growth, and feedback. Using 
mass conservation, we can write 
\begin{equation} \label{eq:delta-m-g}
    \Delta M_{\rm g} = \Delta M_{\rm g, sf} - \Delta M_{\rm *} 
    - \Delta M_{\rm bh} + \Delta M_{\rm g, prev} \,.
\end{equation}

For a galaxy that has entered the satellite stage, we model its star formation 
rate with an exponential form,
\begin{equation} \label{eq:sfr-sat}
    \dot{M}(t) = \dot{M}(t_{\rm inf}) \exp\left(
        - \frac{t - t_{\rm inf}}{\tau_{\rm sat}}
    \right)\,,
\end{equation}
where $t_{\rm inf}$ is the infall time of the galaxy and 
$\tau_{\rm sat} = 4\Gyr/\ln 10$ is the decay timescale. 
This parameterization is motivated by observational results 
\citep[e.g.][]{pengStrangulationPrimaryMechanism2015}. 
The amount of gas is determined according to the `close-box' model with 
instantaneous recycling: gas inflow is cut off as a result of environmental 
effects; the feedback-affected gas is instantaneously recycled into the ISM.
Thus, the amount of remaining gas decreases only by the consumption 
of star formation. The growth of the SMBH is assumed negligible in this 
stage.

Galaxy mergers are modeled by adding a fraction, $f_{\rm merge}$, of the stellar
mass of the satellite galaxy into the central galaxy. The remaining stars 
are assumed to be tidally stripped or scattered into the halo to 
become intracluster light (ICL). The fiducial value, $f_{\rm merge} = 0.1$ 
is taken from Model-II of \citetalias{luEmpiricalModelStar2014}. However,
we note that this value is highly uncertain because of the complicated  
details of mergers and the degeneracy with other model parameters. 
This value also depends on the definition of the galaxy stellar mass in 
observations, i.e. the boundary used to separate galactic stars from 
intergalactic stars and the aperture used in deriving the stellar mass.
To quantify this uncertainty, we bracket $M_*$ by two extremes: one only 
includes stars that have formed in the main branch ($f_{\rm merge} = 0$), 
and the other includes stars that have formed in both the main branch and 
all the side branches ($f_{\rm merge} = 1$). 
The resulted $M_*$-$M_{\rm v}$ relations for these two extremes, shown 
by the green curves in Fig.~\ref{fig:m_gc_vs_mhalo} (ses \S\ref{ssec:gcs-mass}
for details), 
suggest a significant contribution 
by ex-situ stars at $M_{\rm v} \gtrsim 10^{13}\Msun$. Interpretations of our results 
involving $M_*$ should thus take into account this uncertainty.
When counting the total number and mass of GCs in a galaxy (see \S\ref{ssec:impl}), 
we instead include all GCs formed in satellites that have merged into the 
central, based on the fact that observations of GCs usually adopt 
a large aperture \citep[e.g.][]{dornanInvestigatingGCSMRelation2023}.
%For example, the best-fit $f_{\rm merge}$ in Model-I of 
%\citetalias{luEmpiricalModelStar2014} is $0.65$, and in the empirical model of 
%\citet{behrooziUniverseMachineCorrelationGalaxy2019} is $0.400$.

\subsection{Metal enrichment}
\label{ssec:metal}

\begin{figure} \centering
    \includegraphics[width=0.95\columnwidth]{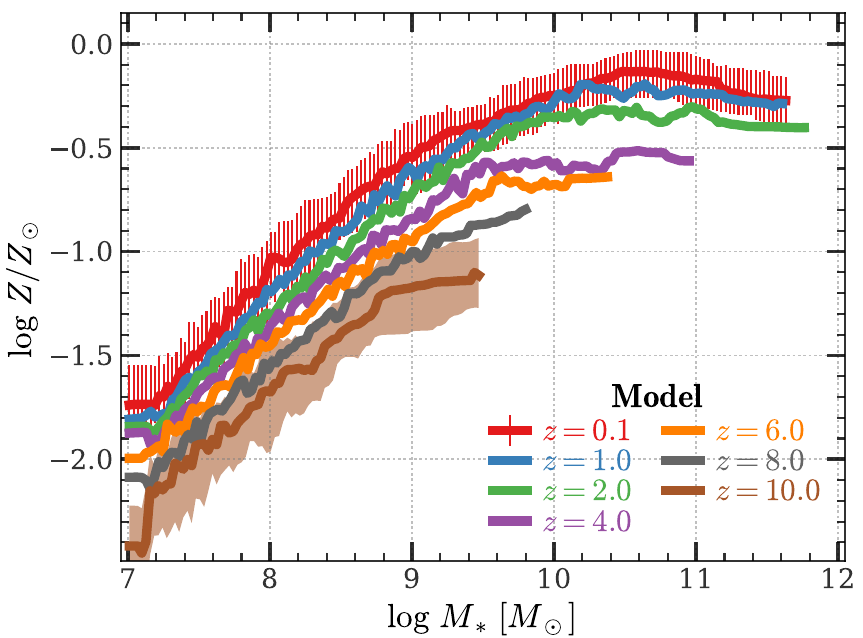}
    \caption{
        Stellar mass-metallicity relation for central galaxies at different 
        redshifts. Each {\bf solid} curve shows the running median. 
        Error bars and shaded area at two redshifts show the 
        1-$\sigma$ range. See \S\ref{ssec:metal} for details.
    }
    \label{fig:mzr}
\end{figure}

Gas-phase metallicity, $Z \equiv M_{\rm Z} / M_{\rm g}$, of galaxies is 
modeled according to the `gas-regulator' scenario 
\citep{lillyGasRegulationGalaxies2013,pengStrangulationPrimaryMechanism2015,
wangGasphaseMetallicityDiagnostic2021}, as our modeling of the SFR is based on 
the amount of available gas and fits in the framework of the gas regulator.
However, certain modifications are needed to account for uncertainties
in the metal enrichment related to the formation of GCs. 
One uncertainty arises from the possibility that high-$z$ galaxies 
may be fed by cold streams, as suggested by numerical simulations 
\citep[e.g.][]{keresHowGalaxiesGet2005,ceverinoHighredshiftClumpyDiscs2010,
danovichFourPhasesAngularmomentum2015,mandelkerColdFilamentaryAccretion2018,
lapinerWetCompactionBlue2023}. 
Cold streams with small cross section can transport pristine gas into the galaxy, 
leading to the formation of metal-poor stars, as suggested by observations
\citep{linMetalenrichedNeutralGas2023}, analytical  
estimations \citep{liFeedbackfreeStarburstsCosmic2024}
and hydrodynamical simulations \citep{mandelkerColdFilamentaryAccretion2018}.
Another uncertainty comes from the expectation that high-$z$ 
galaxies are much more clumpy and that their star formation is more bursty
than in the local Universe, as detailed in \citetalias{moTwophaseModelGalaxy2024}. 
Metal yields from stars at high $z$ may thus be coupled to the ISM and IGM 
in a way very different from that in a more steady and quieter environment
expected at low $z$. One way to deal with these uncertainties is to directly use 
the observed relations of the metallicity with other galaxy properties that can 
be modeled reliably. This approach, however, is limited by the size and quality of 
observational data, the selection bias, and uncertainties in, e.g.,
tracers and calibrations used to infer the metallicity. 
It is also not guaranteed that the observed relation can be applied 
to modeling the GCs, as the environment of GCs may differ 
significantly from the average over an entire galaxy. 
Indeed, the GC catalog produced by post-processing the {\sc Fire-2} zoom-in simulation, 
which is claimed to be able to reproduce the observed stellar mass-metallicity relation up to $z=12$ 
\citep{marszewskiHighRedshiftGasPhaseMass2024},
misses the old (born at $z > 3$), metal-poor population observed in the Milky 
Way \citep{grudicGreatBallsFIRE2023}. 

In this paper, we choose to modify some recipes in the gas-regulator model to account for the
aforementioned uncertainties, while keeping the overall framework unchanged. 
The formulation and calibration used here thus only apply to the modeling of GCs, and may 
not be suitable for other purposes. Following the formulation of \citet{luGalaxyEcosystemsGas2015}, we relate
the change of the oxygen mass, $\Delta M_{\rm Z}$, by the combination of 
inflow, stellar yield, star formation consumption, and feedback ejection.
Thus, the continuity equation for the oxygen mass is
\begin{align} \label{eq:metal-full-form}
    \Delta M_{\rm Z} = & Z_{\rm IGM} \Delta M_{\rm cool} 
        + y_{\rm eff} \frac{\Delta M_*}{1-R} - (\Delta M_* + \Delta M_{\rm bh}) Z \,,
        %- Z_{\rm ej} \Delta M_{\rm g,ej}\,,
\end{align}
where the three terms on the right-hand side are referred to as the inflow,
yield and locked metal masses, respectively.
Thus, $Z_{\rm IGM}$ is the metallicity of the inflow gas, 
$y_{\rm eff}$ is the effective oxygen yield that incorporates the  
intrinsic yield of stars, the escape of decoupled fraction, the 
ejection by stellar feedback, and the recycling due to cooling;
% and $R$ is the fraction of mass returned to the ISM by stellar evolution.
% and $Z_{\rm ej}$ is the metallicity of the gas ejected by feedback. 
Note that gas ejection and recycling 
has been absorbed into the effective yield, $y_{\rm eff}$, % of the ejected gas, 
so that this factor accounts for the net effects of ejection and recycling.
Each of the terms in the above equation has uncertainties.
To proceed, we make the following assumptions:
\begin{enumerate}[topsep=0pt,parsep=0pt,itemsep=0pt]
    \item 
    We set $Z_{\rm IGM}  = 0$ if the metallicity of the inflow gas is much 
    lower than that of the ISM.
    \item 
    A constant returning fraction $R$ (Eq.~\ref{eq:mass-return}) is adopted.
    Instantaneous recycling is also assumed here,
    because the timescale of stellar evolution is much shorter
    than the dynamical timescale of the halo -- roughly $80\%$ of the oxygen is 
    released during the first $10\Myr$ by the Type-II SNe 
    \citep{maiolinoReMetallicaCosmic2019}.
    \item 
    The effective yield is defined as
    \begin{equation}
        y_{\rm eff} = y (1-f_{\rm esc}) f_{\rm mix}\,.
    \end{equation}
    where we adopt the intrinsic oxygen yield $y = 0.0163$ \citep{portinariGalacticChemicalEnrichment1998}, 
    as it is consistent with a broad range of stellar evolution models in 
    the literature. Other models 
    \citep[e.g.][]{henryCosmicOriginsCarbon2000, kobayashiGalacticChemicalEvolution2006}
    give different values, but generally within $0.3\dex$.
    The escaped fraction, $f_{\rm esc}$, depends on the details of wind coupling,
    gas ejection and recycling, and is hard to model.
    Therefore, we choose to model it empirically as
    \begin{equation}
        f_{\rm esc} = \frac{1+\alpha_{\rm esc} x_{\rm v}}{1+ x_{\rm v}}\,,
    \end{equation}
    where $ x_{\rm v} = (V_{\rm max}/V_{\rm esc})^{\beta_{\rm esc}} $ and
    $\alpha_{\rm esc} = z / (z_{\rm esc} + z) $.
    The dependence on $V_{\rm max}$ is motivated by the fact that
    the coupling and ejection of stellar winds in general depend on the the depth of 
    the gravitational potential well of the host halo,
    and the dependence on redshift is motivated by the fact that the overall 
    gas density and clumpiness, and thus the local dynamical timescale and 
    cooling rate, depend on redshift.
    The mixing factor, $f_{\rm mix}$, which characterizes the efficiency of 
    metal mixing between the inflow gas and the existing ISM, is expected to be lower for 
    faster accretion and thus defined as
    \begin{equation}
        f_{\rm mix} = \frac{1}{1+ x_{\rm \gamma}}\,,
    \end{equation}
    where $x_{\rm \gamma} = \left|\gamma / \gamma_0\right|^{\beta_{\rm mix}}$.
    We find that $V_{\rm esc} = 75 \kms$, $\beta_{\rm esc}=2$, 
    $z_{\rm esc} = 5$, 
    $\gamma_0 = 1$, and $\beta_{\rm mix}=2$
    yield a good fit to the observations (see Appendix~\ref{sec:calibration}
    for the calibration).
\end{enumerate}

\noindent
With the above assumptions, Eq.~\eqref{eq:metal-full-form} becomes
\begin{equation} \label{eq:metal-simple-form}
    \Delta M_{\rm Z} = y_{\rm eff} \frac{\Delta M_*}{1-R} - 
    (\Delta M_* + \Delta M_{\rm bh}) Z \,.
\end{equation}
Taking the result of $\Delta M_*$, $\Delta M_{\rm bh}$ and $M_{\rm g}$ from 
the equations described in \S\ref{ssec:star-smbh-formation}, we can
integrate Eq.~\eqref{eq:metal-simple-form} over time to obtain $Z$ for each galaxy.

Fig.~\ref{fig:mzr} shows the stellar mass-metallicity
relation (MZR) at various redshifts from $z=0$ to $10$. 
Due to the redshift dependence of $f_{\rm esc}$ and $f_{\rm mix}$, 
the modeled MZR shows a steady evolution across the entire redshift range.
Low-mass galaxies ($M_* \lesssim 10^{10} \Msun$) become more enriched as their stellar
mass grows, because the star formation efficiency ($M_*/M_{\rm g}$) increases
monotonically with $M_*$, as seen from the blue and green curves in the 
bottom row of Fig.~\ref{fig:history_of_masses}. High-mass galaxies
($M_* \gtrsim 10^{10} \Msun$) shows saturated metallicity, because gas 
accumulation, and thus star formation, are stalled due to AGN feedback and 
inefficient cooling. The assembly of stellar mass for these galaxies
is mainly due to gas-poor (dry) mergers, without significant metal yield.
The slope in logarithmic scale at the low-mass end shows slow evolution,
from $0.6$ at high $z$, to $0.7$ at low $z$. This is a consequence of 
including $f_{\rm mix}$, which suppresses the mixing of low-metallicity 
inflow gas with existing enriched ISM, for halos at high accretion rates.

% The analytic fitting of MZR is now updated by our own results, instead of 
% the previous ones from Choksi, Yingtian Chen and Gnedin.
% Our MZR is oxygen-based, while theirs is iron-based. The index over
% M* is also different (their 0.35 follows a momentum confinement model,
% while our 0.6 is a result of potential confinement).
% The analytical estimation n_{shield, w}, and thus the formation criteria 
% of Pop-II GCs are moderately changed.
In the following, whenever we need an analytical approximation of the 
MZR to give a rough estimate of the results, we will use the following 
relation obtained by fitting our modeled galaxies:
\begin{equation}
     \log \frac{Z}{\Zsun}
     = 0.6 \log (\frac{M_*}{10^{9}\Msun}) - 0.55 \log (1+z) - 0.5 \,,
\end{equation}
For a halo with mass $M_{\rm v}$ and star formation efficiency 
$\epsilon_{\rm v} \equiv M_*/(M_{\rm v}f_{\rm B})$, 
the average metallicity is thus
\begin{align} \label{eq:metal-vs-mhalo-and-z}
    Z/Z_{\rm \odot} = \epsilon_{\rm v,0.1}^{0.6} \times \begin{cases}
        0.029\, M_{\rm v,10}^{0.6}  (1+z)_{10}^{-0.55}  \,,  \\
        0.45\, M_{\rm v,11.5}^{0.6} (1+z)_3^{-0.55}  \,.
    \end{cases}
\end{align}
% Similar relations were used by \citet{liModelingFormationGlobular2014},
% \citet{chenModelingKinematicsGlobular2022} 
% and \citet{chenFormationGlobularClusters2023} in their empirical
% models of globular clusters. 

\bigskip
The model of galaxy formation described in this section focuses on the global properties 
of galaxies. As described earlier, this provides environmental conditions for 
small-scale processes related to the formation of GCs. A consequence of 
this hierarchical strategy is that the spatial distribution of GCs resembles those of 
the halo and galaxies at large scales, with modifications due to refined processes operating 
on small scales. In the next section, we will introduce the model of GC formation
under the constraints provided by the model of galaxies described above. 
Note that this top-down approach does not include the details of 
matter and energy recycling from stars to the gas environment, and thus 
misses the diversity in the feedback produced by different stellar populations.
For example, the IMF in low-metallicity environments is inferred to be top-heavy
\citep{stacyBuildingPopulationIII2016,latifBirthMassFunction2022,klessenFirstStarsFormation2023},
which can alter the evolution paths of individual stars and thus lead to 
different metal yields \citep{portinariGalacticChemicalEnrichment1998}. 
More exotic feedback mechanisms can also be present, such as kilonovae 
\citep{rosswogMultimessengerPictureCompact2013,metzgerKilonovae2017}
and pair-instability supernovae 
\citep{woosleyEvolutionExplosionMassive2002,jankaExplosionMechanismsCoreCollapse2012}.
The inclusion of these populations and related processes may be important in the early 
Universe, and further exploration is needed in this area.

%%%%%%%%%%%%%%%%%%%%%%%%%%%%%%%%%%%%%%%%%%%%%%%%%%%%%%%%%%%%%%%%%%%%%%%%%%%%%%%%

\section{The formation of globular clusters} \label{sec:gc-model}

\begin{figure} \centering
    \includegraphics[width=0.95\columnwidth]{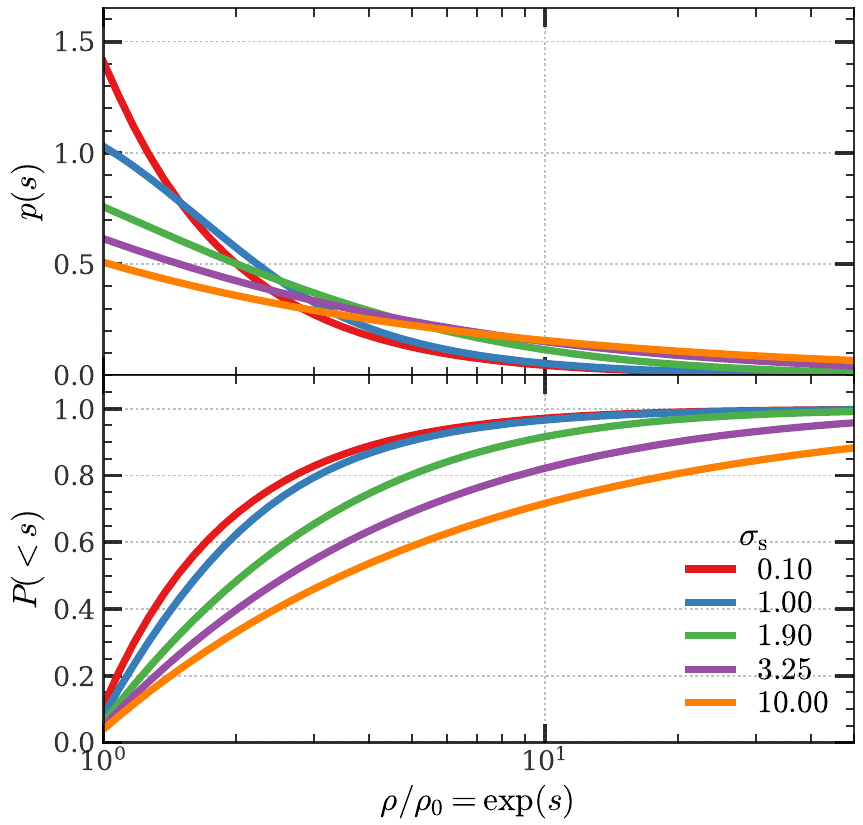}
    \caption{
        The probability distribution of the logarithmic density 
        contrast, $s \equiv \ln(\rho/\rho_0)$, of sub-clouds.
        {\bf Top panel} shows the probability density, while
        {\bf bottom panel} shows the cumulative distribution.
        Curves with different colors are obtained by 
        assuming different log-normal dispersion, $\sigma_{\rm s}$.
        A power-law tail with index $\alpha_{\rm s}$ = 1.5 is adopted 
        in all cases. See \S\ref{ssec:formation-of-sc} for details.
    }
    \label{fig:pdf_of_ln_density}
\end{figure}

A main conclusion of our two-phase model \citepalias{moTwophaseModelGalaxy2024} 
is the formation of sub-structures, referred to as sub-clouds, once the SGC 
becomes self-gravitating. In the fast phase, the SGC is expected to be gas-rich 
and turbulent due to the fast accretion of halo. High-speed winds from SN and AGN 
can also contribute to the turbulence \citep{maSelfconsistentProtoglobularCluster2020,
mercedes-felizDenseStellarClump2024}. Significant compression of the sub-clouds 
is expected if the turbulence is supersonic, which may produce dense sub-clouds 
for the formation of GCs we are interested in here. Despite the complexity of the 
details, it is generally believed that the properties of sub-clouds in a turbulent ISM 
follow `quasi-universal' laws and distributions, as suggested by both observations 
\citep{larsonTurbulenceStarFormation1981,ladaMassSizeRelationConstancy2020} 
and numerical simulations \citep{burkhartRazorsEdgeCollapse2017,
maSelfconsistentProtoglobularCluster2020,grudicModelFormationStellar2021,
appelEffectsMagneticFields2022,buckEscapingMazeStatistical2022,
kiihneFittingMethodsProbability2025}. 
In this section, we sample the sub-cloud population using these universal 
distributions under the constraints given by the global properties of SGCs. 
We then derive the condition for star formation in each sub-clouds, taking into 
account both external and internal feedback effects on the sub-clouds. 
We demonstrate that the condition for a sub-cloud
to form a GC leads to two distinct channels for GC formation.  
Finally, we model the disruption and survival of GCs, which is needed
to make predictions for comparisons with observations.

% Add some discussion about what is an SC, and what is its difference from
% GMC. Refer to \citet{fernandezSlowCoolingLowmetallicity2018} for the motivation
% of two-stage cooling.

\subsection{The formation of sub-clouds} \label{ssec:formation-of-sc}

At formation, an SGC is shock-heated by the release of gravitational 
potential energy to a temperature determined by the virial temperature
of the host halo. Gas cooling in the SGC is efficient, as expected from 
the high gas density and the high rate of atomic cooling expected at the 
relevant temperature. The rapid drop of the gas temperature reduces 
the Jeans mass, and can trigger fragmentation of the SGC into sub-structures. 
The decline of the temperature is expected to be stalled at 
$T_{\rm cool} \approx 10^{4}\Kelvin$ where the cooling rate 
starts to drop precipitously \citep[e.g.][]{smithMetalCoolingSimulations2008,
wiersmaEffectPhotoionizationCooling2009} and the cooling timescale 
becomes long. Thermal motion of gas particles in a cloud fragment 
can thus operate over a sufficient period of time to suppress  
density fluctuations in the fragment, stabilizing it and preventing it 
from further fragmentation. The Jeans mass at $T_{\rm cool}$ therefore 
sets a characteristic mass for the fragmentation of SGCs. In this paper, 
we refer to these self-bound fragments, whose initial temperature 
is $T_{\rm cool}$, as sub-clouds. Each of these sub-clouds is expected 
to feature a thermally-supported internal structure and a mass spectrum 
truncated by the Jeans mass in the high-mass end.
The characteristic temperature, $T_{\rm cool} \approx 10^4 {\rm K}$, 
corresponds to a sound speed of $c_{\rm s} \approx 10\,{\rm km/s}$, and the 
Jeans mass is
\begin{align} \label{eq:jeans-mass}
    M_{\rm J} = v_{\rm eff,10}^3 
     \times \begin{cases}
        \left[ 5.96 \times 10^7 \Msun \right]\, n_{\rm 0.0}^{-1/2} \,,  \\
        \left[ 1.06\times 10^6 \Msun \right]\, n_{\rm 3.5}^{-1/2} \,,
    \end{cases}
\end{align}
where $n_{\rm gas} = 10^{3.5} \perccm \,n_{\rm 3.5} = 1 \perccm\,n_{\rm 0.0}$,
$v_{\rm eff} = ( c_{\rm s}^2 + v_{\rm turb}^2 )^{1/2} = 10\kms\,v_{\rm eff,10}$,
and $v_{\rm turb}$ is the turbulence velocity.
By definition, $v_{\rm turb} < c_{\rm s}$, 
so that $v_{\rm eff} \sim c_{\rm s}$.
We note that the meaning of the sub-clouds so defined should be 
distinguished from that of the GMCs in the literature, with the latter 
usually referring to molecular gas clouds of diverse properties.

The precise mass distribution of sub-clouds depends on the details of how 
SGCs fragment, and therefore on the temperature, turbulence velocity 
and density of individual SGCs, as well as on the 
magnetic and tidal fields they reside in. 
We do not attempt to trace these details here, but adopt an
empirical approach instead.
We assume that the mass function of sub-clouds below the truncation mass 
follows a power-law distribution with an index $\beta_{\rm sc, m}$, 
as expected from scale-free fragmentation of a turbulence-dominated ISM 
\citep{larsonTurbulenceStarFormation1981}.
The value of $\beta_{\rm sc, m}$ is inferred to be $\approx -2$, 
as suggested by the index of the mass function of newly-born massive 
star clusters in the local Universe 
\citep{zhangMassFunctionYoung1999,fallSimilaritiesPopulationsStar2012,adamoStarClustersFar2020} 
and in the cosmic noon \citep{dessauges-zavadskyFirstConstraintsStellar2018}.
The star formation efficiency of massive and dense sub-clouds hosting massive 
and dense star clusters is expected to be close to unity (see \S\ref{ssec:sf-in-scs}), 
and thus the mass spectrum of sub-clouds is expected to be similar to the mass function of 
star clusters. This value of $\beta_{\rm sc, m}$ is also reproduced by high-resolution 
simulations of individual GMCs \citep{grudicModelFormationStellar2021}.
With the consideration above, we model the initial mass function for 
sub-clouds as
\begin{equation} \label{eq:m-sc-pdf}
    \dv{N_{\rm sc}}{M_{\rm sc}} \propto M_{\rm sc}^{\beta_{\rm sc, m}} \exp\left( -\frac{M_{\rm sc}}{M_{\rm sc,t}} \right)\,,
\end{equation}
with $\beta_{\rm sc,m} = -2$. The truncation mass 
$M_{\rm sc,t}=10^{6.5} \Msun$ is set to reflect the Jeans mass given above and 
the mass limit on the most massive star clusters found in observations
(e.g. \citealt{mokConstraintsUpperCutoffs2019}; \citealt{krumholzStarClustersCosmic2019};
see also Fig.~\ref{fig:mass_size}).
The lower and upper bounds of $M_{\rm sc}$ are set to be $M_{\rm sc, min}=10^{4}\Msun$
and $M_{\rm sc, max}=10^{8}\Msun$, respectively. Convergence tests against these choices  
are made in \S\ref{ssec:mass-function}. 

The presence of supersonic shocks (with Mach number $\mathcal{M}$ > 1) can raise the density 
of sub-clouds and significantly change the density distribution of sub-clouds.
As explained in \S3.3 of \citetalias{moTwophaseModelGalaxy2024}, the compressive
nature of supersonic turbulence is critical to make the sub-clouds dense enough
to survive the drag force by ram pressure and collisions with other sub-clouds.
A lack of supersonic turbulence, such as the case where a dynamically-cold 
disk forms and steady star formation ensues, is expected to terminate
the generation of sub-clouds at $T_{\rm cool}$ before gas cooling at lower temperature 
sets a new mass scale, $M_{\rm sc, t}$, that is lower than the original truncation mass.   
Sub-cloud formation in this regime of weak-turbulence may be related to local instabilities, 
such as those induced by spiral arms and SN feedback, and/or to external perturbations, 
such as those from interactions with nearby galaxies.
On the other hand, sub-clouds in strong turbulence are clustered
\citep{liEffectsSubgridModels2020}. They may thus coalesce to form more
massive clouds, raising the value of $M_{\rm sc, t}$, or form massive 
star clusters if rapid star formation can consume the gas before the coalescence  
\citep[see e.g. Fig.~9 of][]{maSelfconsistentProtoglobularCluster2020}. 
Such a variation of $M_{\rm sc, t}$ is evident from observations 
\citep[see \S2.2 of][for a review]{krumholzStarClustersCosmic2019},
and may be modeled by following the evolution of the turbulence in the SGC.
Alternatively, one may start from a parametric model that relates $M_{\rm sc, t}$ 
to the galactic environment, such as the gas density, composition and star 
formation rate density, with necessary calibration from hydrodynamic simulations
and observations. Lacking such a model, we adopt a constant value of $M_{\rm sc, t}$ 
for simplicity.

The density distribution of sub-clouds in supersonic turbulence, coupled with 
non-linear star formation and feedback processes, can modify 
the mass spectrum of sub-clouds and thus that of star clusters formed within them.
Major sources of the turbulence may be summarized as follows:
\begin{enumerate}[topsep=0pt,parsep=0pt,itemsep=0pt]
    \item Global shocks. 
    These arise during the formation of an SGC when 
    the inflow gas reaches the core of the host halo, interacts with itself and collides
    with the existing gas. This is especially important for halos with fast accretion
    and effective cooling, where cold streams penetrate the halo \citepalias[corresponding to 
    the Q1 phase described in][]{moTwophaseModelGalaxy2024}, as suggested by 
    numerical simulations \citep[e.g.][]{ceverinoHighredshiftClumpyDiscs2010,
    danovichFourPhasesAngularmomentum2015,mandelkerColdFilamentaryAccretion2018}.
    A global shock can also arise when angular momentum, turbulence and thermal motion of
    the SGC can no longer support its gravity and it collapses
    \citep[e.g.][]{dekelFormationMassiveGalaxies2009,dekelWetDiscContraction2014,
    zolotovCompactionQuenchingHighz2015,
    tacchellaConfinementStarformingGalaxies2016,
    latifTurbulentColdFlows2022,
    jiReconstructingAssemblyMassive2023,
    dekelEfficientFormationMassive2023}.
    This is expected to be an important source of turbulence for a gas-rich, slow-rotating SGC
    (corresponding to the Q2 phase).
    Gas-rich mergers also produce global shocks and generate gravitational
    instability due to the pile-up of gas and loss of angular momentum.
    Such a global shock can propagate outward across the entire SGC, compress 
    the gas, and produce high-density sub-structures. 
    
    \item Local interactions.
    These occur when a sub-cloud grows via absorbing surrounding gas, or mergers 
    with nearby sub-clouds. The release of gravitational energy can be effectively
    dissipated by the effective cooling above $10^4\Kelvin$, so that this process 
    proceeds rapidly and is roughly isothermal. 
    The process continues until sub-clouds become too 
    small and dense enough for them to collide frequently, or until the 
    driving force of the global turbulence disappears.
     Numerical simulations have shown that such local interactions 
    are critical for the formation of massive and dense sub-clouds
    \citep[e.g.][]{latifTurbulentColdFlows2022,maSelfconsistentProtoglobularCluster2020}.
    
    \item
    Positive feedback. High-speed winds driven by feedback processes 
    can also drive turbulence and compress sub-clouds to high density. 
    Numerical simulations show that quasar winds \citep{mercedes-felizDenseStellarClump2024}
    and SN winds \citep{maSelfconsistentProtoglobularCluster2020} can both  
    lead to the formation of dense stellar clumps reminiscent of     
    YMSCs and GCs. 
    The recent observations by JWST
    \citep{pereira-santaellaExtendedHighionizationMg2024} and ALMA 
    \citep{roman-oliveiraDynamicalModellingOrigin2024}
    also support the presence of stellar feedback as the driver of turbulence.
    This channel of turbulence formation exists as long as star formation and/or AGN 
    is active, and thus may cover the entire history of a galaxy.
\end{enumerate} 
\breakpara
Turbulence driven by the combination of the above sources creates complex density 
structure in the SGC. MHD simulations suggest that the volume density of sub-clouds, 
$\rho_{\rm sc}$, follows a piece-wise distribution combining  
a log-normal piece with a power-law one: 
\begin{equation} \label{eq:ln-density-pdf}
    p(s) = \begin{dcases}
        \frac{1}{\sqrt{2\pi}\sigma_{\rm s}} e^{-\frac{(s-s_0)^2}{2\sigma_{\rm s}} } 
            \,,\ \ \ \ \ \ \ \ \ \ 
        &\text{if}\ s \leqslant s_{\rm t} \,, \\
        A_{\rm s} e^{-\alpha_{\rm s} (s-s_{\rm t}) } 
            \,, 
        &\text{if}\ s > s_{\rm t}  \,,
    \end{dcases}
\end{equation}
where $s = \ln(\rho_{\rm sc}/\rho_0)$ is the logarithmic density contrast, 
$\rho_0$ is the mean density of the SGC, $\sigma_{\rm s}$ 
is a free parameter defining the width of the log-normal distribution,
$s_0 = (-1/2)\sigma_{\rm s}^2$ is the median, and $\alpha_{\rm s}$ is the 
power-law index of the high-density tail. 
The log-normal piece is found to well represent the isothermal
density distribution of a low-density, turbulent ISM 
\citep[e.g.][]{burkhartDensityStudiesMHD2009,myersCharacteristicStructureStarforming2015}.
The power-law tail at the high density, on the other hand, is expected to originate
from compression by self-gravitating collapses and feedback
\citep[e.g.][]{slyzSimulatingStarFormation2005,collinsMassMagneticDistributions2011,
burkhartRazorsEdgeCollapse2017,khullarDensityStructureSupersonic2021,
appelEffectsMagneticFields2022,buckEscapingMazeStatistical2022,appelEffectsMagneticFields2022,
kiihneFittingMethodsProbability2025}. 
The continuity of $p(s)$ and its first derivative at $s=s_{\rm t}$ gives
\begin{align}
    A_{\rm s} &= \frac{1}{\sqrt{2 \pi} \sigma_{\rm s}} e^{-\frac{1}{2} \alpha_{\rm s} \sigma_{\rm s}^2} \,, \\
    s_{\rm t} &= (\alpha_{\rm s}-1/2) \sigma_{\rm s}^2  \,.
\end{align}
So defined, the distribution function has three free parameters, $\rho_0$, $\sigma_{\rm s}$ and $\alpha_{\rm s}$,
which we obtain as follows:
\begin{enumerate}[topsep=0pt,parsep=0pt,itemsep=0pt]
    \item 
    $\rho_0$, the mean density of the SGC, is predicted by the galaxy model 
    described in \S\ref{sec:galaxy-model}: 
    % The normalization, f_r, is dropped, as it is set to 1 everywhere 
    % in this paper.
    \begin{equation}
        % \rho_0 = \frac{M_{\rm g}}{ \frac{4\pi}{3} ( f_{\rm r} f_{\rm gas} R_{\rm v} )^3 } \,.
        \rho_0 = \frac{M_{\rm g}}{ (4\pi/3) ( f_{\rm gas} R_{\rm v} )^3 } \,.
    \end{equation}
    Here, $f_{\rm gas} = M_{\rm g}/M_{\rm v}$ is the gas fraction 
    of the halo, which also determines the size of the SGC by the condition of 
    self-gravitating. 
    % $f_{\rm r}$ is a constant normalization factor, adopted
    % to make adjustments needed to account for uncertainties in the definition of 
    %$R_{\rm v}$ and the boundary of the SGC. We set $f_{\rm r} = 1$ as the fiducial value.    
    \item 
    $\sigma_{\rm s}$, the width of the density distribution, is determined
    by the cascading of turbulence. Numerical simulations 
    \citep[e.g.][]{appelEffectsMagneticFields2022} suggested that it can be 
    parameterized as 
    \begin{equation}
        \sigma_{\rm s}^2 = \ln\left[ 1 + f_{\rm s} \mathcal{M}^2 \right]\,.
    \end{equation}
    Here $f_{\rm s} = b^2 \beta_{\rm m} / (1 + \beta_{\rm m})$, with $b$ accounting for 
    different turbulence modes ($1/3$ for pure solenoidal mode and $1$ 
    for pure compressive mode), and 
    $\beta_{\rm m}$ denotes the ratio between thermal and
    magnetic pressure. The Mach number of the turbulence, $\mathcal{M}$, is 
    estimated as 
    \begin{equation} \label{eq:mach-from-vvir}
        \mathcal{M} = v_{\rm rms,3D} / c_{\rm s}
        = f_{\rm turb} \frac{\sqrt{V_{\rm v}^2 + V_{\rm w}^2}}{c_{\rm s}} \,,
    \end{equation} 
    where $c_{\rm s} \approx 10\kms$ is the sound speed, $f_{\rm turb}$ 
    characterizes the fraction of the kinetic energy initially injected to the turbulent
    ISM, and $V_{\rm w}$ is the speed of the SN wind when it is coupled to the ISM.
    %which should depend on halo accretion and feedbacks (1 if turbulent driving 
    %force is strong, and 0 if the system is steady; may be $>1$ if 
    %the system is not in virial equilibrium).
    %HJ: Since the relative velocity may depend on scale, we may want to treat $f_{\rm burb}$ as a free parameter. If needed 
    %we may also want to a model assuming 
    %\begin{equation}
    %    \mathcal{M} = v_{\rm rms,3D} / c_{\rm s}
    %    = f_{\rm turb} (V_{\rm v} / c_{\rm s})^\tau \,.
    %\end{equation} 
    %This may be checked with hydro simulations like FIRE.   
    Since $f_{\rm s}$ and $f_{\rm turb}$ are degenerate, we absorb $f_{\rm turb}$
    into $f_{\rm s}$, and take $f_{\rm s} = 1$ as the fiducial value,
    % In this paper, the fiducial values are $f_{\rm s} = 1$ and $f_{\rm turb} = 1$,
    based on calibrations using the observed relations of GC systems 
    (see Appendix~\ref{sec:calibration}).
    The fiducial value for the wind speed is set to be $V_{\rm w} = 250\kms$, consistent with 
    the value adopted in \S\ref{ssec:star-smbh-formation} for the 
    SN feedback.
    \item 
    $\alpha_{\rm s}$, the high-density slope, is found to be $1 $--$ 2$ 
    by numerical simulations \citep[e.g.][]{appelEffectsMagneticFields2022, kiihneFittingMethodsProbability2025}. 
    We take $\alpha_{\rm s} = 1.5$, which is consistent with the analytical expectation
    for the pressure-free (free-fall) collapse \citep{girichidisEvolutionDensityProbability2014},
    and for the weak magnetic field and high compression we assumed in setting $f_{\rm s}$ 
    \citep{collinsMassMagneticDistributions2011}.
\end{enumerate}

With these choices, the initial (post-shock) mass and density of a sub-cloud can be obtained 
by sampling the distribution function given by Eqs.~\eqref{eq:m-sc-pdf} and
\eqref{eq:ln-density-pdf}, respectively, as described in \S\ref{ssec:impl}. As an analytical estimation,
the typical density contrast of the post-shock sub-clouds in the high-density 
tail of the supersonic turbulence is
\begin{equation}
    \rho_{\rm sc} / \rho_0 
    \sim e^{s_{\rm t}} 
    \sim e^{ \sigma_{\rm s}^2 }
    \sim 1 + f_{\rm s} \mathcal{M}^2 
    \sim \mathcal{M}^2
    \,.
\end{equation}
Using Eqs.~\eqref{eq:mach-from-vvir} and \eqref{eq:v-vir} for $\mathcal{M}$ and $V_{\rm v}$, respectively,
the Mach number for a halo with $M_{\rm v}$, at redshift $z$, and without 
the presence of SN-induced turbulence ($V_{\rm w} = 0$) is
\begin{align} \label{eq:mach-number}
    \mathcal{M}_1 =
        \begin{cases}
            8.00\,  M_{\rm v, 10}^{1/3} (1+z)_{10}^{1/2} \,,    \\ 
            13.86\,  M_{\rm v, 11.5}^{1/3} (1+z)_{3}^{1/2}  \,,
        \end{cases}
\end{align}
where $c_{\rm s} = 10\kms$ is used. At $z = 9$, $\mathcal{M}$ can be 
high for relatively low-mass halos due to their high densities,  
while at $z=2$ a high value can be achieved in massive halos. 
Thus, at any redshift, the post-shock density of SCs, $\rho_0 \mathcal{M}^2$, can be 
orders of magnitude higher than the mean density of the SGC, 
as long as the driving force of the turbulence is sufficiently strong, 
Meanwhile, the SN-induced turbulence has a Mach number
\begin{equation}\label{eq:mach-number-sn}
    \mathcal{M}_2 \equiv V_{\rm w}/c_{\rm s} = 25\,,
\end{equation} 
which is higher than $\mathcal{M}_1$ over the entire redshift range for all halos 
below the Milky Way (MW) mass. This suggests that supernovae are always important 
sources to create dense sub-clouds, except for cluster-size halos 
($M_{\rm v} \gtrsim 10^{13}\Msun$) expected at low redshift, where $\mathcal{M}_1$ 
becomes very large. The situation is similar to the growth of SMBH described in 
\S\ref{ssec:star-smbh-formation}, where supernovae provide small-scale `positive feedback' 
to enhance the growth of SMBHs, although their large-scale effect 
is to eject gas and thus represents a source of negative feedback for star formation. 

Using Eq.~\eqref{eq:n-sgc} for $n_{\rm sgc}$, the post-shock density without 
SN is thus
\begin{align} \label{eq:n-sc-post-shock}
    n_{\rm sc} 
    & \sim n_{\rm sgc} \mathcal{M}_1^2            \nonumber \\ 
    & = f_{\rm str, 4} f_{\rm gas, 0.04}^{-2} \times
    \begin{cases}
        \left[ 4.24 \times 10^4 \perccm \right] M_{\rm v,10}^{2/3} (1+z)_{10}^4    \,,\\
        \left[ 3.20 \times 10^3 \perccm \right] M_{\rm v,11.5}^{2/3} (1+z)_3^4     \,,
    \end{cases}
\end{align}
and the post-shock density induced solely by SN is
\begin{align} \label{eq:n-sc-post-shock-sn}
    n_{\rm sc} 
    & \sim n_{\rm sgc} \mathcal{M}_2^2            \nonumber \\ 
    & = f_{\rm str, 4} f_{\rm gas, 0.04}^{-2} \times
    \begin{cases}
        \left[ 1.11 \times 10^5 \perccm \right] (1+z)_{10}^3    \,,\\
        \left[ 4.13 \times 10^4 \perccm \right] (1+z)_3^3     \,.
    \end{cases}
\end{align}
As we will show in \S\ref{ssec:gc-channels}, a sub-cloud with $n_{\rm sc} \gtrsim 10^{3.5} \perccm$ 
can remain supernova-free over a free-fall timescale, and may thus be able 
to form a globular cluster. The estimation in Eqs.~\eqref{eq:n-sc-post-shock} 
and \eqref{eq:n-sc-post-shock-sn}
suggests that a gas-rich, dense and turbulent SGC provides the condition    
to form globular clusters, consistent with the results of {\sc Fire-2} zoom-in simulations 
\citep[e.g.][]{maSelfconsistentProtoglobularCluster2020}.

Fig.~\ref{fig:pdf_of_ln_density} shows $p(s)$, the distribution function of 
the logarithmic density contrast and its cumulative function for a number 
of $\sigma_{\rm s}$. In all cases, the low-density, log-normal part ($s < 2$) 
takes the majority ($> 50\%$) of the probability mass, indicating
that stars can form over the entire SGC, albeit with a lower efficiency 
in a more diffuse medium. This is in contrast to the pure feedback-free model of 
\citet{dekelEfficientFormationMassive2023},
which predicts that star formation in massive halos ($M_{\rm v} \approx 10^{10.8}\Msun$) 
at high-z Universe ($z \approx 10$) is concentrated on dense, supernova-free 
sub-clouds. 
%However, the lack of large sample of recent JWST surveys prevent us from distinguishing these two scenarios.
The high-density tail of $p(s)$, on the other hand, is found to 
account for about $20\%$ of the probability mass, assuming the typical Mach 
number at $z \geqslant 2$ in Eqs.~\eqref{eq:mach-number} and \eqref{eq:mach-number-sn} 
(corresponding to $\sigma \approx 3.5$--$7$). As we will show below, sub-clouds
in this regime can reach a star formation efficiency of $100\%$, and
thus, contribute a significant fraction of star formation rate and 
total stellar mass. Recent JWST observations in lensed fields 
\citep[e.g.][]{
vanzellaEarlyResultsGLASSJWST2022,
vanzellaJWSTNIRCamProbes2023,
welchRELICSSmallscaleStar2023,
linMetalenrichedNeutralGas2023,
claeyssensStarFormationSmallest2023,
adamoBoundStarClusters2024,
messaPropertiesBrightestYoung2024,
fujimotoPrimordialRotatingDisk2024,
mowlaFireflySparkleEarliest2024} 
appear to support this prediction.

\subsection{The cooling and fragmentation of sub-clouds} 
\label{ssec:cooling}

\begin{figure} \centering
    \includegraphics[width=0.99\columnwidth]{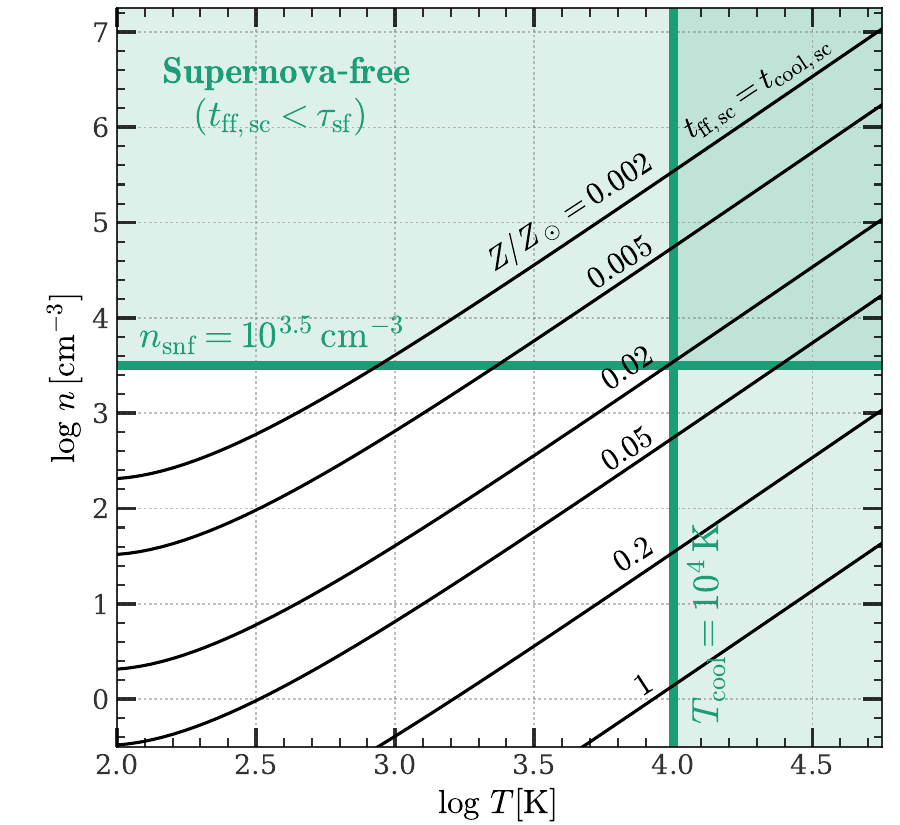}
    \caption{
        The cooling diagram of sub-clouds. {\bf Black} contours show the loci of
        equaling cooling and free-fall timescales, $t_{\rm cool, sc} 
        = t_{\rm ff, sc}$, at different metallicity 
        (see \S\ref{ssec:cooling} for details).
        The {\bf vertical shaded region} indicates
        the regime above the critical temperature
        $T_{\rm cool} = 10^4 \Kelvin$, where cooling
        is effective and fragmentation of SGC is expected to occur to form 
        sub-clouds.
        The {\bf horizontal shaded region} above the `feedback-free' threshold, 
        $n_{\rm snf}=10^{3.5}\perccm$,
        indicates the regime where star formation in sub-clouds is free of 
        (internal) supernova feedback.
        A sub-cloud reaching a density $n_{\rm sc} \geqslant n_{\rm snf}$ thus hosts 
        a GC (see \S\ref{ssec:gc-channels} for details).
        % A complete version 
        %shall be obtained by, e.g. \citet{maioMetalMoleculeCooling2007} or 
        %from the Grackle library (Enzo's cooling implementation).
    }
    \label{fig:cooling_diagram_low_temp}
\end{figure}

\begin{figure*} \centering
    \includegraphics[width=0.99\textwidth]{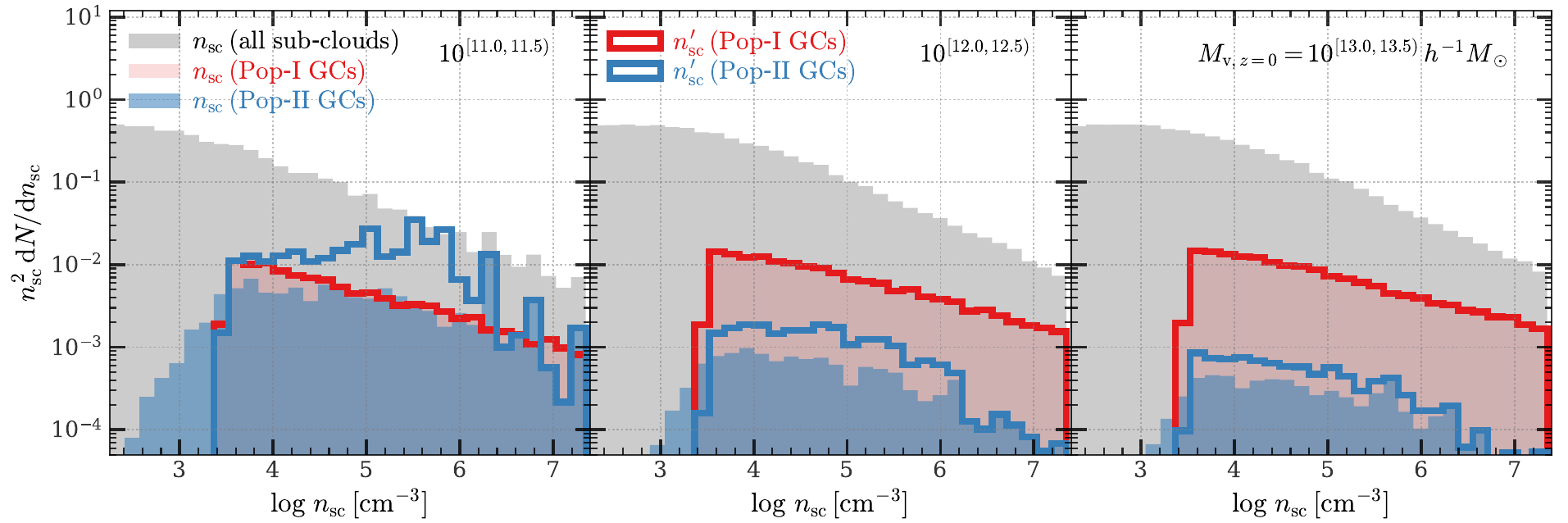}
    \caption{
        Distributions of sub-cloud density. Histograms shown by 
        {\bf shaded areas} and {\bf solid lines} are obtained from $n_{\rm sc}$, 
        the density at formation (post-shock), and $n_{\rm sc}'$, the density 
        at fragmentation (post-cooling), respectively.
        In each panel,
        {\bf grey} histogram shows the result for all sub-clouds,
        while {\bf red} and {\bf blue} histograms show the results for
        Pop-I (metal-rich) and Pop-II (metal-poor) GCs, respectively, 
        selected at the birth time 
        (see \S\ref{ssec:impl} for the selection and classification criteria).
        Different {\bf panels} show the stacked results for 
        central subhalos with different halo mass at $z=0$.
        Sub-clouds formed in the main branches of these subhalos are 
        included in the analysis. See \S\ref{ssec:cooling} for a 
        detailed discussion.
    }
    \label{fig:histogram_n_sc}
\end{figure*}

The initial temperature, $T_{\rm cool} = 10^4 \Kelvin$, set by rapid cooling at the 
formation of sub-clouds, is much higher than that required by star formation.
Thus, other cooling channels, such as those given by metal, dust and 
molecules, determine the subsequent evolution of sub-clouds until the trigger 
of star formation. 

The behavior of how a sub-cloud cools and evolves can be understood by comparing 
its cooling timescale, $t_{\rm cool, sc}$, and its free-fall timescale, $t_{\rm ff}$
\citep[e.g.][]{fernandezSlowCoolingLowmetallicity2018,
mandelkerColdFilamentaryAccretion2018,dekelEfficientFormationMassive2023}.
When cooling is effective, i.e. when $t_{\rm cool, sc} \lesssim t_{\rm ff, sc}$, 
the heating by gravitational collapse is not able to compensate for the cooling loss,
and the growth of small-scale perturbations is no longer 
limited by gas pressure. Sub-clouds with such initial conditions are thus expected to 
fragment and form stars. On the other hand, if $t_{\rm cool, sc} \gtrsim t_{\rm ff, sc}$, 
a global contraction of the sub-cloud can continue for a longer period without being impeded 
by fragmentation and star formation. At given $T$ and $Z$, the cooling timescale $t_{\rm cool, sc} 
\propto n_{\rm sc}^{-1} $ while the free-fall timescale 
$t_{\rm ff, sc} \propto n_{\rm sc}^{-1/2}$. Thus, a contracting sub-cloud 
will eventually enter the regime of effective cooling, and begin to 
fragment once $t_{\rm cool, sc} \sim t_{\rm ff, sc}$.

Following \citet{dekelEfficientFormationMassive2023}, we use the dominant
channel of $[\textsc{Cii}]\,158 \,\mu {\rm m}$ line cooling to estimate the cooling time 
scale for gas at $Z/\Zsun>10^{-3}$. Using results given by 
\citet{krumholzStarFormationAtomic2012}, we have
\begin{equation}\label{eq:t-cool-sc-cii} 
    t_{\rm cool, sc} \approx 0.87\Myr\ n_{\rm sc,3.5}^{-1} Z_{0.02}^{-1} T_4 \exp(\frac{0.009}{T_4}) C^{-1}\,,
\end{equation}
where $Z = 0.02\, \Zsun\, Z_{0.02}$; 
$n_{\rm sc} = 10^{3.5} \perccm n_{\rm sc,3.5}$;
$T = 10^4 \Kelvin\, T_4$; 
a molecular weight $\mu = 1.2$ is used;
$C = \langle n^2 \rangle / \langle n \rangle^2 $ is
the clumpiness factor within the sub-cloud.
The free-fall timescale depends only on the matter density: 
\begin{equation} \label{eq:t-ff}
    t_{\rm ff, sc} 
    = \left( \frac{3\pi}{32 G\rho} \right)^{1/2} 
    = 0.84\Myr\ n_{\rm sc, 3.5}^{-1/2} \,.
\end{equation}
The threshold density for triggering star formation, $n_{\rm sf}$, 
can be estimated using $t_{\rm ff,sc} \approx t_{\rm cool, sc}$, which gives
\begin{equation} \label{eq:n-sf-threshold}
    n_{\rm sf} \approx 3.39 \times 10^3 
    \perccm\ Z_{0.02}^{-2} T_4^2 C^{-2} = n_{\rm sf,1} C^{-2}\,.
\end{equation}

Based on the above arguments, we adopt the following criteria 
for the fragmentation of sub-clouds. If a sub-cloud has an initial density 
$n_{\rm sc} \geqslant n_{\rm sf,1}$, it fragments immediately, followed by star formation 
in individual fragments. On the other hand, if $n_{\rm sc} < n_{\rm sf,1}$, 
the sub-cloud first contracts globally to a density $n_{\rm sc}'$, and then fragments 
to form stars.
Due to the uncertainty in the growth of the clumpiness factor $C$ during the 
contraction of sub-cloud, we randomly sample $n_{\rm sc}'$ in the range  
$[ n_{\rm sc}, n_{\rm sf,1} ]$ from a power-law distribution given by
\begin{equation} \label{eq:sampling-n-sc}
\dv{N_{\rm sc}}{n_{\rm sc}'} \propto n_{\rm sc}^{\prime\beta_{\rm sc, n}}\,,
\end{equation}
where the power-law index is calibrated to be $\beta_{\rm sc, n}=-2.5$ 
(see Appendix~\ref{sec:calibration}).

Fig.~\ref{fig:cooling_diagram_low_temp} shows contours of 
$t_{\rm ff,sc} = t_{\rm cool,sc}$, evaluated using Eqs.~\eqref{eq:t-cool-sc-cii}
and \eqref{eq:t-ff} with $C=1$, at different metallicity. 
The intersection of a contour with the vertical line, $T = 10^4 \Kelvin$,
gives the star formation threshold density, $n_{\rm sf,1}$, as derived
in Eq.~\eqref{eq:n-sf-threshold}. 
The strong metallicity dependence of $n_{\rm sf,1}$ has important 
implications for the formation of star clusters. 
At $z \lesssim 0.02$, $n_{\rm sf, 1}$ is above the 
supernova-free density threshold, $n_{\rm snf}$, indicating that a metal-poor 
sub-cloud, regardless of its initial density, have the possibility to contract to a 
supernova-free cloud before fragmentation, if it is not destroyed by other 
sources of feedback (see \S\ref{ssec:external-feedback} below). 
Thus, a metal-poor galactic environment provides an additional channel for the 
formation of GCs, complementary to that via shock compression. 

The global collapse of a sub-cloud in the metal-poor condition has been 
demonstrated by the simulations of \citet{fernandezSlowCoolingLowmetallicity2018} 
for individual sub-clouds. They found that, with a moderate UV heating 
background, a sub-cloud with $M_{\rm sc} = 10^{6}\Msun$ and 
$Z = 0.01\Zsun$ can collapse globally to form a dense, fragmented core with 
a size of $\approx 5$--$10$ pc, while the low-density, smooth envelop does not 
fragment (see their figure 6). A detailed analysis based on the cooling diagram 
shows that this delayed fragmentation is due to inefficient cooling of metal-poor gas 
(see their figures 2 and 7). A similar conclusion can be found in the simulations of 
\citet{grudicModelFormationStellar2021} for individual sub-clouds, who 
found that the star formation efficiency on sub-cloud scales increases with 
increasing sub-cloud surface density, and that the fraction of stars formed 
in bound clusters is higher in their low-metallicity runs.

Grey shades in Fig.~\ref{fig:histogram_n_sc} show the distributions of 
sub-cloud density at formation ($n_{\rm sc}$), for all sub-clouds formed 
in the main branches of central subhalos at $z=0$ with different halo masses.
Depending on internal properties and environments, sub-clouds can have 
different fates: some will be destroyed by external feedback, while others 
will continue to contract, fragment and form stars at $n_{\rm sc}'$. 
Only the most massive and/or densest sub-clouds can shield 
themselves from feedback and form sufficient amounts of stars to qualify as
(the progenitors of) globular clusters. These will be discussed thoroughly 
in \S\S\ref{ssec:external-feedback}, \ref{ssec:sf-in-scs} and 
\ref{ssec:gc-channels}, and be synthesized into a coherent model 
in \S\ref{ssec:impl}. Briefly, the aforementioned conditions, 
together with supersonic turbulence and metal-poor environment, give rise to two 
distinct formation channels that lead to two populations of GCs: 
Pop-I (the metal-rich population) and Pop-II (the metal-poor population).

The red and blue shades in Fig.~\ref{fig:histogram_n_sc} show the 
distributions of gas density at formation for sub-clouds that will eventually
evolve into Pop-I and Pop-II GCs, respectively, while the solid lines show the
distributions of the gas density at fragmentation for the two populations.
At $10^{3.5} \perccm$, the lower limit for a sub-cloud to qualify  
as (the progenitor of) a globular cluster, only
$< 10\%$ of sub-clouds are massive enough to form globular clusters,
mainly due to the bottom-heavy nature of the sub-cloud mass 
function (Eq.~\ref{eq:m-sc-pdf}).
A significant fraction of metal-poor sub-clouds with low initial density 
can contract to higher density to reach the condition 
for globular cluster formation, as clearly suggested
by comparing the blue histograms for $n_{\rm sc}$ and $n_{\rm sc}'$.
The effects of sub-cloud contraction appear to be more significant for
low-mass halos, as their metallicity is lower.
In contrast, metal-rich sub-clouds hosting globular clusters can fragment immediately 
after formation. The difference in cooling and fragmentation between the two populations 
of sub-clouds turns out to be the key to understanding the two channels of GC 
formation and the observed bimodality in the GC population, as to be
discussed in \S\ref{ssec:gc-channels}.

\subsection{Destruction of sub-clouds by external feedback}
\label{ssec:external-feedback}

Once formed, sub-clouds are subject to various external feedback effects produced 
by other stars in the galaxy. One of the main feedback sources is the shock wave 
associated with the energy injection from stellar evolution and SNe,
which we collectively refer to as `wind feedback'.
Depending on the state of the gaseous structure that is affected 
by the wind feedback, the effect of such feedback can be either negative or positive.
For dense sub-clouds where cooling is effective and an isothermal state can be maintained, 
shocks are to compress the gas. 
On the other hand, for the diffuse component or low-density sub-clouds 
where gas cannot cool effectively, shocks may disperse gas and disrupt sub-clouds
\citep[see e.g. \S8.3 of][]{moGalaxyFormationEvolution2010,hopkinsStellarFeedbackGalaxies2012,rosenMassiveStarBorn2022}.
The effect of the SN feedback is also scale-dependent. At the galactic scale, 
its effect is to reduce the amount of star-forming gas and prevent 
star formation, as modeled by the $F_{\rm sn}$ factor in 
\S\ref{ssec:star-smbh-formation}. Consequently, the total number of sub-clouds 
survived is also reduced. In terms of individual sub-clouds,
the high-density ones are compressed to higher density by shocks,
and such positive feedback is already described in \S\ref{ssec:formation-of-sc}.
Here we focus on the negative side of the feedback.  

As suggested by \citet{dekelEfficientFormationMassive2023}, wind feedback
can ablate or destroy sub-clouds. Following their arguments, we use $t_{\rm crush,sc}$, the timescale for the 
shock to crush a sub-cloud, to estimate the time needed for the shock 
to destroy the sub-cloud. This timescale is given by \citet{kleinHydrodynamicInteractionShock1994}: 
\begin{equation} \label{eq:t-crush-sc}
    t_{\rm crush,sc} = 2 \frac{R_{\rm sc} \rho_{\rm sc}^{1/2}}{V_{\rm w} \rho_{\rm w}^{1/2}}\,.
\end{equation}
Here, $V_{\rm w}$ and $\rho_{\rm w}$ are, respectively, the wind velocity and density when it 
hits the sub-cloud. These two quantities are related by 
\begin{equation} \label{eq:rho-wind}
    \rho_{\rm w} = \frac{{\rm \Phi}\, \eta_{\rm w}}{4 \pi R_{\rm sgc}^2 V_{\rm w}}\,,
\end{equation}
where $\Phi$ is the star formation rate of the galaxy, and $\eta_{\rm w}$ is the mass loss 
fraction of stars. Using the Starburst99 library \citep{leithererStarburst99SynthesisModels1999}, 
and assuming a wind-lasting time of $\approx 10\Myr$ and a specific loss rate of 
$10^{-8}{\rm yr}^{-1}$, \citet{dekelEfficientFormationMassive2023} estimated that  
$\eta_{\rm w} \approx 0.1$. The wind velocity is estimated to be $10^{3.5} \kms$, 
which is the typical value of an SN-driven wind before it is coupled with the ISM. 
The radius of the SGC, as discussed in \S\ref{ssec:collapse-gas}, is 
$R_{\rm sgc} = f_{\rm gas} R_{\rm v}$. 
For the sub-cloud to survive the wind disruption and proceed to the 
regime of collapse and fragmentation,
the cooling timescale in the turbulence mixing layer around the sub-cloud 
has to be shorter than the crushing timescale: 
$t_{\rm cool,sc} < t_{\rm crush,sc}$. Substituting Eq.~\eqref{eq:t-cool-sc-cii}
and \eqref{eq:t-crush-sc} into this condition, we obtain 
the density threshold for a sub-cloud to survive the wind feedback:
\begin{align}\label{eq:n-shield-w}
    n_{\rm shield,w}
        & = 94.71 \perccm (\eta_{\rm w, 0.1} \Phi_1)^{3/7} \times  \nonumber \\
        & \ \ \ \ \ \ \ \ \  (Z_{\rm 0.02}R_{\rm v,25} f_{\rm gas,0.04})^{-6/7} 
        M_{\rm sc,6}^{-2/7}\,,
\end{align}
where $\eta_{\rm w} = 0.1 \eta_{\rm w,0.1}$,
$R_{\rm v} = 25\Kpc\, R_{\rm v,25}$
and $\Phi = 1\Msun/{\rm yr}\, \Phi_1$.
Using the equation of $\dot{M}_{\rm v}$ in Eq.~\eqref{eq:m-vir-dot}, 
assuming a star formation efficiency $\epsilon_{\rm v} = M_*/(f_{\rm B}M_{\rm v}) = 0.1$, 
a mass returning fraction given by Eq.~\eqref{eq:mass-return}, 
the mean $Z(M_{\rm v},z)$ relation in Eq.~\eqref{eq:metal-vs-mhalo-and-z}, and $R_{\rm v}$ in Eq.~\eqref{eq:r-vir},
the shielding density can be approximated as
\begin{align}\label{eq:n-shield-w-estimate}
    n_{\rm shield,w} = 
    & \eta_{\rm w, 0.1}^{3/7} 
    (f_{\rm gas,0.04})^{-6/7} 
    M_{\rm sc,6}^{-2/7} 
    \epsilon_{\rm v,0.1}^{-0.086}
    \times \nonumber\\ 
    & \ \ \ \ \ 
    \begin{cases}
        \left[ 233.96 \perccm \right] M_{\rm v,10}^{-0.31} (1+z)_{10}^{2.4} \,,\\
        \left[ 4.49 \perccm \right] M_{\rm v,11.5}^{-0.31} (1+z)_{3}^{2.4} \,.
    \end{cases}
\end{align}
This depends weakly on halo mass, but as fast as $n_{\rm sgc}$ on redshift 
(see Eq.~\ref{eq:n-sgc}). At $z=2$ and $z=9$, it is much smaller than $n_{\rm sgc}$ and 
thus than the post-shock tail of $n_{\rm sc}$ (see Eqs.~\ref{eq:n-sc-post-shock} 
and \ref{eq:n-sc-post-shock-sn}), 
indicating that only low-density, 
low-mass sub-clouds are destroyed. 

Another source of feedback is the UV radiation of massive stars, which can heat and unbound sub-clouds.
The shielding length of an SC is obtained by equaling the \textsc{Hii} recombination rate 
and the photon flux \citep{dekelEfficientFormationMassive2023}, as 
\begin{equation}
    \Delta R = \frac{f_{\rm OB}\nu_{\rm ion} \tau_{\rm OB} \Phi}{4 \pi R_{\rm sgc}^2 n_{\rm sc}^2 \alpha_{\rm rec}}\,, 
\end{equation}
where $f_{\rm OB} = 0.01 \Msun^{-1}$ is the number of massive (O/B) stars per 
solar mass of formed stars, $\tau_{\rm OB} = 10\Myr$ is their lifetime, 
$\nu_{\rm ion} = 10^{49}{\rm s}^{-1}$ is the UV photon rate per massive star,
and $\alpha_{\rm rec}=4\times 10^{-13}{\rm cm}^3{\rm s}^{-1}$ is the recombination rate 
per unit density. For a sub-cloud to survive the UV radiation, its size, 
$R_{\rm sc}$, has to be larger than the shielding length, $\Delta R$, namely
$R_{\rm sc} > \Delta R$. This gives the density threshold for a sub-cloud to survive 
the UV radiation:
\begin{align} \label{eq:n-shield-r}
    n_{\rm shield,r} = 
        & 8.27 \perccm\, (\tau_{\rm 10} \Phi_1)^{3/5} \times \nonumber \\
        & \ \ \ \ \ \ (R_{\rm v,25} f_{\rm gas,0.04})^{-6/5} M_{\rm sc,6}^{-1/5} \,,
\end{align}
where $\tau_{\rm OB} = 10\Myr\,\tau_{10} $.
This value of $n_{\rm shield,r}$ is usually smaller than $n_{\rm shield, w}$, meaning that UV radiation
is not a main source to disperse sub-clouds.

For both the wind feedback and radiation feedback, the $\Phi/R_{\rm sgc}^2$-dependence in the 
shielding density suggests that the survival of sub-clouds is preferred in low-mass, gas-rich halos, 
because
\begin{equation}
    \frac{\Phi}{R_{\rm sgc}^2} 
    \sim \Sigma_{\rm SFR} 
    \sim \Sigma_{\rm gas}^{\gamma} \sim \left(\frac{M_{\rm sgc}}{R_{\rm sgc}^2}\right)^\gamma
    \sim f_{\rm gas}^{-\gamma} M_{\rm v}^{\gamma/3}
    \,,
\end{equation}
where $\gamma \approx 1.5$ is the Kennicutt-Schmidt index. 
The negative dependence on $M_{\rm sc}$ and $Z$ indicates that  
sub-clouds with larger mass and higher metallicity have a better 
chance of surviving against the feedback effects. The effects of the $Z$-dependence 
and the $\Sigma_{\rm SFR}$-dependence cancel each other, weakening the dependence  
on $M_{\rm v}$, as shown by Eq.~\eqref{eq:n-shield-w-estimate}.

The effect of stellar feedback on star formation has been confirmed by 
\citet{appelEffectsMagneticFields2022} using MHD simulations, where the 
inclusion of stellar feedback (outflow and heating) significantly increases 
the amount of diffuse gas below the mean density, reduces the rate of mass 
transfer from the log-normal peak to the power-law tail, and thus reduces SFR.

Other sources of feedback external to sub-clouds include the cosmic UV background, 
which is anticipated to suppress star formation in halos with mass below 
$\approx 10^9 \Msun$ after the epoch of reionization, and the feedback from accreting SMBHs.
As their effects have already been included by the two-phase model in the star formation law 
of the entire galaxy (\S\ref{ssec:star-smbh-formation}) to set the normalization of the 
total amount of star formation, we do not include them in the sub-cloud model. 

For SNe inside a sub-cloud, the timescale for them to disperse the sub-cloud   
is too short for cooling to take place. This sets a time window for star formation
and a key criterion for the formation of GCs, as we will describe in \S\ref{ssec:sf-in-scs} 
and \S\ref{ssec:gc-channels}, respectively. 
The combination of these positive and negative effects will become 
clearer when we join the pieces to construct a complete model in \S\ref{ssec:impl}. 
The operation of these processes is also demonstrated by high-resolution zoom-in simulations. 
For example, as shown in figure 3 of \citet{maSelfconsistentProtoglobularCluster2020} using the 
{\sc Fire-2} zoom-in simulations, as soon as SNe are triggered by the formation of a massive 
star cluster, most of the surrounding gas is dispersed, leaving a number of high-density clumps 
produced by the compression of wind feedback.

\subsection{Star formation in sub-clouds} 
\label{ssec:sf-in-scs}

\begin{figure} \centering
    \includegraphics[width=0.99\columnwidth]{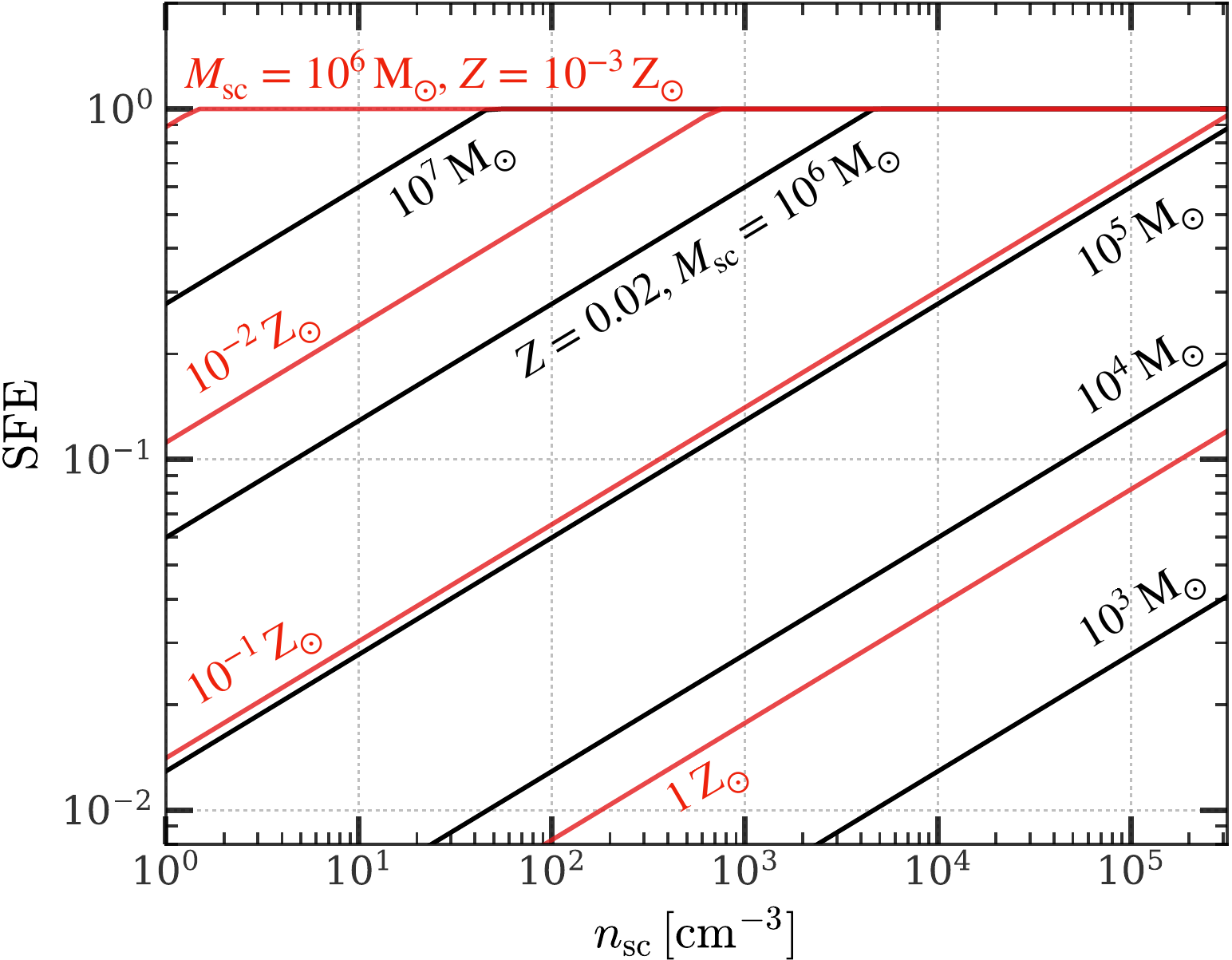}
    \caption{Star formation efficiency (SFE) of sub-clouds as a function of their density
    ($n_{\rm sc}$), adopting wind feedback and $\epsilon_{\rm w, max}=1$. 
    {\bf Black} lines show the cases with different sub-cloud mass ($M_{\rm sc}$)
    and a fixed metallicity, $Z = 0.02\Zsun$.
    {\bf Red} lines show the cases with different metallicity and a fixed
    sub-cloud mass, $M_{\rm sc} = 10^6\Msun$.
    See \S\ref{ssec:sf-in-scs} for details.
    }
    \label{fig:SFE}
\end{figure}

Once a sub-cloud survives the external feedback and contracts to a density 
so that $t_{\rm cool, sc} \sim t_{\rm ff, sc}$, it is expected to fragment and 
form stars. The newly formed stars then produce internal feedback, 
which suppresses star formation within the sub-cloud.
As estimated by \citetalias{moTwophaseModelGalaxy2024}, supernova feedback 
can affect a self-gravitating, gaseous system with an escaping velocity 
$V_{\rm esc} \lesssim 200 \kms$ \citep[see also e.g.][]{hopkinsWhatCausesFormation2023}. 
This escaping velocity is much higher than that of a sub-cloud, implying that 
a sub-cloud will be dispersed as soon as the formation of its first generation of stars.
Thus, the delay time of supernova feedback, namely the lifetime of massive stars, sets 
the time window for star formation in sub-clouds.
Based on the Starburst99 library \citep{leithererStarburst99SynthesisModels1999}
for individual starbursts,
\citet[see their figure 1]{dekelEfficientFormationMassive2023} found 
that this time window is about $\tau_{\rm sf} \approx 1\Myr$, quite independent of 
the metallicity. Within this time window, star formation in sub-clouds is regulated by 
other channels of stellar feedback, as we discuss below.

Stellar winds generated by massive stars can affect the surrounding gas in the sub-cloud
and reduce the subsequent star formation. Here, we adopt a metallicity-dependent wind 
energy deposition, 
\begin{equation}
    E_{\rm w} = e_{\rm w} \eta_{\rm w} \tau_{\rm sf}  \left(\frac{Z}{\Zsun}\right)^{0.9} M_{\rm cls}   \,.
\end{equation}
Here, $M_{\rm cls}$ is the stellar mass of the star cluster formed within the 
sub-cloud;
the normalization $e_{\rm w} \approx 10^{40}\, {\rm erg\,s^{-1}\,(10^6\Msun)^{-1}}$ 
is computed by \citet{dekelEfficientFormationMassive2023} based on the same
library in the estimation of $\tau_{\rm sf}$;
$\eta_{\rm w} \approx 0.1$ is the wind coupling factor \citep{guptaHowRadiationAffects2016}. 
The metallicity dependence comes mainly from the metallicity-dependent 
threshold for Wolf-Rayet stars that dominate the wind power, while the power-law index 
0.9 was suggested by \citet{hirschiVeryLowmetallicityMassive2007}. 
For the wind to be effective in destroying the sub-cloud, 
$E_{\rm w}$ must be comparable to the binding energy of the sub-cloud, 
$E_{\rm b} \approx (1/2) G M_{\rm sc}^2 / R_{\rm sc}$.  
Equaling these two energies leads to a star formation efficiency, 
$\epsilon_{\rm w} \equiv M_{\rm cls}/M_{\rm sc}$, that determines the fraction of 
the gas in the sub-cloud to be converted into stars under the effect of stellar winds:
\begin{align}\label{eq:eps-star-by-wind}
    \epsilon_{\rm w} 
    &= \frac{G M_{\rm sc}}{2 R_{\rm sc} (Z/\Zsun)^{0.9} 
        \tau_{\rm sf} \eta_{\rm w} e_{\rm w}}   \nonumber\\
    &= 3.81 \frac{M_{\rm sc, 6}}{R_{\rm sc, 12} 
        Z_{0.02}^{0.9} \tau_{\rm sf, 1} \eta_{\rm w, 0.1} }  \nonumber\\
    &= \epsilon_{\rm w,max} M_{\rm sc, 6} R_{\rm sc, 12}^{-1}
        Z_{0.02}^{-0.9}
    \,,
\end{align}
where $M_{\rm sc} = 10^6 \Msun\,M_{\rm sc,6}$, 
$R_{\rm sc} = 12\pc\,R_{\rm sc,12}$,
$\tau_{\rm sf} = 1\Myr\, \tau_{\rm sf,1}$, 
$\eta_{\rm w} = 0.1\, \eta_{\rm w, 0.1}$,
and $\epsilon_{\rm w,max} \equiv 3.81/( \tau_{\rm sf,1}\eta_{\rm w, 0.1} ) \sim 1$ 
is the maximum star formation efficiency.
Fig.~\ref{fig:SFE} shows the star formation efficiency of sub-clouds as a 
function of their density, regulated by the wind feedback.
At $Z/\Zsun \approx 0.02$, the star formation efficiency reaches 
unity for a sub-cloud with mass of $10^6 \Msun$ and a size of $12\pc$,
indicating that a low-metallicity sub-cloud free of internal supernova feedback 
is also free of internal wind feedback. A reduction of density or mass, or 
an increase of metallicity, decreases the resistance of sub-cloud to wind feedback
and thus lowers the star formation efficiency.
Assuming that $\tau_{\rm sf}$ and $\eta_{\rm w}$ are independent of 
$M_{\rm sc}$, $R_{\rm sc}$ and $Z$, the star formation efficiency scales as
\begin{equation}\label{eq:eps-star-by-wind-scaling}
    \epsilon_{\rm w} \propto M_{\rm sc} R_{\rm sc}^{-1} Z^{-0.9} 
    \propto M_{\rm sc}^{2/3} Z^{-0.9}\,.
\end{equation}

Photo-ionization from stellar radiation also imparts momentum/pressure on the gas 
within a sub-cloud, which can also disperse the gas and reduce the star formation efficiency
\citep[e.g.][]{liEffectsSubgridModels2020}. Following the argument of 
e.g. \citet{fallStellarFeedbackMolecular2010}, the specific momentum injection rate is 
\begin{equation}
    \dot{p}_{\rm r} = \frac{1}{c}  \left< \frac{L_*}{M_*} \right> \beta_{\rm boost} \,,
\end{equation}
where $\left< L_*/M_* \right> \approx 10^3 L_\odot/M_\odot$, corresponding
to the peak emission of a blackbody radiation with $T \approx 40,000\,{\rm K}$
for OB stars.
The boosting factor $\beta_{\rm boost}$ takes into account the effect of 
multiple scattering. Following \citet{marinacciSimulatingInterstellarMedium2019},
we set $\beta_{\rm boost} = 1 + \kappa_{\rm IR} \Sigma_{\rm sc}$, 
where $\Sigma_{\rm sc} = M_{\rm sc}/( \pi R_{\rm sc}^2 )$ 
is the surface density of the sub-cloud and 
$\kappa_{\rm IR} = 10 (Z/\Zsun)\,{\rm cm^2\,g^{-1}}$ is the infrared opacity. 
Equaling the total momentum injection rate, $\dot{p}_{\rm r} M_{\rm cls}$, to the gravity, 
$G M_{\rm sc}^2 / R_{\rm sc}^2 $, we obtain the star formation efficiency
$\epsilon_{\rm r}$ for a radiation-regulated sub-cloud, as
\begin{equation} \label{eq:eps-star-by-radiation}
    \epsilon_{\rm r} 
    = \frac{\pi G}{\frac{1}{c}  \left< \frac{L_*}{M_*} \right> (\Sigma_{\rm sc}^{-1} + \kappa_{\rm IR})}
    = 2.16 \frac{\Sigma_{\rm sc, 3.5}}{1 + 0.13 \Sigma_{\rm sc, 3.5} Z_{0.02}}
    \,.
\end{equation} 
At $Z/\Zsun \approx 0.02$, $\epsilon_{\rm r}$ is 
proportional to $\Sigma_{\rm sc,3.5}$ and thus we have  
\begin{equation}
    \epsilon_{\rm r} \propto \Sigma_{\rm sc} \propto M_{\rm sc}^{1/3}\,.
\end{equation}
This efficiency reaches unity at $\Sigma_{\rm sc} \approx 10^{3.2} \Msunperpcsq$, 
comparable to  the condition for sub-clouds with $\epsilon_{\rm w} \approx 1$.
At high metallicity, the momentum injection from multiple scattering can effectively 
reduce the star formation. 
A more general form of Eq.~\eqref{eq:eps-star-by-wind-scaling} was originally 
proposed by \citet{fallStellarFeedbackMolecular2010}, and subsequently confirmed in
hydrodynamical simulations and observations. 
In particular, \citet{grudicWhenFeedbackFails2018} 
showed that the formula is accurate for a mixture of feedback processes, 
both momentum-driven and energy-driven, including radiation pressure, 
photo-ionization, stellar winds, etc. 
The formula is also consistent with various observations of 
proto-clusters/sub-clouds in the MW and the Large Magellanic Cloud
\citep[LMC;][]{mokFeedbackFormingStar2021}.

Based on the above discussion, we can parameterize the sub-cloud-scale star formation 
efficiency as 
\begin{equation}
    \epsilon_{\rm sf} \equiv M_{\rm cls} / M_{\rm sc} = \epsilon_{\rm sf, max} M_{\rm sc,6}^{\beta_{\rm sf}} 
    Z_{0.02}^{\gamma_{\rm sf}} \,,
\end{equation}
where $\epsilon_{\rm sf, max} \approx 1$ is the maximum star formation efficiency,
which is reached at $M_{\rm sc} \approx 10^{6} \Msun$ and $Z \approx 0.02 \Zsun$.
The values of $\gamma_{\rm sf}$ and $\beta_{\rm sf}$ depend on the dominant channel of stellar 
feedback. For example, $(\beta_{\rm sf},\gamma_{\rm sf})=(2/3,-0.9)$ in the wind-regulated regime,
while $(\beta_{\rm sf},\gamma_{\rm sf})=(1/3,0)$ in the radiation-regulated, metal-poor regime. 
If these two channels operate in multiplication, $(\beta_{\rm sf},\gamma_{\rm sf})=(2/3+1/3,-0.9)$.

By definition, the size of a sub-cloud with density $n_{\rm sc}$ is
\begin{equation} \label{eq:sc-size-mass}
    R_{\rm sc} 
      = \left[ \frac{M_{\rm sc}}{(4\pi / 3) \mu m_{\rm p} n_{\rm sc}} \right]^{1/3}
      =  13.65\,{\rm pc}\ M_{\rm sc, 6}^{1/3} n_{\rm sc, 3.5}^{-1/3} \,,
\end{equation}
where the default values of $M_{\rm sc}$ and $n_{\rm sc}$ are set by the 
Jeans mass (Eq.~\ref{eq:jeans-mass}) and supernova-free density, respectively. 
Thus, the typical spatial extent of GC-forming sub-clouds is $10 \pc$.

The half-stellar-mass radius of a star cluster, $R_{\rm cls}$,
is determined by $R_{\rm sc}$, the size of its parent sub-cloud and we model it as 
\begin{equation}\label{eq:gc-size}
    R_{\rm cls} = f_{\rm r,cls} R_{\rm sc}\,,
\end{equation}
where $f_{\rm r,cls} \lesssim 1$ depends on the density profile of the cluster.
For an isothermal sub-cloud, $f_{\rm r,cls} = 1/2$, which we find to be able to 
reproduce observed sizes of young massive star clusters and globular clusters
(see \S\ref{ssec:size-mass} below).
Such a similarity between cluster size and sub-cloud size is expected 
if star formation in sub-cloud follows a scale-free fragmentation process 
in which the free-fall timescale is the same throughout the sub-cloud.

Combining the above equations for sub-cloud size-mass relation, 
star formation efficiency, and the definition of star-cluster size, 
we obtain the size-mass relation of star clusters at birth:
\begin{align} \label{eq:gc-size-mass-vs-feedbacks}
    R_{\rm cls} & = 6.83\, {\rm pc}\,n_{\rm sc,3.5}^{-1/3} \left( \frac{M_{\rm cls,6}}{\epsilon_{\rm sf,max} Z_{0.02}^{\gamma_{\rm sf}}} \right)^{\frac{1}{3(\beta_{\rm sf}+1)}}  
    \nonumber \\
    &= 6.83\, {\rm pc}\, n_{\rm sc,3.5}^{-1/3} \begin{cases}
        M_{\rm cls,6}^{1/5} Z_{0.02}^{0.18} \ \ \ \ \ \ \ \ \ & \text{(wind)} \,, 
        \\
        M_{\rm cls,6}^{1/4} & \text{(radiative)}  \,,
        \\
        M_{\rm cls,6}^{1/6} Z_{0.02}^{0.18} & \text{(wind + radiative)} 
        \,,
    \end{cases}
\end{align}
where we have defined $M_{\rm cls} = 10^6 \Msun\,M_{\rm cls,6}$.
In both cases, star clusters have roughly constant size, as the dependence on stellar mass and metallicity is weak. 
This is indeed seen in observations, where star clusters spanning orders of 
magnitude in masses have a narrow size range of $1$--$10\pc$
\citep[e.g.][]{krumholzStarClustersCosmic2019,brownRadiiYoungStar2021}.

%%%%%%%%%%%%%%%%%%%%%%%%%%%%%%%%%%%%%%%%%%%%%%%%%%%%%%%%%%%%%%%%%%%%%%%%%%%%%%%% 

\subsection{The fate of sub-clouds and the two channels of 
globular cluster formation} 
\label{ssec:gc-channels}

\begin{figure} \centering
    \includegraphics[width=0.99\columnwidth]{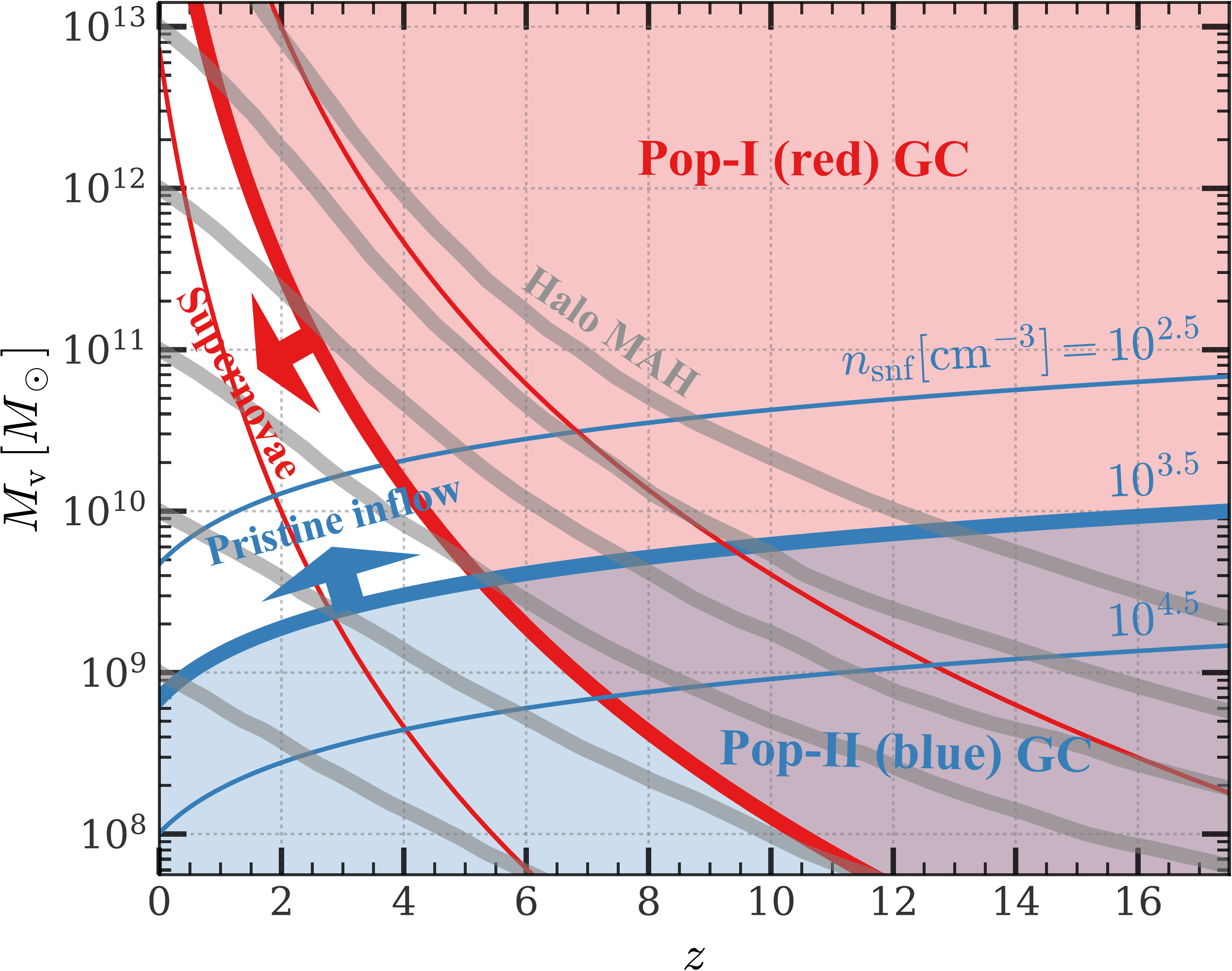}
    \caption{
        The criteria for the formation of Pop-I and Pop-II GCs. 
        {\bf Red} and {\bf blue shaded areas} indicate the halo mass and redshift 
        where Pop-I and Pop-II GCs can actively form, according to 
        Eqs.~\eqref{eq:criteron-pop-i} and \eqref{eq:criteron-pop-ii}, respectively,
        assuming a supernova-free density threshold 
        $n_{\rm snf} = 10^{3.5} \perccm$.        
        Other red and blue curves are the criteria boundaries with alternative 
        choices of $n_{\rm snf}$.
        In the presence of supernovae and/or metal-poor inflow, the criteria 
        for GC formation can be further relaxed, as indicated by the red and blue
        arrows.
        {\bf Gray curves} are median mass assembly histories 
        of halos with different masses at $z=0$, 
        generated using {\sc Diffmah} \citep{hearinDifferentiableModelAssembly2021}.
        Note that this figure only shows the environment expected for active 
        GC formation. The random sampling of halo assembly history, 
        sub-cloud density and metallicity can lead to the formation of GCs in 
        unusual environments. See \S\ref{ssec:gc-channels} for the details
        of this figure.
    }
    \label{fig:gc-criteria}
\end{figure}

For a sub-cloud ready to fragment, a combination of the free-fall time 
scale, $t_{\rm ff,sc}$, and the star formation time window set by supernova 
feedback, $\tau_{\rm sf}$, leads to an interesting consequence.
If the density of the subsub-cloud is $n_{\rm sc} \gtrsim 10^{3.5} \perccm$, 
the free-fall timescale (Eq.~\ref{eq:t-ff}) will be so short that 
$t_{\rm ff, sc} \lesssim \tau_{\rm sf}$.
Because $t_{\rm ff,sc}$ sets the timescale for a sub-cloud to fragment, 
the sub-cloud with such high density is expected to be free of supernova feedback. 
Meanwhile, as suggested by Eqs.~\eqref{eq:eps-star-by-wind} and 
\eqref{eq:eps-star-by-radiation}, feedback from stellar wind and radiation within 
such a sub-cloud is ineffective, leading to a star formation efficiency of about unity.
Thus, according to the size-mass relation of sub-clouds (Eq.~\ref{eq:sc-size-mass}), 
a sub-cloud with gas mass $\sim 10^6 \Msun$ and size $\sim 10 \pc$
converts nearly all its gas into stars, producing a massive star cluster
with stellar mass $\sim 10^6 \Msun$ and size $\sim 10 \pc$. 
Such a star cluster falls into the mass and size range of globular clusters,
and can thus be viewed as the progenitor of a globular cluster.
In the remainder of this paper, we will refer to the density, 
$n_{\rm snf} = 10^{3.5} \perccm$ as the `supernova-free' density threshold, and use it 
as a condition for the formation of globular clusters.
%For a detailed analytical description, refer to \citet{dekelEfficientFormationMassive2023}.

The star formation threshold density $n_{\rm sf,1}$ obtained 
in Eq.~\eqref{eq:n-sf-threshold} is sensitive to metallicity. In the high-redshift 
Universe where pristine inflow dominates the gas accretion, the metallicity of sub-clouds can be
as low as $Z \approx 0.02 \Zsun$ so that the density at the triggering of 
fragmentation can reach $n_{\rm snf}$. 
Thus, the low-metallicity environment naturally gives rise to the formation
of globular clusters, regardless of the initial density produced by shocks. 
On the other hand, for galactic environments where gas is already enriched,
sub-clouds with low initial densities can fragment before they reach $n_{\rm snf}$.
This will lead to the formation of star clusters under the influence of supernova 
feedback, making them less dense, less massive, and more fragile to the tidal 
field of the host galaxy. Such a star cluster is expected to become an open cluster or 
to disperse and join the diffuse stellar component. In metal-enriched galaxies,
therefore, a globular cluster can form only if the initial density of the parent 
sub-cloud is $\gtrsim n_{\rm snf}$.

The above discussion leads to two channels of globular cluster formation, 
depending on the order of three densities: the star formation threshold
($n_{\rm sf,1}$); the supernova-free density ($n_{\rm snf}$); and the initial
density of the sub-cloud ($n_{\rm sc}$). We thus define two types of globular clusters as follows:
\begin{itemize}[topsep=0pt,parsep=0pt,itemsep=0pt]
    \item {\bf Pop-I GC} which forms in a sub-cloud with $n_{\rm sc} \geqslant n_{\rm snf} > n_{\rm sf,1}$. 
    A globular cluster of this type relies on compression by supersonic shocks for 
    its progenitor sub-cloud to reach the supernova-free threshold and to fragment. 
    \item {\bf Pop-II GC} which forms in a sub-cloud with $n_{\rm sf,1} \geqslant n_{\rm snf}$. 
    A globular cluster of this type forms by contracting towards the star formation
    threshold that is above the feedback-free threshold. Shock compression can help 
    to set a high initial density, but is not necessary. Due to the uncertainty 
    in the growth of the clumpiness factor as described around Eq.~\eqref{eq:sampling-n-sc},
    a sub-cloud with $n_{\rm sf,1} \geqslant n_{\rm snf}$ may fragment at a 
    density $n_{\rm sc}'$ below $n_{\rm snf}$, and thus fail to form a Pop-II GC.
\end{itemize}

Using Eq.~\eqref{eq:n-sf-threshold} for $n_{\rm sf, 1}$, the critical density
$n_{\rm sf, 1} = n_{\rm snf}$ that separates the two types of GCs 
immediately yields a critical gas-phase metallicty, 
$Z_{\rm crit} = 10^{-1.68}\Zsun$.
Our definition of GCs resembles the definition of 
Pop-I and Pop-II stars, in the sense that Pop-I GCs are metal-rich and Pop-II 
GCs are metal-poor. Observationally, Pop-I GCs appear red in color, 
while Pop-II GCs appear blue. Other phenomenological classifications of GCs, 
such as those based on in- and ex-situ origins, can be derived by integrating 
the star formation history and merger history of a given system (galaxy/halo).
Our classification of GCs is based on the properties of sub-clouds prior 
to the star formation. This should not be confused with the Type-I/II
GCs classified by the observed stellar populations 
\citep[e.g.][]{miloneHubbleSpaceTelescope2017,simioniStatisticalAnalysisGalactic2020}
that reflect the star formation and enrichment history of a GC 
\citep[see \S4 of][for a review]{miloneMultiplePopulationsStar2022}. 

To gain some intuition on the condition of globular cluster formation, here we 
provide a simple analytical estimation. In the absence of `positive feedback'
from supernovae, the post-shock density of sub-clouds at the high-density tail
is given by Eq.~\eqref{eq:n-sc-post-shock}, and the Pop-I 
GC formation criterion, $n_{\rm sc} \geqslant n_{\rm snf}$, becomes
\begin{equation} \label{eq:criteron-pop-i}
    M_{\rm v,11.5}^{2/3} (1+z)_3^4 f_{\rm gas, 0.04}^{-2} f_{\rm str, 4} 
    \geqslant 0.99\, n_{\rm snf,3.5} \,,
\end{equation}
where $n_{\rm snf} = 10^{3.5}\perccm n_{\rm snf, 3.5}$.
If supernovae dominate the generation of turbulence, the condition can be 
obtained by using Eq.~\eqref{eq:n-sc-post-shock-sn} as
\begin{equation}
    (1+z)^3 f_{\rm str, 4} f_{\rm gas,0.04}^{-2} \geqslant 2.07\, n_{\rm snf,3.5} \,.
\end{equation}
Thus, supernovae can effectively enhance the formation of Pop-I GCs 
in halos of any mass over a wide range of redshift. At $z \lesssim 1$, 
the cold stream disappears ($f_{\rm str} \approx 1$) in most halos and supernova-driven globular
cluster formation ceases. On the other hand, using the mean $Z(M_{\rm v}, z)$ 
relation in Eq.~\eqref{eq:metal-vs-mhalo-and-z} and the metallicity-dependent
star formation threshold in Eq.~\eqref{eq:n-sf-threshold},
the Pop-II GC formation criterion $n_{\rm sf,1} \geqslant n_{\rm snf}$ becomes 
\begin{equation} \label{eq:criteron-pop-ii}
    M_{\rm v,10}^{-1.2} (1+z)_{10}^{1.1} \epsilon_{\rm v,0.1}^{-1.2} T_{\rm 4}^2 
    \geqslant 1.96\,  n_{\rm snf,3.5} \,.
\end{equation}
In the presence of metal-poor inflows, such as that from cold streams and 
mergers with low-mass galaxies, the above criterion can be further relaxed
(see \S\ref{ssec:metal-bimodal}).

Fig.~\ref{fig:gc-criteria} shows the conditions for the formation of Pop-I and Pop-II 
GCs in the $(z,\,M_{\rm v})$ plane, where fiducial values, 
$n_{\rm snf} = 10^{3.5}\perccm$, $\epsilon_{\rm v}=0.1$, 
$f_{\rm gas}=0.04$, $f_{\rm str}=4$ and $T=10^{4}{\rm K}$ are used. 
Both conditions have a positive dependence on redshift and thus favor globular cluster
formation in high-redshift halos. However, at a given redshift, 
they have opposite dependence on halo mass: the Pop-I GCs preferentially form 
in high-mass halos owing to the high Mach number (strong turbulence), 
while Pop-II GCs preferentially form in low-mass halos owing to the low 
metallicity (slow cooling and delayed fragmentation). 
At $z \approx 5$, roughly the peak redshift for the formation of dynamically 
hot systems \citepalias[see e.g. figure 12 of][]{moTwophaseModelGalaxy2024}, 
the boundaries of two criteria start to diverge, leaving halos of intermediate 
mass less efficient in forming GCs.
This divergence has direct implications for the GC frequency expected 
in halos of different masses, as we will discuss in \S\ref{ssec:gcs-mass} and 
Fig.~\ref{fig:m_gc_vs_mhalo}. At $z\approx 0$, the low density of the Universe, 
the enrichment of the environment (galaxy), and the disappearance of cold stream 
prevent the formation of GCs in most galaxies. This leaves supernova compression 
and/or galaxy-galaxy merger as the open channel for GC formation.   

% {\color{red} Notes by: HJ
% (i) The halo population may be formed in low-mass progenitors at high redshift where
% compression by turbulent shocks is small. They have low metallicity so that allow SCs to grow 
% directly to high density to form GCs without much external compression. 
% (ii) At lower-$z$ where the density of SGCs becomes low and the metallicity is high, large 
% compression is needed to form GCs. This can only happen for sub-clouds already in more massive SGCs 
% where compression by turbulent shocks is larger.}

\subsection{Dynamical evolution}
\label{ssec:dyn-evol}

After formation, dynamical evolution of a star cluster changes its structure,
causes continuous mass loss and may eventually disrupt it. 
Sources that drive the dynamical evolution include two-body relaxation, 
tidal stripping, and tidal shocks, all of which may depend on the tidal environment 
and the internal composition and structure of the star cluster. 
As most GCs are born early, the long time available for dynamical effects 
to operate can significantly change the number of GCs survived 
today. For example, \citet{harrisDarkMatterHalos2015} used the observed constancy
of GCS mass-halo mass ratio to speculate that only $1/3$-$1/5$ proto-GCs can survive
to the present day. \citet{choksiOriginsScalingRelations2019} modeled the evolution 
of GCs within external, Milky Way-like tidal field by adopting a disruption timescale of 
$5\Gyr$ for clusters with $M_{\rm cls} = 2\times 10^5 \Msun$. Consequently,
most of such GCs born at $z \gtrsim 2$ are disrupted. 
\citet{chenModelingKinematicsGlobular2022} 
reported a more detailed modeling of the disruption timescale by tracing 
tidal fields in a hydrodynamical simulation around individual GCs   
sampled using a semi-analytical method. Their results suggested that $\approx 90\%$ 
of the GCs are disrupted, with significant dependence on their ages and 
distances to the galactic center (see their figure 13). \citet{rodriguezGreatBallsFIRE2023}
performed a more detailed modeling, based on the bulk properties of GMCs 
resolved by the {\sc Fire-2} simulation of a Milky Way-mass galaxy, a semi-analytical cluster 
sampler calibrated by simulations of individual GMCs 
\citep{grudicModelFormationStellar2021,grudicGreatBallsFIRE2023}, and 
a star-by-star cluster evolution code with assumptions on the cluster structure 
and tidal field. Their results highlight the complex interplay between the
internal stellar population, metallicity and structure of the star 
cluster, and the varying tidal environment of the host galaxy.

Due to the large sample and dynamic range that we aim to cover, 
the spatially resolved tidal environment is not available for individual star 
clusters. Following \citet{chenFormationGlobularClusters2023} and 
\citet{chenCatalogueModelStar2024}, we adopt a semi-analytical 
approach, which was originally introduced by 
\citet{fallDynamicalEvolutionMass2001}, 
extended by \citet{mclaughlinShapingGlobularCluster2008},
and calibrated by the high-resolution N-body simulations of 
\citet{gielesMasslossRatesStar2023}, with modifications according to the 
SGC properties from our model. For a star cluster with mass $M_{\rm cls}(t)$ 
moving in an orbit with radius $R_{\rm orb}(t)$ and $V_{\rm orb}(t)$ around 
the host galaxy, the total mass loss rate can be modeled as 
\begin{equation} \label{eq:tidal-disruption-rate}
    \dv{m(t)}{t} = 
    - \frac{ \nu_{\rm 0} \Omega_{\rm tid} (t) }{250} 
    \left[ \frac{M_{\rm cls}(t_0)}{2\times 10^5 \Msun} \right]^{\alpha_{\rm tid}}
    m ^{\beta_{\rm tid}} (t)\,.
\end{equation}
where $m(t) \equiv M_{\rm cls}(t) / M_{\rm cls}(t_0)$ is the 
scaled mass and $t_0$ is the birth time of the cluster, 
$\nu_{\rm 0} \sim 1$ is a free parameter
that needs to be calibrated by observations 
due to the unresolved tidal environments in the simulation, 
such as those producing tidal shocks. We choose $\nu_0 = 0.6$
as the fiducial value (see Appendix~\ref{sec:calibration} for the 
calibration). 
We set $\alpha_{\rm tid} = -2/3$ and $\beta_{\rm tid} = -1/3$ 
using calibrations by the N-body simulation of \citet{gielesMasslossRatesStar2023}, 
which also takes into account the effects of stellar black holes. 
Since their calibrations assume a static potential and may thus miss 
potential non-linear correlations between the two-body relaxation and 
tidal disruption, we modify their formulation of tidal frequency, 
$\Omega_{\rm tid}$, to be time-dependent, as
\begin{equation} \label{eq:tidal-frequency}
    \Omega_{\rm tid} (t)
    = \frac{V_{\rm orb}(t)}{R_{\rm orb}(t)} 
    = \frac{V_{\rm v}(t)}{u_{\rm r} f_{\rm gas} R_{\rm v}(t)}\,,
\end{equation}
where all the normalization constants have been absorbed into $\nu_{\rm 0}$,
and $u_{\rm r} \sim U[0.0, 1] $ is a uniformly distributed random 
number describing the distribution of the effective orbital radius of the star 
clusters within the host galaxy. The above formulation relies on the following facts. 
Firstly, the SGC is self-gravitating, so that $V_{\rm v}$ and $f_{\rm gas} R_{\rm v}$ give 
its typical circular velocity and radius, respectively, as discussed in 
\citetalias{moTwophaseModelGalaxy2024} and \citetalias{chenTwophaseModelGalaxy2024a}.
Secondly, the initial distribution of star clusters within the host galaxy
is nearly isothermal, so that the approximations that $V_{\rm orb} = V_{\rm v}$ and 
$R_{\rm orb} = u_{\rm r} f_{\rm gas} R_{\rm v}$ may be valid. This is consistent 
with the GC evolution model of \citet[see their figure~10]{rodriguezGreatBallsFIRE2023} 
based on the post-processing of the {\sc Fire-2} simulation and a cluster Monte Carlo
code, where the effective tidal strength is found to be well described by the 
expectation of an isothermal sphere. 
Finally, the upper bound of $u_{\rm r}$ comes from 
observations that GCs can extend to large radius, such as $0.1 R_{\rm vir}$ or even beyond 
\citep[e.g.][]{
    karthaSLUGGSSurveyGlobular2014,
    alabiSLUGGSSurveyMass2016,
forbesMetallicityGradientsGlobular2018,
dornanInvestigatingGCSMRelation2023}.

Combining Eqs.~\eqref{eq:v-vir} and \eqref{eq:r-vir} for halo virial velocity
and radius, respectively, and substituting the time variable with 
$H = \dot{a}/a$ and $a=1/(1+z)$, Eq.~\eqref{eq:tidal-disruption-rate} 
can be analytically integrated from the initial redshift $z_0$ to arbitrary 
redshift $z<z_0$:
% critical for the numeric implementation
\begin{align} \label{eq:tidal-integration}
    M_{\rm cls}(z) & = M_{\rm cls}(z_0) (1-R) \times  \nonumber \\
    & \left[
        1 - 
        \frac{ (1-\beta_{\rm tid}) \nu_0 }{u_{\rm r} f_{\rm gas,0.04}} 
        M_{\rm cls, 5.3}^{\alpha_{\rm tid}}(z_0) \ln \left(\frac{1+z_0}{1+z} \right) 
    \right]^{\frac{1}{1-\beta_{\rm tid}}}\,,
\end{align}
where
$M_{\rm cls}(z_0) = 2 \times 10^5 \Msun\,M_{\rm cls,5.3}(z_0)$,
$f_{\rm gas} = 0.04 f_{\rm gas,0.04}$,
and $R$ is the returned fraction of mass (Eq.~\ref{eq:mass-return}).
Note that the change of tidal environment in mergers is difficult to model. 
We ignore such effects in our modeling and apply Eq.~\eqref{eq:tidal-integration} 
to all star clusters, regardless of whether or not they have merged into other galaxies. 
This is a rough approximation, but may be reasonable given that tidal disruption is more 
significant at higher $z$, so that most of the mass loss is expected to occur in the 
galaxy where star clusters were born \citep[e.g.][]{kruijssenGlobularClusterFormation2014}.

To gain some intuition of the disruption, we define the disruption timescale 
for a star cluster as $\tau_{\rm tid} = |m/\dot{m}|_{z=z_0}$, and by using 
Eq.~\eqref{eq:tidal-disruption-rate}, we obtain
\begin{align} \label{eq:tidal-time-scale}
    \tau_{\rm tid} & = u_{\rm r} f_{\rm gas,0.04} 
        M_{\rm cls,5.3}^{-\alpha_{\rm tid}} (z_0) \nu_0^{-1} H^{-1}(z_0)  
        \nonumber \\
    & = u_{\rm r} f_{\rm gas,0.04}
        M_{\rm cls,5.3}^{-\alpha_{\rm tid}} (z_0) \nu_0^{-1} \times 
    \begin{cases}
        \left[ 0.82 \Gyr \right] (1+z)_{10}^{-3/2} \\
        \left[ 5.00 \Gyr \right] (1+z)_{3}^{-3/2} \,.
    \end{cases}
\end{align}
The dependence of $\tau_{\rm tid}$ on $M_{\rm cls}$ has a power-law index of
$2/3$ in our fiducial choice, close to the analytically-estimated and observed 
values, $0.62$ and $0.60$, respectively, reported by 
\citet{lamersDisruptionTimeScales2005}.
The dependence of $\tau_{\rm tid}$ on the tidal frequency 
(Eq.~\ref{eq:tidal-disruption-rate}) implies a dependence of $\tau_{\rm tid}$
on the ambient gas density, $\tau_{\rm tid} \sim \rho_{\rm gas}^{-1/2}$,
consistent with N-body simulations and observations shown
by \citet{lamersDisruptionTimeScales2005}.
The disruption timescale is proportional to the Hubble timescale, and does 
not explicitly depend on halo mass, indicating that the GCS mass-halo mass
relation is not distorted by the dynamical evolution while the normalization 
decreases as the formation of GCs stops.
The short disruption timescale at $z = 9$ indicates that nearly all GCs formed 
at such high redshift are eventually disrupted. The relatively long disruption 
timescale at $z = 2$ suggests that massive GCs born with $M_{\rm cls, 5.3} \gtrsim 1$ 
at $z=2$ can survive until $z=0$.
At $z=0$, the disruption timescale is about $3\Gyr$ for a cluster
with $u_{\rm r} = 0.5$, $f_{\rm gas} = 0.04$ and $M_{\rm cls} = 10^4\Msun$,
falling into the range found by \citet{lamersDisruptionTimeScales2005}
using observational data.
Variations of GC disruption among halos due to their differences in 
GC formation history are allowed in Eqs.~\eqref{eq:tidal-frequency}
and \eqref{eq:tidal-integration} through the gas fraction $f_{\rm gas}$ and time variables 
$t$ and $z_0$. Note that we have ignored substructures of galaxies, 
such as clumps, spiral arms and bars. These should be included in the future
with detailed modeling of the galactic structure.

%%%%%%%%%%%%%%%%%%%%%%%%%%%%%%%%%%%%%%%%%%%%%%%%%%%%%%%%%%%%%%%%%%%%%%%%%%%%%%%%

\section{Model predictions and comparison with observational data}
\label{sec:results}

% A history map, such as
% those presented by \citet{el-badryFormationHierarchicalAssembly2019},
% is likely useful.

In this section, we combine the physical recipes described in \S\ref{sec:galaxy-model} and \S\ref{sec:gc-model},
and implement them into (sub)halo merger trees taken from a dark-matter-only simulation. 
The product of the implementation is the galaxy, including its gaseous, stellar and SMBH 
components, and globular clusters, for each subhalo in the simulation.
A variety of summary statistics are provided for comparisons with currently 
available observations and other GC models. Other predictions are provided to be 
tested by future observations. 
% A number of case studies are presented to demonstrate the use of the model to 
% unveil the formation histories of individual galaxies.

The simulation used here is the TNG100-1-Dark run, conducted as a part of the 
IllustrisTNG project \citep{pillepichFirstResultsIllustrisTNG2018, 
nelsonIllustrisTNGSimulationsPublic2019}.
The cosmology adopted by the simulation is consistent with that used in 
this paper (see \S\ref{sec:intro}).
The run has a periodic box with a side length of $75\mpc$, 
$1820^3$ dark matter particles, each with a mass of $6.0\times 10^6 \msun$,
a Plummer equivalent gravitational softening length varying from $1 \kpc$
at high $z$ to $0.5 \kpc$ at low $z$. A total of 100 
snapshots spanning from redshift $z=20.0$ to $0$ have been saved. Halos are 
identified using the friends-of-friends (\softwarenamestyle[FoF]) algorithm
with a scaled linking length of $0.2$ \citep{davisEvolutionLargescaleStructure1985}. 
Subhalos are identified using the \softwarenamestyle[Subfind] 
algorithm \citep{springelPopulatingClusterGalaxies2001,
dolagSubstructuresHydrodynamicalCluster2009}, and subhalo merger trees are 
constructed using the \softwarenamestyle[SubLink]
algorithm \citep{springelCosmologicalSimulationCode2005,
boylan-kolchinResolvingCosmicStructure2009, rodriguez-gomezMergerRateGalaxies2015}. 
The lower limit for FoF halo mass is about $2\times 10^8 \msun$,
and the most massive halo at $z=0$ in the simulation volume 
has $M_{\rm v} = 2.47\times 10^{14}\msun$.
The main progenitor of a subhalo is defined as the one with the most massive 
history among all progenitors \citep{deluciaHierarchicalFormationBrightest2007,
rodriguez-gomezMergerRateGalaxies2015}. The central subhalo of a FoF halo 
is defined as the one with the most massive history among all subhalos within the halo. 
The main branch of an FoF halo corresponds to the main branch of its central subhalo. 
The specific run we choose ensures a balance between statistical 
robustness and resolution in halo assembly histories. 

\subsection{Numerical implementation of the model} 
\label{ssec:impl}

\begin{center}
\begin{table*} 
\caption{List of the model components, their fiducial parameters
and the sources of the values. See \S\ref{ssec:impl} for a list of
the steps of the numerical implementation. The parameters calibrated 
in this paper are detailed in Appendix~\ref{sec:calibration}.
}
\begin{tabular}{ >{\centering\arraybackslash}m{5cm} | m{4cm} | m{6.5cm} }
    \hline
    \rowcolor{Gainsboro!60}
        \makecell{Model Component} 		 
        &
        \makecell{Fiducial Parameters}
        &
        \makecell{Sources}
        \\
    \hline
        Halo assembly and galaxy formation \linebreak
        (\S\S\ref{ssec:halo-assembly}, \ref{ssec:collapse-gas},
        \ref{ssec:star-smbh-formation})
        &
        -
        &
        Calibration in \citetalias{moTwophaseModelGalaxy2024} and 
        \citetalias{luEmpiricalModelStar2014}
        \\
    \hline
        Metal enrichment (\S\ref{ssec:metal})
        &
        $V_{\rm esc} = 75 \kms$, $\beta_{\rm esc}=2$, 
        $z_{\rm esc} = 5$, 
        $\gamma_0 = 1$, $\beta_{\rm mix}=2$
        &
        This paper
        \\
    \hline 
        \multirow{4}{*}{
            Sub-cloud formation (\S\ref{ssec:formation-of-sc})
        }
        &
        $\beta_{\rm sc,m}=-2$, $M_{\rm sc,t} = 10^{6.5}\Msun$;
        &
        Star cluster mass function by e.g. \citet{krumholzStarClustersCosmic2019};
    \\
        &
        % $f_{\rm r} = 1$, 
        % $f_{\rm turb} = 1$;
        $f_{\rm s} = 1$;
        &
        This paper;
    \\
        &
        $V_{\rm w} = 250\kms$;
        &
        The same value as \S\ref{ssec:star-smbh-formation}
        for the galactic-scale model;
    \\
        &
        $\alpha_{\rm s} = 1.5$
        &
        \citet{girichidisEvolutionDensityProbability2014} and 
        \citet{kiihneFittingMethodsProbability2025}
    \\
    \hline
        Sub-cloud fragmentation
        (\S\ref{ssec:cooling})
        &
        $\beta_{\rm sc,n} = -2.5$
        &
        This paper
        \\
    \hline
        
        \multirow{2}{*}{
            Star formation within sub-cloud
            (\S\ref{ssec:sf-in-scs})
        }
        &
        $\epsilon_{\rm w, max} = 1$;
        &
        \citet{dekelEfficientFormationMassive2023};
        \\
        &
        $f_{\rm r,cls} = 1/2$
        &
        Isothermal assumption
        \\
    \hline
        GC formation
        (\S\ref{ssec:gc-channels})
        &
        $n_{\rm snf} = 10^{3.5}\perccm$
        &
        \citet{dekelEfficientFormationMassive2023}
        \\
    \hline
        \multirow{2}{*}{
            Star cluster dynamical evolution
            (\S\ref{ssec:dyn-evol})
        }
        &
        $\alpha_{\rm tid}=-2/3$, $\beta_{\rm tid}=-1/3$;
        &
        \citet{gielesMasslossRatesStar2023};
        \\
        &
        $\nu_0 = 0.6$
        &
        This paper
        \\
    \hline
\end{tabular}
\label{tab:parameters}
\end{table*}
\end{center}

The numerical algorithm takes each subhalo merger tree as the input unit, 
and processes it by the steps summarized below.
\begin{enumerate}[parsep=6pt]
    \item {\bf Decomposition of halo assembly history:} The subhalo merger 
    tree is decomposed into a set of disjoint branches. 
    Physical quantities in each branch are smoothed by a running kernel,
    with the kernel size given by the dynamical timescale, $t_{\rm dyn}$.
    Each branch is decomposed into two stages, 
    a central stage and a satellite stage.
    The mass assembly history, $M_{\rm v}(z)$, during the central stage is
    fitted to a parametric function. The specific halo growth rate, 
    $\gamma(z)$, derived from the parametric fitting, and a threshold, 
    $\gamma_{\rm f}$, are used to separate 
    the central stage into two phases: 
    a fast phase with $\gamma(z) \geqslant \gamma_{\rm f}$, 
    and a slow phase with $\gamma(z) < \gamma_{\rm f}$. 
    The details of this step are described in \S\ref{ssec:halo-assembly}.
    
    \item {\bf The amount of cooled gas}: At each time step in a branch,
    the total amount of available gas, $\Delta M_{\rm g,avail}$, 
    is given by Eq.~\eqref{eq:delta-m-g-avail}. The amount of gas
    cooled down to the SGC, $\Delta M_{\rm g,cool}$, 
    is modified by the cooling factor, $F_{\rm cool}$, 
    given by Eq.~\eqref{eq:delta-m-g-cool}.

    \item {\bf Star formation and SMBH growth:} 
    Taking into account the feedback from AGN and supernova,
    the amount of star-forming gas, $\Delta M_{\rm g, sf}$, is a fraction 
    $F_{\rm sn} F_{\rm agn}$ (Eqs.~\ref{eq:f-sn} and \ref{eq:f-agn}) 
    of the cooled gas. The amount of formed stars, $\Delta M_*$, is given by 
    Eq.~\eqref{eq:delta-m-star}, from which SFR is also obtained. 
    The growth of SMBH, $\Delta M_{\rm bh}$, 
    is given by Eq.~\eqref{eq:delta-m-bh}. 
    The amount of gas remaining in the SGC, $\Delta M_{\rm g}$, is obtained 
    by the conservation of mass (Eq.~\ref{eq:delta-m-g}). 
    Star formation in the satellite stage is assumed to follow 
    an exponential decay (Eq.~\ref{eq:sfr-sat}), and the gas is modeled 
    assuming a `close-box'. Galaxy mergers are modeled by assuming that a 
    fraction, $f_{\rm merge}$, of the stellar mass of the infall
    galaxy is added to the main branch.

    \item {\bf Metal enrichment:}
    Metal yield and recycling are modeled by following the idea of the 
    gas regulator,
    with mixing and escaping fractions modified for the gaseous component 
    that hosts star clusters. The details are described in \S\ref{ssec:metal}.
    To account for the missed fluctuations, a Gaussian random number with 
    $\sigma_{\log Z} = 0.1\dex$ is added to the metallicity $Z$.

    \item {\bf Sampling of sub-clouds: }
    A set of sub-clouds is randomly sampled, with initial mass 
    $M_{\rm sc}$ following the distribution given by Eq.~\eqref{eq:m-sc-pdf}, 
    and with the initial density $n_{\rm sc}$ following the distribution given by
    Eq.~\eqref{eq:ln-density-pdf}. $M_{\rm sc}$ and $n_{\rm sc}$ are assumed 
    to be independent of each other at formation, but subsequent processes, e.g. 
    external feedback and star formation, can lead to correlations.
    The size of a sub-cloud, $R_{\rm sc}$, is obtained from $M_{\rm sc}$ and 
    $n_{\rm sc}$. The metallicity of a sub-cloud is assumed to follow 
    the galactic average given by the previous step, 
    with an additional $0.3\dex$ random fluctuation added to 
    account for variations among sub-clouds.
    Because the timescale involved in sub-cloud generation and dispersal
    can be much shorter than the time interval between two adjacent snapshots 
    saved by the simulation, we use the amount of stars to constrain
    $N_{\rm sc}$, the number of sub-clouds to be sampled at each snapshot. 
    This is achieved by requiring that the total amount of stars 
    formed in all sub-clouds to be equal to that of the entire galaxy:
    $\sum_{i \leq N_{\rm sc}} M_{\rm cls, i} = \Delta M_*$, where
    $M_{\rm cls}$ is given by Step (viii) and $\Delta M_*$ is modeled 
    in Step (iii).

    \item {\bf Cooling and fragmentation: } 
    The threshold density, $n_{\rm sf, 1}$, 
    at which a sub-cloud is able to fragment and form stars, 
    is obtained according to Eq.~\eqref{eq:n-sf-threshold}. 
    If the initial density is so low that $n_{\rm sc} < n_{\rm sf, 1}$, 
    the fragmentation and star formation are delayed to a later time
    when the sub-cloud cools and contracts to a density $n_{\rm sc}'$ sampled 
    from the distribution function given by Eq.~\eqref{eq:sampling-n-sc}. 
    On the other hand, if $n_{\rm sc} \geqslant n_{\rm sf, 1}$, 
    the sub-cloud fragments immediately.
    For convenience, we set $n_{\rm sc}' = n_{\rm sc}$.

    \item {\bf External feedback: } A sub-cloud can survive the feedback from
    previous generations of stars within the galaxy if the initial 
    density of the sub-cloud is high enough to shield against the feedback.
    In this paper, we only include the wind feedback (Eq.~\ref{eq:n-shield-w}), 
    as its effect is estimated to be more significant than the 
    radiative feedback (Eq.~\ref{eq:n-shield-r}) 
    on sub-cloud scales. If a sub-cloud is not shielded, we remove 
    it from the set of sub-clouds.

    \item {\bf Star formation in sub-clouds:} 
    The fraction of gas converted into stars in each sub-cloud,
    and thus the mass of the resulted star cluster, $M_{\rm cls}$,
    is determined by the stellar feedback within the sub-cloud
    (Eqs.~\ref{eq:eps-star-by-wind} and \ref{eq:eps-star-by-radiation}). 
    We adopt the wind feedback in the fiducial model, 
    while leaving other choices optional.
    With alternative choices of feedback, the power-law index of the size-mass 
    relation of star clusters differs within a narrow range of $1/6$--$1/4$, 
    and the dominant channel of feedback is still unconstrained by the 
    current observations. The fiducial parameters are chosen so that 
    the star formation efficiency in a sub-cloud is $\epsilon_{\rm w} = 1$
    at $M_{\rm sc} = 10^{6}\Msun$, $R_{\rm sc} = 12\pc$ and $Z = 0.02$,
    consistent with both our theoretical expectations in \S\ref{ssec:sf-in-scs} 
    and the observed truncation mass and size of young massive clusters 
    and globular clusters \citep[e.g.][]{krumholzStarClustersCosmic2019}.

    \item {\bf Selection of globular cluster:}
    At formation, a star cluster is classified as a (progenitor of) GC if 
    (i) the star formation in the sub-cloud is supernova-free, 
    i.e. $n_{\rm sc}' \geqslant n_{\rm snf}$, and (ii) the stellar mass 
    $M_{\rm cls} \gtrsim M_{\rm gc,min} = 10^4 \Msun$. 
    The mass limit set by the second criterion is roughly the threshold 
    adopted to classify an observed star cluster as a GC
    \citep[e.g.][]{kruijssenFormationAssemblyHistory2019},
    and is also adopted by other models \citep[e.g.][]{chenFormationGlobularClusters2023}.
    We have checked that this lower bound of mass is 
    sufficient to robustly obtain all the statistics of GCs, because the peak of 
    the cumulative GC mass function is much larger than this lower bound (see 
    \S\ref{ssec:mass-function}).
    As discussed in \S\ref{ssec:gc-channels}, a GC is 
    classified as Pop-I if $n_{\rm sf,1} < n_{\rm snf}$, and otherwise 
    classified as Pop-II. These two populations directly match 
    the metal-rich (red) and metal-poor (blue) populations, respectively,
    found in observations. Once a side-branch galaxy is merged into the 
    main branch, all of the side-branch GCs are added to the merger 
    remnant. The strong tidal field in the merger process
    can strip GCs from the infalling galaxy and spread them over 
    a large range of galactocentric distance, as to be discussed in 
    \S\ref{ssec:spatial-dist}. GCs formed in the main and 
    side branches are referred to as the in-situ and ex-situ populations, 
    respectively.

    \item {\bf Dynamical evolution:} The change of stellar mass of a star 
    cluster in the galactic tidal field is modeled according to 
    the time integration, from the birth redshift $z_0$ to the observed 
    redshift $z$, given by Eq.~\eqref{eq:tidal-integration}, with 
    a randomly sampled effective orbital radius within the SGC.
    Consequently, a star cluster born as a GC, i.e. 
    $M_{\rm cls}(z_0) \geqslant M_{\rm gc,min}$,
    can gradually lose its mass, and begin to be classified as a 
    non-GC (e.g. open) star cluster or a diffuse stellar component, 
    once $M_{\rm cls}(z) < M_{\rm gc,min}$.
\end{enumerate}

The above pipeline involves a number of random sampling procedures
(e.g. those for $M_{\rm sc}$, $n_{\rm sc}$, $n_{\rm sc}'$), which are all difficult to carry out directly. This is due to 
several reasons. Firstly, the ranges of quantities to be sampled are wide,
and the probability densities vary significantly over these ranges. 
For example, the distribution of $M_{\rm sc}$ has a power-law index 
$\approx -2$ (Eq.~\ref{eq:m-sc-pdf}), implying  that the number of 
sub-clouds per unit logarithmic mass differs by four 
orders of magnitude between $10^4$ and $10^8 \Msun$. The consequence 
of this is that sub-populations are highly imbalanced, potentially leading to 
large statistical fluctuations for the rare population (e.g. GC)
when the sample size of dark matter halos is limited. 
Secondly, the number of sub-clouds to sample is rather large even for a single halo. 
For example, the total number of sub-clouds to sample is about $1.5 \times 10^7$
for a subhalo merger tree rooted in a present-day Milky Way-size subhalo
with $M_{{\rm v},z=0}\approx 10^{12} \msun$, and more than $2 \times 10^8$ for a cluster halo 
of $10^{14} \msun$ (see Fig.~\ref{fig:b-sample-scale}). This presents a great challenge for computation, especially
when we want to generate samples of GCs on cosmological scales.

For the sake of statistical robustness and computational efficiency, we
design a `balanced' sampling algorithm (referred to as {\sc BSampling} hereafter)
to generate sub-clouds. 
Briefly, the algorithm defines a set of proposal 
distributions $q^{(j)}(x^{(j)})$, one for each property $x^{(j)}$ to be sampled, 
as replacements of the original distributions $p^{(j)}(x^{(j)})$.
Once a sub-cloud, represented by a set of its properties 
${\bf x}_i = (x_{i}^{(1)},\,x_{i}^{(2)},\,...)$, 
is drawn from the proposal distributions, a weight 
\begin{equation}\label{eq:bsamp-weight-sc}
    w_i \equiv \prod_j \frac{
        p^{(j)}(x_i^{(j)})
    }{
        q^{(j)}(x_i^{(j)})   
    }
\end{equation}
is assigned to it to compensate for the bias between the proposal 
and original distributions. The weights of sub-clouds are normalized 
by $N_{\rm sc}$, the total number of sub-clouds to be sampled,
as $\sum_i w_i = N_{\rm sc}$,
so that a weight $w_i > 1$ represents multiple sub-clouds, 
therefore reducing the computational cost. Summary statistics 
derived from the sampled sub-clouds need to incorporate 
these weights.
If $q^{(j)}$ is carefully chosen so that it is more uniform than $p^{(j)}$, 
the statistical fluctuation can be reduced.
If a real, un-weighted sample of sub-clouds 
is needed for, e.g. mock observations, 
it can be generated by resampling $N_{\rm sc}$ times 
from the discrete values $\{{\bf x}_i\}$,
according to probabilities given by $\{ w_i / \sum_i w_i \}$.
The idea here is motivated by the importance sampling technique in
Monte Carlo integrations, and by the sampling-importance-resampling (SIR)
technique in the discrete approximation of continuous distributions
\citep[see e.g. chapter 11 of][]{bishopPatternRecognitionMachine2006}.
A formal description of this sampling technique and the specific 
choice for this paper are detailed in Appendix~\ref{sec:bsampling}. 

Table~\ref{tab:parameters} lists the parameters adopted in this paper
for each component of the model. The free parameters that cannot be 
fixed via theoretical arguments, hydrodynamical simulations or 
direct observations need to be calibrated by summary statistics 
of observations, which are detailed in Appendix~\ref{sec:calibration}.
In the remainder of this paper, we present the results of the model
and discuss their physical implications.

\subsection{Mass and age distributions} \label{ssec:mass-function}

\begin{figure*} \centering
    \includegraphics[width=\textwidth]{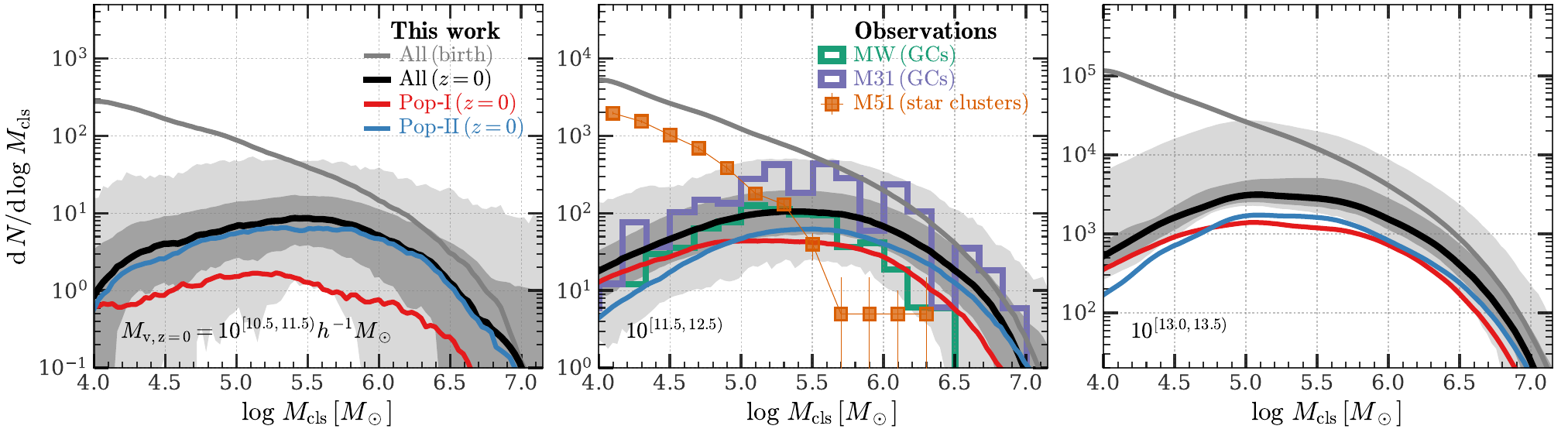}
    \caption{
        Mass functions of GCs. 
        Different {\bf panels} show the mass functions of GCs in central 
        galaxies with different halo masses at $z=0$. Both in-situ 
        and ex-situ GCs are included.
        In each panel, 
        {\bf grey} curve shows the birth-time mass function of 
        all GCs formed in the history. 
        {\bf Black} curve shows the $z=0$ 
        mass function of GCs survived after the mass loss due to dynamical 
        evolution is incorporated, with dark shade showing 
        the $1$-$\sigma$ range among individual 
        galaxies and light shade showing the minimum to maximum. 
        {\bf Red} and {\bf blue} curves show the Pop-I and Pop-II sub-components, 
        respectively, of the black curve.
        See \S\ref{ssec:impl} for the details of modeling and sample selection.
        In the center panel, the histograms show the observed GC mass 
        functions for MW ({\bf green}) and M31 ({\bf purple}), 
        compiled by \citet{rodriguezGreatBallsFIRE2023}. 
        {\bf Orange} markers show the mass function of star clusters
        in M51, obtained from the star-cluster catalog of LEGUS
        \citep{messaYoungStarCluster2018a}.
        See \S\ref{ssec:mass-function} for a detailed discussion of this figure.
    }
    \label{fig:mass_func}
\end{figure*}

\begin{figure*} \centering
    \includegraphics[width=1\textwidth]{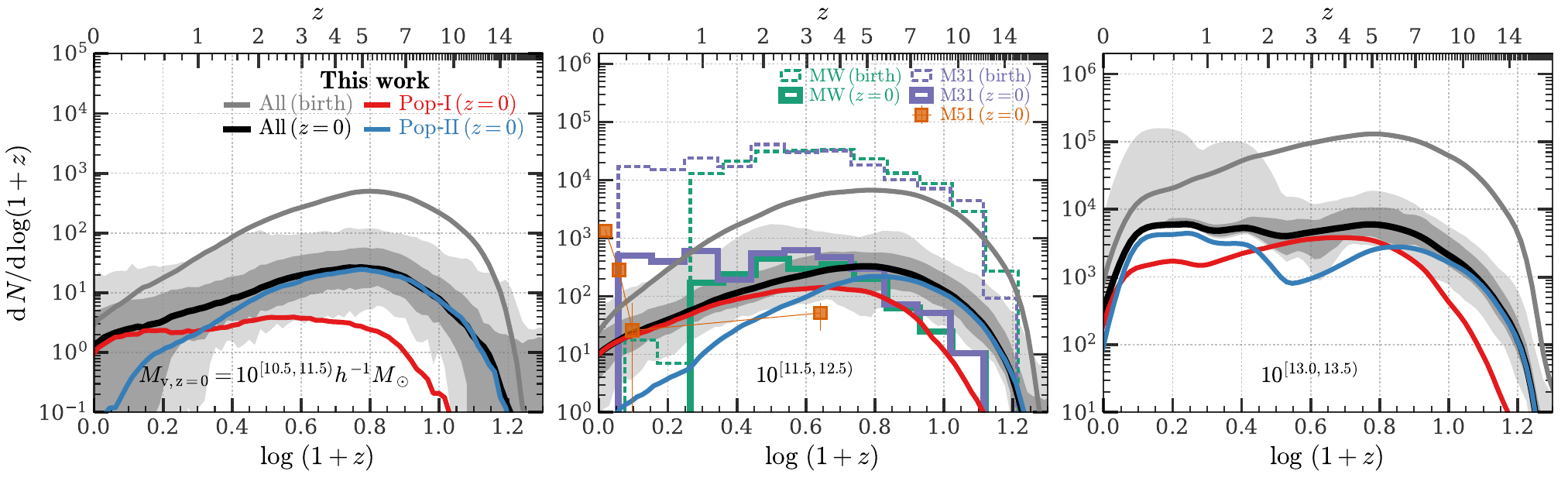}
    \caption{
        Age distributions of GCs.
        Different {\bf panels} show the age distributions of GCs in central 
        galaxies with different halo masses at $z=0$. Both in-situ 
        and ex-situ GCs are included.
        In each panel, 
        {\bf grey} curve shows the distribution of the birth time of 
        all GCs formed in the history. 
        {\bf Black} curve shows that 
        of GCs survived until $z=0$ after the mass loss due to 
        dynamical evolution is incorporated,
        with dark shade showing 
        the $1$-$\sigma$ range among individual 
        galaxies and light shade showing the minimum to maximum.
        {\bf Red} and {\bf blue} curves show the Pop-I and Pop-II sub-components, 
        respectively, of the black curve.
        See \S\ref{ssec:impl} for the details of modeling and sample selection.
        In the center panel, the histograms show the results of 
        the empirical model of \citet{chenCatalogueModelStar2024} for
        MW ({\bf green}) and M31 ({\bf purple}). 
        Their results for all and survived GCs are shown by 
        {\bf dashed} and {\bf solid} histograms, respectively.
        {\bf Orange} markers show the age distribution of star clusters
        more massive than $10^5\Msun$ in M51, obtained from the star-cluster 
        catalog of LEGUS \citep{messaYoungStarCluster2018a}.
        See \S\ref{ssec:mass-function} for a detailed discussion of this figure.
    }
    \label{fig:gc_history}
\end{figure*}

Fig.~\ref{fig:mass_func} shows the mass functions of GCs in central 
galaxies at $z=0$ in three different ranges of halo mass. Here, 
the mass function is defined as $\dd{N}/\dd{\log M_{\rm cls}}$, 
the number of GCs per unit logarithmic mass, and the result is 
averaged over all central galaxies in the given range of halo mass.
Both the in-situ and ex-situ components are included.
The birth-time mass function, obtained by selecting GCs and using 
statistics based on their masses at formation (see \S\ref{ssec:impl} for the 
selection criteria), is shown by the grey curve.
The slope of the mass function in the logarithmic scale 
is a constant of $\approx 0.7$ at $M_{\rm cls} \lesssim 10^6 \Msun$, 
due to the power-law form, with an index of $\beta_{\rm sc,m} = -2$, 
of the sub-cloud mass function (Eq.~\ref{eq:m-sc-pdf}), and the 
decreasing star formation efficiency with decreasing sub-cloud mass
(Eq.~\ref{eq:eps-star-by-wind-scaling}).
The steepening of the mass function at the high-cluster-mass end is due 
to the exponential truncation of the sub-cloud mass function.
The present-day mass function, obtained by taking into account 
the effects of dynamical evolution (\S\ref{ssec:dyn-evol}), 
is shown by the black curve. The difference between the present-day mass function 
and that at the birth time highlights the effect of mass loss. 
As GCs with lower mass have shorter tidal disruption timescales 
(Eq.~\ref{eq:tidal-time-scale}), the mass loss is quicker 
for lower-mass GCs, leading to greater suppression of the mass function 
towards lower cluster mass. The dynamical evolution also changes 
the shape of the mass function, from a monotonical form at birth to
a unimodal form at the present day.

The peak of the unimodal distribution is around $10^5 $--$ 10^{5.5} \Msun$,
with moderate dependence on the host halo mass.
Such an unimodal distribution also ensures that the lower limit of sub-cloud mass, 
$M_{\rm sc,min}=10^4 \Msun$ (Eq.~\ref{eq:m-sc-pdf}), 
and the lower limit of GC mass, $M_{\rm gc,min}=10^4 \Msun$ (\S\ref{ssec:impl}),
are sufficient to obtain the summary statistics for the observed population of GCs.

The red and blue curves in each panel of Fig.~\ref{fig:mass_func} show
the mass functions of Pop-I and Pop-II GCs, formed via the two channels 
defined in \S\ref{ssec:gc-channels}, respectively.
The mass functions of the two populations are quite similar
in shape, both being unimodal. 
The mass function of Pop-II GCs appears more top-heavy, which
is because this population on average is born earlier, 
and thus has a longer time for mass loss and a higher probability 
for the low-mass ones to be disrupted.
The amplitudes of the mass function of both populations increase
with increasing halo mass, as we will discuss in more detail
in \S\ref{ssec:gcs-mass}.
Pop-I is sub-dominant in central galaxies 
with $M_{{\rm v}, z=0} < 10^{11.5} \msun$, but becomes 
comparable to Pop-II in more massive halos. This is a direct
outcome of the more enriched ISM in galaxies that reside in more 
massive halos, as shown in Fig.~\ref{fig:mzr} and \S\ref{ssec:metal}.

In the center panel of Fig.~\ref{fig:mass_func}, we also include 
the observed mass functions of GCs in the Milky Way (green curve) and  
M31 (purple curve), compiled by 
\citet[see their figure 3]{rodriguezGreatBallsFIRE2023} 
using catalogs from
\citet{harrisCatalogParametersGlobular1996,harrisMassiveStarClusters2010}
and \citet{caldwellStarClustersM312011}, respectively.
The observed GC mass function of M31 has an amplitude 
about $0.5\dex$ higher than that of the MW, and appears
more top-heavy. This may be explained partly by the more massive halo 
of M31, and partly by the fact that it has more recent mergers that can 
bring in ex-situ GCs as well as increase the in-situ  
formation of GCs \citep[see the age distribution below; see also, e.g. \S4.3 of][]{chenCatalogueModelStar2024}.
The mass function of GCs predicted by our model (black curve) falls between 
those of MW and M31 over the entire range of $M_{\rm cls}$, indicating that 
the model is capable of reproducing the average GC mass function for MW-size 
galaxies. The observed mass functions of MW and 
M31 are well within the range of the model prediction (light shade),
indicating that they are not outliers in the ensemble of MW-sized galaxies
predicted by our model.

Fig.~\ref{fig:gc_history} shows the age distributions of GCs 
in the same sets of central galaxies as in Fig.~\ref{fig:mass_func}.
In each panel, histograms of the birth redshift are shown for star clusters 
born as GCs (grey curve) and for the ones that can survive to the present.
For comparison, in the middle panel, we also show the results of the 
empirical model of \citet{chenCatalogueModelStar2024} for MW (green curves) and 
M31 (purple curves), obtained by calibrating the model with observations 
and by populating GCs in appropriate halos in simulations.
The results shown here are the averages over their three best-match realizations
for both MW and M31.
Their adopted halo masses for MW and M31 are about $10^{12.1} \Msun$ and
$10^{12.25}$, respectively, which gives a slightly larger number of GCs in M31 than in MW. 
Their selections of MW-like and M31-like halos are observationally motivated, 
so that there is no major merger in the past $10\Gyr$ for the MW case
and there is at least one major merger in the past $6\Gyr$ for the M31 case. 
This, combined with their merger-triggered GC formation scenario,
predicts a larger and younger population of GCs in M31 than in MW,
as seen from the excess of young GCs in M31 at $z \lesssim 1$.
The fraction of disrupted GCs predicted by their model is significantly larger than ours, 
although the predicted number of GCs at $z=0$ is similar in both models. 
This highlights the degeneracy between GC formation and disruption in the modeling, 
which are respectively controlled by the parameters $(p_2, p_3)$ and $\kappa$
in \citet[see their \S2.2]{chenCatalogueModelStar2024}, 
and by $n_{\rm snf}$ (\S\ref{ssec:gc-channels}) and $\nu_0$ (\S\ref{ssec:dyn-evol}) 
in our model.

A key difference between our model and theirs is that we include   
the Pop-II (metal-poor) channel which elevates GC formation in metal-poor 
galaxies, mainly low-mass galaxies at high $z$. 
The consequence is the enhanced GC formation at high redshift,
which produces an age distribution peaked at $z\approx 5$, 
roughly the peak redshift for cosmic bulge formation \citepalias{moTwophaseModelGalaxy2024}, 
instead of at $z\approx 2$, the peak redshift of the cosmic star formation history
\citep[e.g.][]{madauCosmicStarFormation2014}.
Thus, a stronger connection between GCs and bulges is predicted 
by our model in comparison to other models and early observations.
The new GC candidates in the galactic bulge discovered recently by 
Gaia \citep{palmaAnalysisPhysicalNature2019,camargoThreeCandidateGlobular2019,
minnitiIntriguingGlobularCluster2021} 
appears to support the prediction of our model. 
The empirical model of GC formation by 
\citet[see also \citealt{valenzuelaGalaxyArchaeologyWet2024}]{valenzuelaGlobularClusterNumbers2021} 
also supports the presence of a distinct channel in the early Universe.
Metal-poor GCs in present-day central galaxies 
hosted by halos below the MW-halo mass are dominated by the in-situ component, and thus 
have systematically older age than the metal-rich counterparts, as seen in the left and center panels. 
On the other hand, for cluster-size halos, the age distribution of metal-poor GCs 
in their central galaxies is bimodal, as seen from the right panel of Fig.~\ref{fig:gc_history}.
The additional metal-poor, low-$z$ mode in our model is due to the ex-situ 
GCs brought in by satellite galaxies merged into the central, 
and it is consistent with the inference 
by, e.g., \citet{harrisPhotometricSurveyGlobular2023} from observations 
of bright elliptical galaxies.

The difference between MW and M31 in the mass and age distributions of GCs 
indicates the diversity of GC formation in galaxies with different states and
assembly histories. To clearly see this, we consider another galaxy, 
M51 (NGC 5194), a spiral galaxy interacting with a massive companion, 
NGC 5195. The stellar mass of M51 is inferred to be $\approx 10^{10.6}\Msun$
\citep{leroyStarFormationEfficiency2008,weiSpatiallyresolvedStellarPopulation2021}, 
slightly smaller than that of MW. We take the star-cluster candidates
of the LEGUS survey for M51 \citep{messaYoungStarCluster2018a}, 
remove the contamination by selecting only those with machine-learning
class 1 or 2, and exclude those with age less than $10\Myr$ to avoid 
unbound systems. The mass distribution of the star clusters so obtained 
is shown in the center panel of Fig.~\ref{fig:mass_func}.
We further restrict the sample to massive clusters with 
$M_{\rm cls} \geqslant 10^5\Msun$ so that they are age-complete
\citep[see Fig.~5 of][]{messaYoungStarCluster2018a}, and show the age 
distribution of these clusters in the center panel of 
Fig.~\ref{fig:gc_history}.
At $M_{\rm cls} \gtrsim 10^5\Msun$, the cluster mass function of M51
is much lower than those of MW, M31 and the model prediction.
The age distribution shows that these massive clusters are mostly born
at $z < 0.5$ by recent star formation, and none of them are as old as
those produced at $z \approx 5$, the peak of the GC-formation activity
predicted by our model. Such a situation can be created if M51 forms 
very late, and is hosted by a halo with very stable dynamical environment
in the early stage. The lack of galactic-scale turbulence thus suppresses the 
formation of old and massive GCs, and large amounts of GCs are produced 
only later by gas compression through, e.g. disk instabilities and/or 
external interactions (see the discussion in \S\ref{ssec:formation-of-sc}).
Even considering the extreme ranges of the mass and age distributions
predicted by our model (light shades in Figs.~\ref{fig:mass_func} and \ref{fig:gc_history}), 
the observed distributions of M51 are still outliers.
This highlights the importance of explicitly modeling effects of galaxy 
interactions on the formation of GCs. Such interactations can complement and 
even dominate the source of turbulence in the ISM, and may produce a 
density distribution different from that based on the assumption of the
single Mach number given by Eq.~\eqref{eq:mach-from-vvir}.

%Fig.~\ref{fig:gc-disruption-fraction} shows the distribution of metallicity, 
%birth redshift, and mass of GCs at formation and at present.

\subsection{The size-mass relation} \label{ssec:size-mass}

\begin{figure} \centering
    \includegraphics[width=0.99\columnwidth]{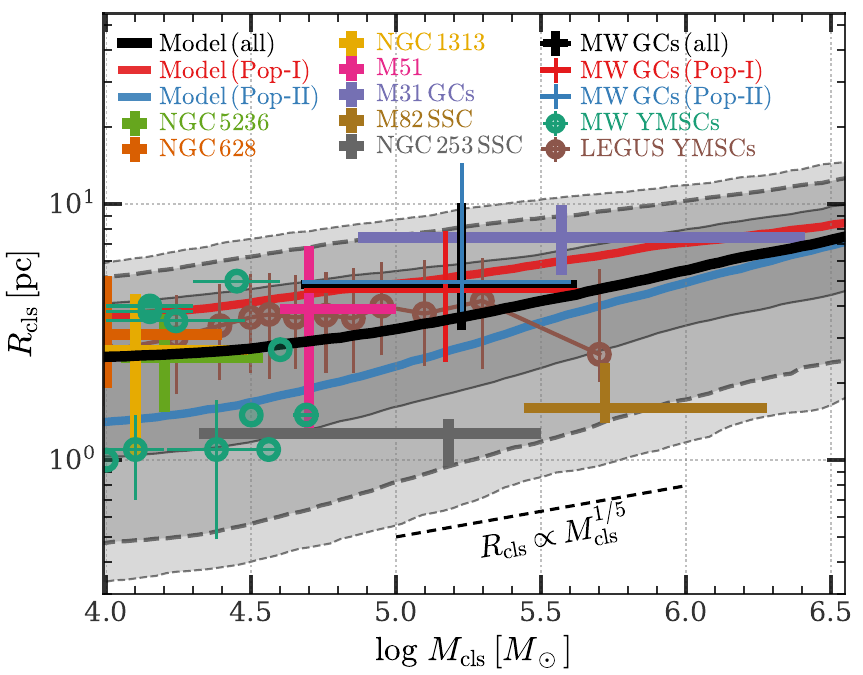}
    \caption{
        Size-mass relation of GCs. 
        {\bf Thick black} curve shows the model prediction
        of the birth-time median size at a given mass 
        by stacking all GCs formed in the histories of central galaxies with 
        $10^{11.5} \leqslant M_{{\rm v}, z=0}/(\msun) < 10^{12.5}$.
        Both in-situ and ex-situ GCs are included in the analysis.
        Here $R_{\rm cls}$ is defined as the half-mass radius of GC.
        Shaded areas, from inner to outer, 
        encompass 1, 2 and 3-$\sigma$ ranges, respectively.
        {\bf Red} and {\bf blue} curves show relations of 
        Pop-I and Pop-II GCs, respectively.
        Star clusters compied by \citet{krumholzStarClustersCosmic2019} 
        from observations for individual galaxies also shown:
        Each {\bf cross marker} represents the median and 1-$\sigma$ range of 
        all observed massive ($M_* \gtrsim 10^4 \Msun$) star clusters in a 
        given galaxy; each {\bf green circle} represents a single YMSC in MW,
        with the error bar indicating the uncertainty for this star cluster.
        For MW GCs, we compile the catalog of 
        \citet[2010 edition]{harrisCatalogParametersGlobular1996}
        and \citet{baumgardtCatalogueMassesStructural2018}, and show 
        the results for all, Pop-I and Pop-II of them by 
        {\bf black}, {\bf red} and {\bf blue cross markers}, respectively.
        {\bf Brown circles} with error bars are running median and 1-$\sigma$ 
        uncertainty for a large sample of YMSCs in the LEGUS survey 
        obtained by \citet{brownRadiiYoungStar2021}.
        The observed size-mass trend of GC is very weak and the 
        scatter of GC size is large even in a single galaxy. 
        The power-law index of our prediction, due to the adoption of 
        wind feedback, is about $1/5$, as indicated by the {\bf black dashed} line
        (see Eq.~\ref{eq:gc-size-mass-vs-feedbacks}), consistent with 
        the observation of \citet{brownRadiiYoungStar2021} for YMSCs.
        The present-day size-mass relation of GCs is expected to be
        similar if GCs evolve along their birth-time relation.
        See \S\ref{ssec:size-mass} for a detailed discussion 
        of this figure.
    }
    \label{fig:mass_size}
\end{figure}

In addition to marginal distributions of individual GC properties, 
a number of joint distributions have also been shown in the literature to 
characterize the interconnection among different physical processes. 
One of such relations is the GC size-mass relation, or more formally the 
size distribution conditioned on the mass, $p(R_{\rm cls} | M_{\rm cls})$.
The importance of this relation can be understood by applying 
the virial theorem to individual bound star clusters.
At a given GC mass, the binding-energy distribution of GCs 
at the birth-time depends on their stellar density profiles which are related to the 
histories of gas dissipation (cooling). 
The subsequent dynamical processes of a GC, such as mass segregation, 
binary hardening, black-hole heating, core collapse, evaporation and tidal 
heating, can cause changes in the distribution of stars and 
exchanges between the cluster and its environment \citep[see][for a review]{portegieszwartYoungMassiveStar2010}.
The histories of these processes can thus leave imprints on the size of a GC,
and shape the final size-mass relation of GCs.
% and energy injection, as well as to other GC properties through these processes.
%Examples of this argument have been shown for the inference of 
%the cooling and fragmentation of SGC in the formation 
%of dynamically hot galaxies \citep{chenTwophaseModelGalaxy2024a}
%the energy injection from mergers with bound orbits in the size expansion of early-type galaxies
%\citep{shenSizeDistributionGalaxies2003MassSizeRelation},
%and the the energy release from stellar feedback in the early, bursty 
%stage in the formation of cored dark matter halo profile
%\citep{el-badryBreathingFIREHow2016}, while here the size-mass 
%relation of GCs is used to infer the stellar feedback at sub-cloud scale.

Fig.~\ref{fig:mass_size} shows the birth-time size-mass relation of GCs, 
both in-situ and ex-situ, in central galaxies with $M_{{\rm v}, z=0} \in 10^{[11.5,\,12.5]} \msun$.
Results for other halo masses are similar and not shown here.
As discussed in \S\ref{ssec:sf-in-scs}, the amount of gas converted into stars before 
the dispersal of a sub-cloud by internal stellar feedback is determined by the 
balance between feedback energy and binding energy. Thus, the size-mass relation 
of GCs is a direct reflection of the gravity-feedback relation.
As our fiducial model adopts the wind feedback, the resulted 
GC size-mass relation is expected to have a power-law index of
$\approx 1/5$ (Eq.~\ref{eq:gc-size-mass-vs-feedbacks}), 
smaller than the $1/3$ power-law index expected from the assumption 
of a constant density for all GCs. This is seen from the median relation shown by the 
black solid curve in Fig.~\ref{fig:mass_size} at $M_{\rm cls} \gtrsim 10^{5.5}\Msun$. 
At lower $M_{\rm cls}$, the effect of metallicity becomes important, 
as seen from the divergence of the red (Pop-I) and blue (Pop-II) curves,  
and the median relation for all GCs becomes flatter and has a 
larger scatter. 
This is expected, as massive stars formed in metal-rich ISM evolve faster, 
strengthening the feedback and lowering the mass of Pop-I GCs relative 
to that of Pop-II GCs formed in sub-clouds with comparable mass and density.

Symbols in Fig.~\ref{fig:mass_size} show the sizes and masses of massive star 
clusters compiled by \citet{krumholzStarClustersCosmic2019} 
and obtained by \citet{brownRadiiYoungStar2021} from the HST/LEGUS survey.
For MW GCs, we take the catalog of \citet[2010 edition]{harrisCatalogParametersGlobular1996} 
for metallicity, and incorporate the size and mass measurements 
from \citet{baumgardtCatalogueMassesStructural2018}. 
We use a two-component Gaussian mixture model to fit the metallicity 
distribution of MW GCs, classify them into Pop-I and Pop-II,
and show the sizes and masses separately for all, Pop-I and Pop-II GCs using 
three markers with different colors.
Note that classifications of star clusters are not clean-cut and may not 
carry particular physical significance
\citep[see discussion in, e.g. \S1 of][]{krumholzStarClustersCosmic2019}.
YMSCs traditionally refer to massive bound star clusters with young age 
($M_{\rm cls} \gtrsim 10^4 \Msun$, age$< 100\Myr$);
GCs in the MW are considered old (age $\gtrsim 6\Gyr$; see Fig.~\ref{fig:gc_history})
but those in M31 are not; star clusters in some other galaxies do 
not even have a clear classification. Despite of the wide range of age, 
nearly all the observed sizes and masses of massive clusters fall well within 
the 2-$\sigma$ range of our model prediction. The amplitude and slope 
(in logarithmic scale) of the size-mass relation obtained by \citet{brownRadiiYoungStar2021}
for LEGUS YMSCs are very close to our model prediction, thus providing
support to our model. Given the large scatter in the observed cluster size at 
given cluster mass and the significant galaxy-galaxy variation, it is still
difficult to use the observed relation to constrain different feedback channels.

Another uncertainty arises from the dynamical evolution of GCs,
which can not only change their masses, as modeled in \S\ref{ssec:dyn-evol}, 
but also modifies their sizes. As our model does not track the change of the 
GC size during the dynamical evolution, it is unclear whether or not the present-day 
size-mass relation is different from that at the birth time. Some studies have 
suggested a `two-phase' scenario for the size evolution in the tidal field 
\citep[e.g.][]{gielesDistinctionStarClusters2011}: an early `expansion-dominated'
phase before the cluster becomes Roche lobe filling, and a 
later `evaporation-dominated' phase. The starting point and the duration 
of the two phases thus depend on the initial conditions for cluster formation. 
Using N-body simulations, \citet[see their figure 2]{kruijssenModellingFormationEvolution2011} 
found that, during the second phase, 
the evolution paths of individual star clusters in the size-mass plane 
follows the observed size-mass relation at $z=0$. 
Thus, if a population of GCs are born on the local size-mass relation, 
as suggested by our analysis, they are expected to follow the observed 
relation after the dynamical evolution. 
We note that the birth-time size-mass relation of star clusters, as well as other 
initial conditions for their subsequent dynamical evolution, 
is set up by a complex interplay of stellar physics, gas magnetohydrodynamics, 
and radiative transfer, which is not well understood and constrained 
theoretically and observationally \citep[see \S3 of][for a review]{portegieszwartYoungMassiveStar2010}.
Our model that star-clusters are born in the observed size-mass relation 
and evolve along it thus only represents a possible solution. Whether this is a 
coincidence due to model simplification or a more general result remains to be determined.
Taking MW GCs as an example, the sizes and masses of Pop-I and Pop-II GCs
have no obvious difference in median, but the sizes of 
Pop-II GCs appear to be skewed towards larger values, as seen from the 
error bar. This is potentially due to the older age, and thus 
the longer time for dynamical evolution, of Pop-II GCs.
Alternatively, it may originate from the different tidal environments 
of the two populations. In such cases, dynamical evolution is required to drive 
them off the birth-time size-mass relation which lies below the observed one.
\citet[see their figure~2 and \S3]{arcaseddaDRAGONIISimulationsEvolution2024} 
used N-body simulations and reported such a possible path for the size-mass 
evolution: star clusters are initially small and under-filling the Roche lobes, 
but expand to the observed range of sizes within a timescale of $\lesssim 1\Gyr$ 
due to internal dynamical effects.

\subsection{Connection to host galaxies and halos} 
\label{ssec:gcs-mass}

\begin{figure} \centering
    \includegraphics[width=0.99\columnwidth]{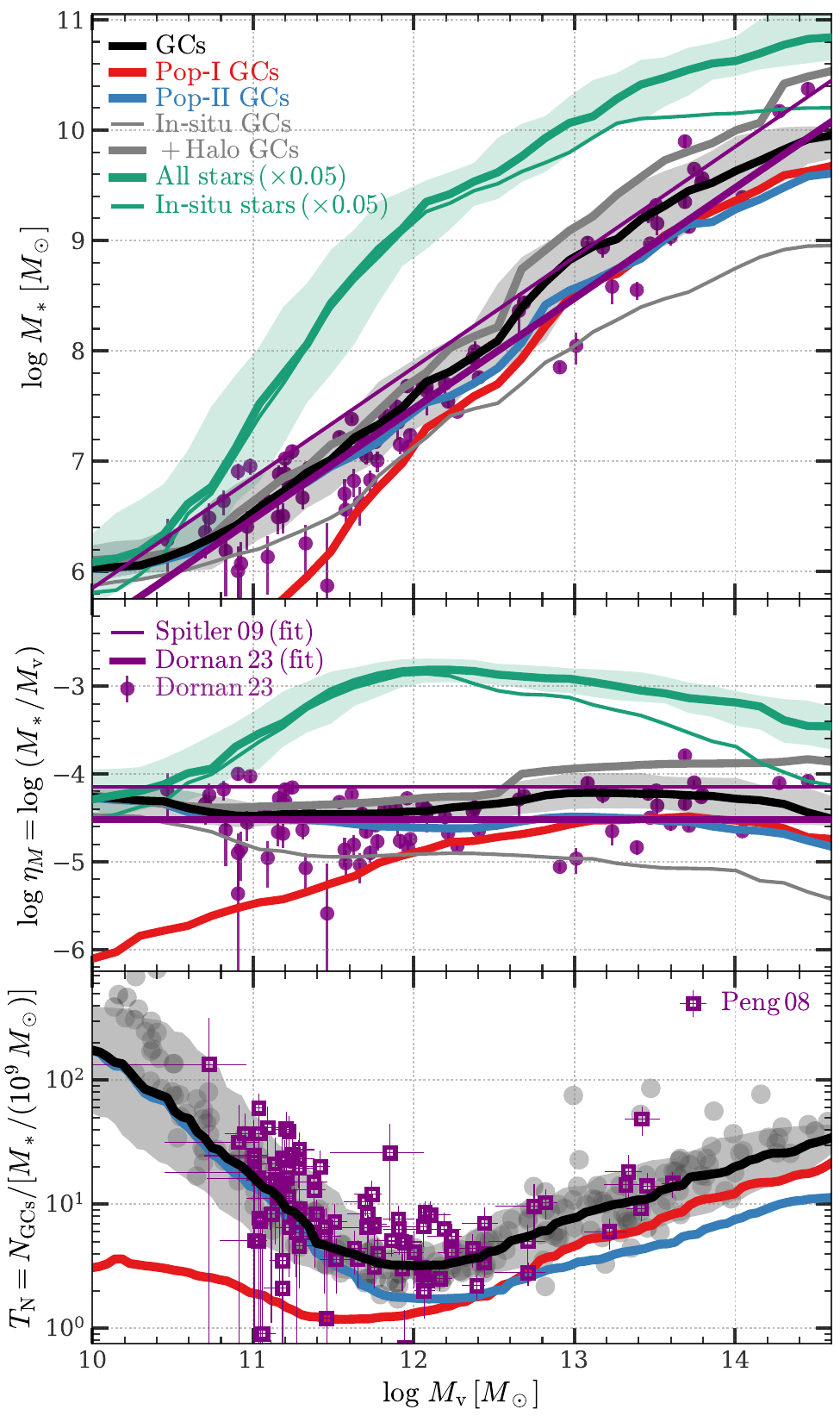}
    \caption{
        Total mass and number of GCs in galaxies at $z=0$. All results 
        are shown as a function of host halo mass, $M_{\rm v}$.
        {\bf Top panel} shows $M_*^{\rm (GCs)}$, the total mass of GCs
        within individual galaxies.
        {\bf Middle panel} shows $\eta_M$, the ratio of 
        $M_*^{\rm (GCs)}$ and $M_{\rm v}$.
        {\bf Bottom panel} shows GC frequency, $T_{N}$, defined as the 
        number of GCs, $N_{\rm GCs}$, per $10^9 \Msun$ stellar mass of 
        the host galaxy.
        {\bf Thick black curve} is obtained by 
        including all (in-situ + ex-situ) GCs within the central 
        galaxy.
        {\bf Red} and {\bf blue curves} are obtained by only including 
        Pop-I and Pop-II galactic GCs, respectively. 
        {\bf Thin grey curve} is obtained by only including
        in-situ galactic GCs.
        {\bf Grey dots} show the results for galactic GCs of 
        individual galaxies.
        {\bf Thick grey curve} is obtained by including 
        both galactic GCs and halo GCs (in satellite galaxies).
        For reference, 
        {\bf green curves} show $M_*$, the stellar mass 
        of all stars (GC and non-GC), 
        {\bf thin} for in-situ only ($f_{\rm merge} = 0$) and {\bf thick} 
        for in-situ plus ex-situ ($f_{\rm merge} = 1$),
        both scaled by 5\% for clarity.
        {\bf Shaded areas} represent the 1-$\sigma$ ranges of the results.
        For comparison, {\bf purple dots} show the observational results 
        by \citet{dornanInvestigatingGCSMRelation2023}
        for individual galaxies, and {\bf thick purple line}
        shows their linear fit, $M_*^{\rm (GCs)} = 3.0\times 10^{-5} M_{\rm v}$.
        {\bf Thin purple line} shows the linear fit, 
        $\log M_{\rm v} = 4.15 + \log M_*^{\rm (GCs)}$, obtained by 
        \citet{spitlerNewMethodEstimating2009}.
        {\bf Purple squares} show the results obtained
        by \citet{pengACSVirgoCluster2008} for individual galaxies
        from HST/ACS Virgo Cluster Survey,
        where we have converted their stellar mass to halo mass by the 
        fitted relation obtained from our modeled galaxies. See 
        \S\ref{ssec:gcs-mass} for a detailed discussion of this figure.
    }
    \label{fig:m_gc_vs_mhalo}
\end{figure}

The systematic change in the distribution function of GCs 
with different host halo mass, as discussed in \S\ref{ssec:mass-function}, 
indicates a strong connection between the GC population and the host galaxy and halo. 
An intriguing observational fact is the linear relation between the total mass of 
the GC population, $M_*^{\rm (GCs)}$, and the host halo mass, $M_{\rm v}$, discovered 
by \citet{blakesleeDependenceGlobularCluster1997,blakesleeGlobularClustersDense1999}.
Follow-up studies have extended this relation to galaxies with wide ranges of masses 
\citep[e.g.][]{mclaughlinEfficiencyGlobularCluster1999,
spitlerNewMethodEstimating2009,
harrisCatalogGlobularCluster2013,
hudsonDarkMatterHalos2014,
harrisDarkMatterHalos2015,
harrisGalacticDarkMatter2017,burkertHighprecisionDarkHalo2020,
forbesGlobularClustersComa2020,dornanInvestigatingGCSMRelation2023},
and found that it is obeyed by galaxies of different masses, from dwarfs to 
massive ellipticals. A key question for modeling the GC population is 
how such a relation is established by physical processes.

% Here we add a superscript (GCs) to M_* to indicate the mass of GCs. 
% $M_*$ without the superscript is the total stellar mass of the galaxy (GC + non-GC).
The top panel of Fig.~\ref{fig:m_gc_vs_mhalo} shows $M_*^{\rm (GCs)}$,
the mass of the GC population obtained by summing up the masses of individual GCs 
within the galaxy, as a function of halo mass, $M_{\rm v}$.
Here, the analysis includes central galaxies at $z=0$ and the GCs survive in them. 
For convenience, the middle panel shows $\eta_M$, the GC mass fraction 
defined as the ratio between  $M_*^{\rm (GCs)}$ and $M_{\rm v}$.
For galactic GCs (in-situ + ex-situ; shown by the black curve), a linear 
relation, $M_*^{\rm (GCs)} \propto M_{\rm v}$, is seen over the 
entire range of halo mass considered here. 
The 1-$\sigma$ residual, as shown by the grey shaded area, 
is about $0.3\dex$, comparable to the 
root mean square error of $0.35\dex$ reported by 
\citet[see their \S5]{dornanInvestigatingGCSMRelation2023}.
Including only the in-situ component
decreases $M_*^{\rm (GCs)}$ by a halo-mass-dependent factor, 
from negligibly small at $M_{\rm v} \approx 10^{10}\Msun$, to about 
$0.5\dex$ at $10^{12}\Msun$ and about $1\dex$ at $ \gtrsim 10^{14}\Msun$. 
%The ex-situ contribution predicted by our model
%is slightly larger than that obtained by \citet[see their figure 3]{choksiOriginsScalingRelations2019} 
%from an empirical model, probably due to bent shape of the  
%$M_*^{\rm (GCs)}$-$M_{\rm v}$ relation they 
Including additionally halo GCs (formed in satellite galaxies) increases $M_*^{\rm (GCs)}$ 
by a halo-mass-dependent factor, from negligibly small at $M_{\rm v} \lesssim 10^{12.5}\Msun$,
to about $0.5\dex$ at $\gtrsim 10^{14}\Msun$, consistent with the results for the 
intracluster GCs obtained by 
\citet{spitlerNewMethodEstimating2009} and \citet{hudsonDarkMatterHalos2014}.
Separating galactic GCs into Pop-I and Pop-II components reveals the effects of 
metal enrichment on the two channels of GC formation, 
as already discussed in \S\ref{ssec:mass-function} using 
the mass function: below $M_{\rm v} \approx 10^{12.5}\Msun$, 
Pop-II GCs are dominating, while Pop-I GCs become comparable in abundance 
to and then slightly more abundant than Pop-II GCs above this halo mass.
The difference between the two populations in the $M_*^{\rm (GCs)}$-$M_{\rm v}$ relation
found here is very similar to the result obtained by 
\citet[see their figure 7]{harrisDarkMatterHalos2015}, and we 
will quantify this in more detail in \S\ref{ssec:metal-bimodal}.

The measurement of the $M_*^{\rm (GCs)}$-$M_{\rm v}$ relation in observations
has to deal with a number of systematic uncertainties in, e.g.,
the selection of GCs, sample incompleteness, etc.  
%the contamination by diffuse light,
%the mask of satellite galaxies,
%the integration boundary of the GC system,
%the average mass of mass-to-light ratio of GCs,
%the halo mass estimation
%and the fitting strategy,
%each of them relying on a calibration or some assumptions, 
%which may not be homogeneous among different studies.
Collecting data published in the literature, we found that observational 
results generally agree with each other within about $0.3\dex$, 
and the internal uncertainties originated from errors of individual 
data points 
%and the residual of the fitting after 
%other assumptions are fixed,
are at a similar level \citep[see e.g. figure 8 of][]{dornanInvestigatingGCSMRelation2023}.
We therefore take two representative results from \citet{spitlerNewMethodEstimating2009}
and \citet{dornanInvestigatingGCSMRelation2023},
and show them in the first two panels of Fig.~\ref{fig:m_gc_vs_mhalo}
by the thin and thick purple lines, respectively. For comparison, 
we also show results for individual galaxies obtained by \citet{dornanInvestigatingGCSMRelation2023} 
using purple dots. 
The result produced by the model, shown by the black curve, matches the observational 
data well over the full range of the halo mass. This is partly because of our 
calibration with observations (see Appendix \ref{sec:calibration}),
and partly because of the two underlying mechanisms in the model
that shape the linearity between $M_*^{\rm (GCs)}$ and $M_{\rm v}$, as detailed below.

The origin of the linear relation between $M_*^{\rm (GCs)}$ and $M_{\rm v}$
can be understood by comparing different curves in Fig.~\ref{fig:m_gc_vs_mhalo} produced by our model.
Firstly, we note that, as shown by the green curves in the first two panels,
the total stellar mass, $M_*$, is a non-linear function of $M_{\rm v}$
\citep[see also observations of, e.g.][]{yangEVOLUTIONGALAXYDARK2012,
behrooziAVERAGESTARFORMATION2013}, with $M_*/M_{\rm v}$ varying  
by more than $1\dex$ over the range of halo mass considered here.
Thus, the relation between $M_*^{\rm (GCs)}$ and $M_{\rm *}$ must 
be non-linear, as clearly shown by the GC frequency,
defined as the number of GCs per $10^9 \Msun$ of the host galaxy,
versus $M_{\rm v}$ in the bottom panel for galactic GCs produced by our 
model and obtained by \citet{pengACSVirgoCluster2008} from observational data.
The frequency of Pop-II (metal-poor) GCs closely follows that 
of the total population, with only small deviations above the MW mass. 
Pop-I is very different in that it follows the frequency of the total 
population only at the high end of the halo mass.
The proximity of the Pop-II GCs to the total population suggests that 
the key to understanding the linear relation is to understand such a relation 
for Pop-II GCs.
At $M_{\rm v} \lesssim 10^{12} \Msun$ where the in-situ Pop-II
component dominates, the condition for a halo to host a significant
number of Pop-II GCs is given by Eq.~\eqref{eq:criteron-pop-ii}
and shown by the blue shaded area in Fig.~\ref{fig:gc-criteria}.
The negative correlation between $M_{\rm v}$ and $z$, which defines
the boundary of this condition, leads to a narrower time window for 
a more massive halo to host GCs. It turns out that the metal enrichment 
of the ISM, which is positively correlated with galaxy mass and halo mass (Fig.~\ref{fig:mzr}), 
plays an important role in blocking the Pop-II channel of GC formation in massive halos
and in shaping the linear relation between $M_*^{\rm (GCs)}$ and $M_{\rm v}$.
At $M_{\rm v} \gtrsim 10^{12} \Msun$ where mergers become 
more frequent, this proportionality is preserved by the central limit theorem, 
as Pop-II GCs in satellite galaxies follow this linear relation and so does their 
summation. Thus, the origin of the observed $M_*^{\rm (GCs)}$-$M_{\rm v}$ relation
is a combination of {\em the initial linear relation set up by the metal enrichment 
process and the central limit theorem that preserves it.}

A similar conclusion for the importance of the central limit theorem
in shaping the  $M_*^{\rm (GCs)}$-$M_{\rm v}$ relation was reached by 
\citet[see also \citealt{valenzuelaGlobularClusterNumbers2021}]{boylan-kolchinGlobularClusterdarkMatter2017} using a simple additive model for 
the formation of Pop-II GCs. In their model, an initial linear relation is 
assumed to populate tiny progenitor halos with GCs at $z \approx 6$, 
and the growth of a GC is followed by mass acquisition along its merger tree.
Alternative explanations have also been proposed, e.g. by assuming a larger
gas fraction and an enhanced GC formation rate, for halos with lower mass
\citep[e.g.][]{choksiFormationGlobularCluster2018,
choksiOriginsScalingRelations2019,
el-badryFormationHierarchicalAssembly2019,
chenModelingKinematicsGlobular2022,
chenFormationGlobularClusters2023,chenCatalogueModelStar2024}.
Future observations to measure the gas fraction, density and metallicity, 
the reconstruction of the formation histories of GCs 
in low-$z$ galaxies (see e.g. Fig.~\ref{fig:gc_history}), 
and to directly identify GC-forming galaxies at high $z$ 
\citep[e.g.][]{vanzellaEarlyResultsGLASSJWST2022,
vanzellaJWSTNIRCamProbes2023,
welchRELICSSmallscaleStar2023,
linMetalenrichedNeutralGas2023,
claeyssensStarFormationSmallest2023,
adamoBoundStarClusters2024,
messaPropertiesBrightestYoung2024,
fujimotoPrimordialRotatingDisk2024,
mowlaFireflySparkleEarliest2024}, will be critical to distinguish different scenarios.

\subsection{The metallicity dichotomy} 
\label{ssec:metal-bimodal}

\begin{figure*} \centering
    \includegraphics[width=0.95\textwidth]{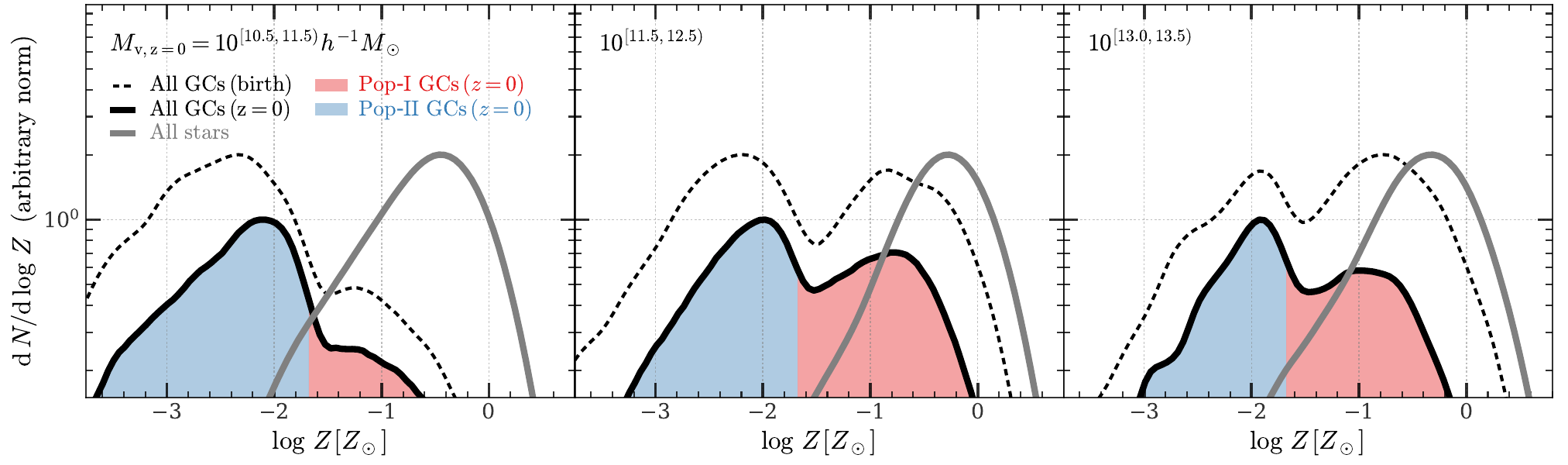}
    \caption{
        Metallicity distribution of GCs. {\bf Three panels} show 
        the results for GCs in central galaxies with 
        different halo masses at $z=0$. All galactic GCs (in-situ and ex-situ)
        are included in the analysis.
        {\bf Black solid} curve, {\bf red} shade and {\bf blue} shade
        show the distributions of all, Pop-I and Pop-II GCs, respectively,
        survived until $z=0$. {\bf Black dashed} curve is obtained 
        by including all GCs formed in the history.
        For comparison, {\bf grey} curve shows the metallicity distribution
        of all stars (GC and non-GC), obtained by using all sub-clouds 
        formed in the history, weighted by their stellar masses.
        Distributions are arbitrarily normalized for clarity.
        See \S\ref{ssec:metal-bimodal} for the detailed description
        of this figure.
    }
    \label{fig:metal_dist}
\end{figure*}

\begin{figure*} \centering
    \includegraphics[width=0.95\textwidth]{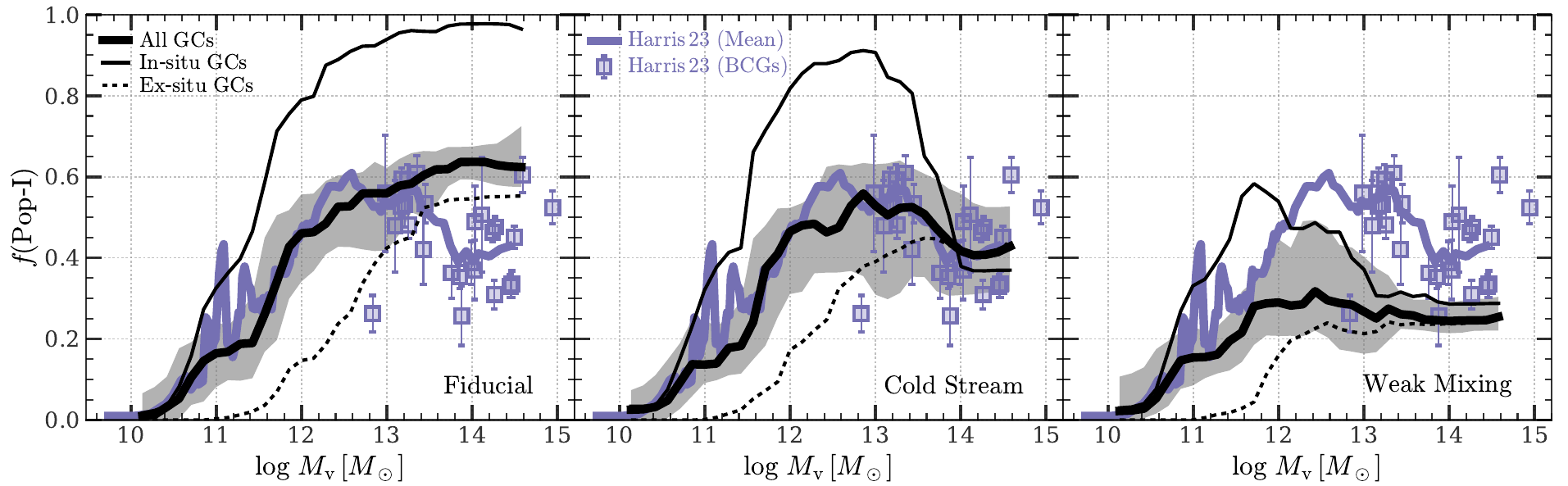}
    \caption{
        Pop-I GC fraction, $f(\text{Pop-I})$, as a function of halo mass at $z \approx 0$.
        {\bf Three panels} show the results for three model variants, 
        respectively: {\bf left} for the fiducial, {\bf center} for 
        a variant that mimics the pristine inflow for stream-fed 
        galaxies in massive halos at high redshift, 
        and {\bf right} for a variant with more relaxed criteria
        for a galaxy to be stream-fed.
        In all panels,
        {\bf black curves} show the median results of our model: 
        {\bf thick} for all galactic GCs, with a shaded area indicating 
        the 1-$\sigma$ range, 
        and {\bf thin} and {\bf dotted} for the in-situ
        and ex-situ components, respectively.
        {\bf Purple curve} is the running mean compiled by 
        \citet{harrisPhotometricSurveyGlobular2023} from observations,
        while {\bf purple squares} represent the results for individual
        brightest cluster galaxies. 
        The observed drop of $f(\text{Pop-I})$ at high halo mass, 
        as our inference, is the fossils of the cold streams
        at high redshift. See \S\ref{ssec:metal-bimodal} for details.
    }
    \label{fig:f_red_vs_mhalo}
\end{figure*}

To demonstrate the relative contribution to the GC formation of the two channels,
in Fig.~\ref{fig:metal_dist} we show the metallicity distribution of galactic GCs 
in central galaxies with different halo masses at $z=0$.
The distribution for all survived GCs (black solid curve) is clearly bimodal, 
which is a consequence of the two channels of GC formation.
For low-mass halos ($M_{\rm v} \sim 10^{11} \Msun$), Pop-II GCs (blue curve) dominates
the total number, while Pop-I GCs (red curve) are just starting to appear in the 
metal-rich end of the distribution. For MW-size halos ($M_{\rm v} \sim 10^{12}\Msun$), 
The Pop-I channel becomes important, and the two populations are nearly 
equal in number. For more massive halos ($M_{\rm v} \gtrsim 10^{13} \Msun$),
the relative contribution from the two channels appears to be stabilized, 
which is an outcome of the slightly increased Pop-I GC fraction at birth 
(black dashed curve) due to more enriched ISM, as compared to
MW-size halos, and the stronger effect of tidal disruption due to 
the longer time for mass loss.
For all halos and for both Pop-I and Pop-II, the metallicity of GCs is 
significantly lower than that of the total stellar component (grey curve), 
which reflects the preference of GC formation in a high-density environment 
in the early Universe (see Eqs.~\ref{eq:criteron-pop-i}
and \ref{eq:criteron-pop-ii}). Such a bimodal distribution of GCs has been 
found by observations in nearly all galaxies, using either direct metallicity 
measurements from spectra \citep[e.g.][]{fahrionFornax3DProject2020a}, or 
conversion from color indices \citep[e.g.][]{harrisPhotometricSurveyGlobular2023}. 

To quantity the systematic change in the contribution of the two channels
for halos with different masses,
in Fig.~\ref{fig:f_red_vs_mhalo}, we show the fraction of Pop-I (metal-rich) GCs, 
defined as the number ratio between Pop-I GCs and the total galactic GCs:
\begin{equation}
    f(\text{Pop-I}) = \frac{
        N_{\rm GCs}(\text{Pop-I})
    }{
        N_{\rm GCs}
    }\,,
\end{equation}
as a function of halo mass. Both in-situ and ex-situ components are included in the analysis, 
and the results are shown for central galaxies at $z=0$.
Our model predicts a monotonically increasing Pop-I fraction with 
increasing halo mass, regardless of the parameter choices. 
The fiducial model (left panel, black thick curve) is calibrated by the 
observation of \citet[purple curve]{harrisPhotometricSurveyGlobular2023} at 
$M_{\rm v} \lesssim 10^{12.5}\Msun$ (see Appendix \ref{sec:calibration}),
and thus matches the observation in this mass range. The Pop-I
fraction is near zero at $10^{10} \Msun$;  increases rapidly 
to about $0.5$ at $10^{12} \Msun$; becomes saturated and slowly increases 
towards higher halo mass. This is consistent with the inferences from  
the effects of metal enrichment on the mass functions (\S\ref{ssec:mass-function}),
on the masses of GC systems (\S\ref{ssec:gcs-mass}), and 
on the metallicity distributions (Fig.~\ref{fig:metal_dist}). 
The reason for the dominance of Pop-II GCs at $M_{\rm v} < 10^{11}\Msun$ 
can also be inferred from the GC formation criteria shown in
Fig.~\ref{fig:gc-criteria}, where the mass assembly histories
of halos with $M_{{\rm v}, z=0} \lesssim 10^{11} \Msun$
do not intersect with the Pop-I channel (the region covered only by red).
Similar results have been obtained by, 
e.g. \citet[see their figure 7]{choksiOriginsScalingRelations2019}
and \citet[see their figure 9]{el-badryFormationHierarchicalAssembly2019}.
In more detail, we separate galactic GCs into in-situ and ex-situ
components, and show their Pop-I fractions separately by the 
thin and dotted black curves, respectively.
These two components show similar monotonic behavior to the total 
population in the Pop-I fraction, but with different 
amplitude and different halo mass for saturation.
This difference can be explained by the fact that progenitors of 
satellites in general have lower halo mass than those of centrals at a given 
redshift, and their metal enrichment is systematically delayed, 
as seen by comparing the three panels of Fig.~\ref{fig:history_of_masses}.

The observational results from \citet{harrisPhotometricSurveyGlobular2023} 
cover a wide range of halo mass, with the BCGs highlighted by individual 
purple squares. At $M_{\rm v} \approx 10^{12.5}\Msun$, 
the observed Pop-I fraction begins to decline
and drops to about $0.4$ at $10^{14}\Msun$, while the result of fiducial model
continues to increase.
This significant discrepancy can arise from
a number of sources. The first possibility is the systematic error in the
observation. As many of the BCGs are quite distant, individual GCs within them 
are point sources even viewed by HST, and have to be distinguished 
from individual stars by, e.g., the color-magnitude diagram with 
trade-offs between completeness and contamination.
The crowded environment in the inner regions of BCGs can also hide their GCs,
and thus statistical inferences have to rely on the identification of the 
turnover point of the GC luminosity function and assumptions on the 
completeness curve. These systematics combined may lead to an underestimate of the
Pop-I fraction in BCGs. The second possibility is the difference in the 
definition of Pop-I GCs between our model and the observation.
The two populations in each BCG are obtained by
\citet{harrisPhotometricSurveyGlobular2023} 
using a bi-Gaussian fitting to the distribution of the stellar 
$[{\rm Fe/H}]$ converted from a color index. 
In our model, on the other hand, the populations are classified 
using a physical criterion on the gas-phase oxygen abundance of 
the sub-clouds that host GCs (see \S\ref{ssec:gc-channels}). 
This difference in definition is not likely to be the main reason for 
the discrepancy, however, as the Pop-I fraction is related to the normalization of the two 
components of a bimodal distribution, rather than the metallicity of 
individual GCs. A linear conversion between oxygen and iron abundances for the alpha 
enhancement, such as that adopted by, e.g. 
\citet[see their appendix]{chenCatalogueModelStar2024}, does not change 
the Pop-I fraction. We have also tried to use a bimodal fitting, 
instead of our physical criterion, to classify the two populations, and 
found no significant difference in the result.
The final possibility is that the observed drop of the Pop-I fraction is
physical and reflects a deficiency of our model, as we discuss in the following.

One simplification of our model, as discussed in \S3.4 of 
\citetalias{moTwophaseModelGalaxy2024}, is the assumption of spherically symmetric
accretion of gas into the SGC. Galaxies in the real Universe, however, 
can be fed by filamentary inflows of cold streams. The thin, cold streams of 
pristine gas are expected to penetrate the IGM, with only little mixing with the 
enriched outflow \citep[see e.g. an analytical estimation by][]{liFeedbackfreeStarburstsCosmic2024}.
High resolution numerical simulations 
\citep[e.g.][]{mandelkerColdFilamentaryAccretion2018} 
suggest that dense clumps can form within such streams, leading to 
even less mixing with the enriched gas.
Thus, stream-fed galaxies may have a higher fraction of Pop-II GCs 
originated from cold-stream inflows. To test the effects of such 
accretion on GC formation, we design a suite of variants to the fiducial model
by lowering the metallicity of sub-clouds in stream-fed galaxies. Formally, we define 
a galaxy to be stream-fed when it is in the fast phase of halo accretion, and if
\begin{equation} \label{eq:stream-fed-criteria}
    \eta_{\rm s}(M_{\rm v}, z) \equiv \frac{ M_{\rm v} }{10^{10.8}\Msun} 
    \left(\frac{1+z}{10}\right)^{6.2} \geq \eta_{\rm s,0}\,,
\end{equation}
where $\eta_{\rm s, 0}$ is a free-parameter.
The requirement for fast accretion is motivated by hydrodynamic
simulations, which found that cold gas streams are seeded by 
dark matter filaments outside the virial radius of fast-accreting halos 
\citep[e.g.][]{danovichFourPhasesAngularmomentum2015}. 
The parameterization of $\eta_{\rm s}$ is derived by 
\citet{dekelEfficientFormationMassive2023} to identify galaxies that 
are `feedback-free', and is used by \citet{liFeedbackfreeStarburstsCosmic2024} 
to model `feedback-free' galaxies.
From their arguments, a galaxy with $\eta_{\rm s} \gtrsim 1$ is purely fed by 
cold streams, and a lower value of $\eta_{\rm s}$ leads to a galaxy with more 
spherical accretion. For a stream-fed galaxy, we lower the metallicity of 
each sub-cloud by randomly assigning it a metallicity, $Z_{\rm sc}$, 
drawn from the distribution
\begin{equation} \label{eq:pdf-Z-sc}
    \dv{N_{\rm sc}}{Z_{\rm sc}} \propto Z_{\rm sc}^{\beta_{\rm sc, Z}}\,,
\end{equation}
in the range of $Z_{\rm sc} \leq Z$, where $Z$ is the metallicity of the
SGC predicted by the fiducial model (\S\ref{ssec:metal}), 
and $\beta_{\rm sc,Z}$ is a free-parameter.

With this modification, we define two variants as follows. The first one,
referred to as the `Cold Stream' variant, has 
$\eta_{\rm s,0} = 0.2$ and  $\beta_{\rm sc, Z} = -0.35$.
This variant is obtained on the basis of the fiducial model whose calibration
is presented in Appendix \ref{sec:calibration}, with additional tuning 
of $\eta_{\rm s, 0}$ and $\beta_{\rm sc, Z}$ so that the Pop-I fraction
matches those of the BCGs at $M_{\rm v} \gtrsim 10^{12.5}\Msun$
obtained by \citet{harrisPhotometricSurveyGlobular2023}.
The second variant, referred to as the `Weak Mixing' variant, has
$\eta_{\rm s, 0}=0.02$, a more relaxed criterion for 
a galaxy to be stream-fed, and the same value of $\beta_{\rm sc,Z}$ 
as the first variant. The results of these two variants
for the Pop-I GC fraction are shown in the central and right panels of 
Fig.~\ref{fig:f_red_vs_mhalo}, respectively. 
The `Cold Stream' variant appears to match the observations well 
owing to parameter tuning. An inspection of $f(\text{Pop-I})$ for the 
in-situ and ex-situ components suggests that cold streams mainly affect 
the in-situ formation in massive halos, which is a consequence of the
mass and redshift dependence of $\eta_{\rm s}$. The `Weak Mixing' variant, 
on the other hand, produces too many metal-poor sub-clouds in low-mass halos, 
and thus under-predicts the Pop-I fraction in most halos.

Given the small sample size and the large uncertainty in identifying 
$f(\text{Pop-I})$ GCs from individual BCGs, it is still premature to make 
definite conclusions on the origin of the observed drop of the Pop-I fraction
at high halo mass. Observations of GCs at higher redshift, such as those made by JWST 
at $z \approx 0.3$ \citep{harrisPhotometricSurveyGlobular2023,
harrisJWSTPhotometryGlobular2024}, or at even higher redshift via strong 
lensing \citep{vanzellaEarlyResultsGLASSJWST2022,
vanzellaJWSTNIRCamProbes2023,
welchRELICSSmallscaleStar2023,
linMetalenrichedNeutralGas2023,
claeyssensStarFormationSmallest2023,
adamoBoundStarClusters2024,
messaPropertiesBrightestYoung2024,
fujimotoPrimordialRotatingDisk2024,
mowlaFireflySparkleEarliest2024},
will provide more clues.
Gas dynamics and metallicity measured via multiple tracers 
\citep[e.g.][]{wangDiscoveryStronglyInverted2019,wangEarlyResultsGLASSJWST2022,zhangInspiralingStreamsEnriched2023,
linMetalenrichedNeutralGas2023,venturiGasphaseMetallicityGradients2024} have unveiled from a small number of 
systems that pristine gas inflows can invert the radial gradient of 
the metallicity profile. A larger sample of such measurements is needed to 
quantify the importance of the stream-fed population of galaxies.

% (i) gas-rich mergers is thought to be one important channel to 
% induce the formation of GCs, but more likely metal-rich GCs in massive halos, 
% because the ex-situ GCs (first panel) carried by satellite galaxies 
% have already been enriched, with a red fraction larger than 0.5.
% (ii) random fluctuation of metallicity in the galactic gas is hard 
% to produce a massive sub-cloud ($M_{\rm sc} \sim 10^6 \Msun$) 
% with so low metallicity ($Z \sim 0.02 \Zsun$).
% Simulation work also reports the fragmentation and star formation 
% of metal-poor inflow filaments \citep{mandelkerColdFilamentaryAccretion2018}. 

\subsection{Spatial distribution}
\label{ssec:spatial-dist}

\begin{figure} \centering
    \includegraphics[width=0.95\columnwidth]{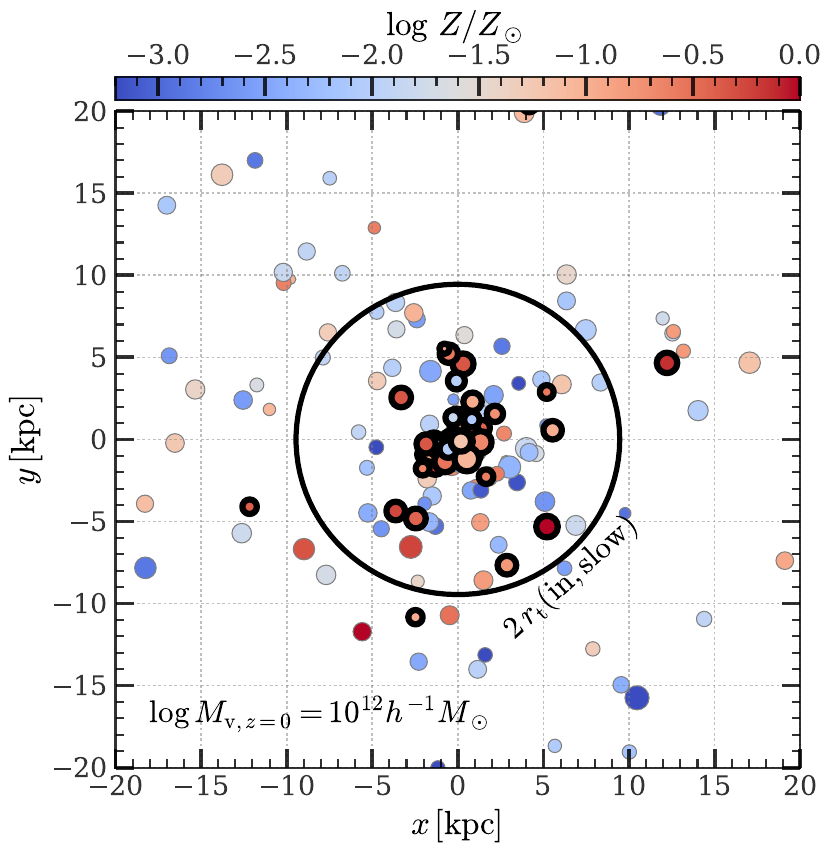}
    \caption{
        Projected spatial distribution of GCs within a MW-size galaxy
        with $M_{\rm v} = 10^{12} \msun$ at $z=0$.
        All galactic GCs 
        predicted by our model and resampled with the {\sc BSampling}
        method (see \S\ref{ssec:impl}) are shown. 
        Each {\bf dot} represents 
        a survived GC, with {\bf bold} and {\bf thin} edges indicating 
        their in-situ and ex-situ origins, respectively.
        The {\bf size} of each circle is proportional to $R_{\rm cls}$, 
        the half-mass radius of the GC (see \S\ref{ssec:sf-in-scs}).
        Metallicity of each GC is {\bf color-coded} according to the
        color bar on the top.
        The {\bf big circle} indicates two times the exponential truncation 
        radius of the in-situ, slow-phase GC system of this galaxy. 
        See \S\ref{ssec:spatial-dist} for a detailed discussion 
        of this figure.
    }
    \label{fig:spatial-example}
\end{figure}

\begin{figure*} \centering
    \includegraphics[width=\textwidth]{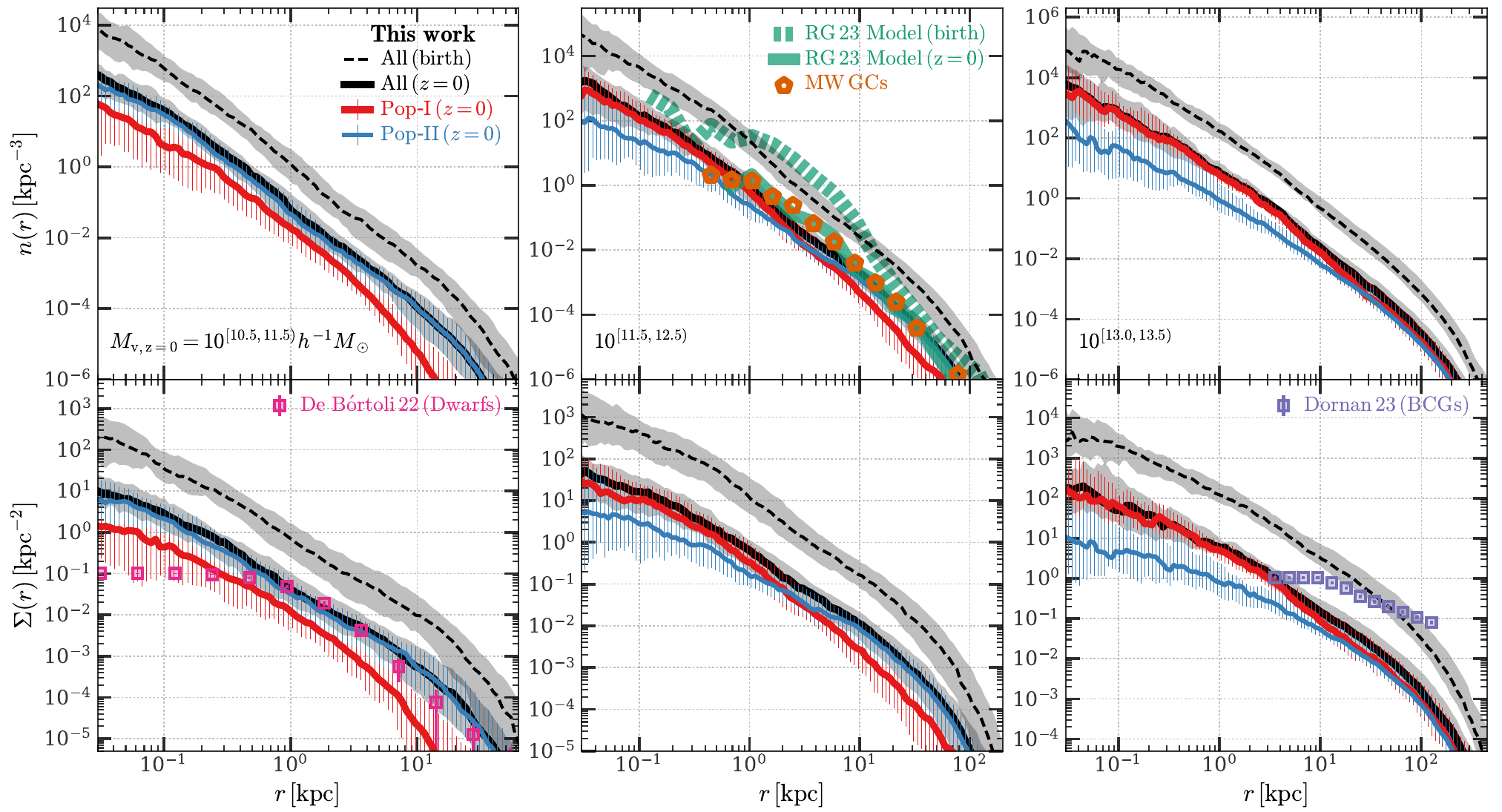}
    \caption{
        {\bf Upper row}: number density profile of galactic GCs.
        {\bf Lower row}: surface number density profile of galactic GCs.
        {\bf Three columns} show 
        the results for central galaxies with different halo masses 
        at $z=0$. Both in-situ and ex-situ components are included.
        In each panel, {\bf solid black}, {\bf red} and {\bf blue} curves are for 
        all, Pop-I and Pop-II GCs, respectively, all including 
        only survived GCs.
        {\bf Dashed black} curve includes all GCs at formation 
        (see \S\ref{ssec:impl} for the selection).
        Each curve shows the median of GC systems, 
        with error bars or shaded areas indicating the 1-$\sigma$ 
        range.
        {\bf Green} curves are obtained by 
        \citet{rodriguezGreatBallsFIRE2023} using the model of GC formation
        developed by \citet{grudicGreatBallsFIRE2023} and a star-by-star 
        model of cluster evolution, applied to a MW-size galaxy in 
        the {\sc Fire-2} simulation. The solid and dashed curves include 
        their GCs at birth and survived at $z=0$, respectively.
        {\bf Orange} markers are obtained by their compilation of MW GCs 
        based on the catalog of \citet[2010 edition]{harrisCatalogParametersGlobular1996}.
        {\bf Pink} markers are staked results for 7 sets of dwarf galaxies,
        taken from the observations by 
        \citet[including those obtained earlier by \citealt{casoScalingRelationsGlobular2019}]{debortoliScalingRelationsGlobular2022}.
        {\bf Purple} markers are stacked results for 4 BCGs, taken from 
        the observations by \citet{dornanInvestigatingGCSMRelation2023}.
        See \S\ref{ssec:spatial-dist} for a detailed 
        discussion of this figure.
    }
    \label{fig:profile}
\end{figure*}

Complementary to the age distribution (\S\ref{ssec:mass-function})
that describes when GCs formed, the spatial distribution of GCs
in their host galaxies and halos contains information about 
their birthplaces and provides additional insights into the conditions and mechanisms of GC formation.
In observations, this information can be obtained as long as resolved images are available.
To predict the spatial distribution from the first principle is more challenging, as it requires precise modeling of the host environment, 
including small-scale 
perturbations from close encounters of density clumps, and precisely tracking the orbits of individual GCs.
One approach is to attach each newly formed star cluster to a neighbor young stellar particle that serves as an environment tracer
\citep[e.g.][]{
pfefferEMOSAICSProjectSimulating2018,
chenModelingKinematicsGlobular2022,
rodriguezGreatBallsFIRE2023}. 
This approach, however, can only be applied to the `post-processing'
models based on high-resolution hydrodynamical simulations, because 
it needs stellar particles to trace the orbits.
Such details are not available in the large-volume, 
dark-matter-only simulations used here.
Meanwhile, the tracer-based method cannot deal with 
dynamical friction and orbital decay of star clusters
in a self-consistent way, and analytical approximations are 
still necessary.

% The difficulty is mainly due to: (i) Star clusters generally do not follow 
% circular orbits, and have a distribution of ellipticity. As tidal 
% disruption is a non-linear process, the effective orbit of a star cluster
% in the tidal disruption is not a simple time average of its orbit.
% (ii) The orbit can evolve, if the mass distribution of the host galaxy is 
% not uniform or evolves with time. This is particularly significant in 
% a halo during the fast accretion phase, where the orbits of star clusters 
% in the central galaxy can be significantly perturbed.
% (iii) Star clusters in a satellite can be stripped away, become intracluster
% component or mixed with the central galaxy. This is most 
% significant near the core of the halo, where tidal force is strong.

Here we adopt a semi-empirical approach to assign a spatial location 
to each modeled star cluster, with the formulation and 
parameters motivated physically and calibrated by observations. We note that
the purpose here is to statistically reproduce the profile of GCs in 
their host galaxies at a given redshift, instead of tracking 
the evolution of individual orbits.
For each galaxy at the given redshift $z$, we partition its survived GCs 
into three components: (A) the in-situ component formed during the fast phase;
(B) the in-situ component formed during the slow phase; (C) the 
ex-situ component accreted from merged companion galaxies.
A GC is assigned a galactocentric distance, $r$, according to the number density profile 
\begin{equation} \label{eq:n-r-gc}
    n(r) \propto \frac{1}{r^2} \exp\left(
            - \frac{ r }{ r_{\rm t} }
        \right)\,,
\end{equation}
where GCs belonging to the same component share a common parameter 
$r_{\rm t}$ that describes the characteristic size of the 
distribution of the GC component. 
The isothermal form, $1/r^2$,
is motivated by the expectation that the fragmentation of an SGC 
is associated with effective cooling that keeps the temperature to be 
$10^4\Kelvin$ at the formation of dense sub-clouds (\S\ref{ssec:formation-of-sc}).
The exponential truncation ensures that the distribution 
of galactic GCs has a finite extent.
Once $r$ is sampled for a GC, the angular location is randomly 
drawn from a uniform distribution on a unit sphere.

The typical value of $r_{\rm t}$ for each of the three components
can be obtained by assuming that GCs follow the profile of 
diffuse stellar components. 
For the in-situ GC system formed during the fast-phase, its 
characteristic size is expected to be the size of the stellar 
bulge, which is discussed in \citetalias{chenTwophaseModelGalaxy2024a} 
and can be modeled to be about $ 0.01 R_{\rm f}$, where 
$R_{\rm f}$ is the halo virial radius at the end of the 
fast phase. 
We thus set $r_{\rm t}^\text{(in,fast)} = 0.01 R_{\rm f}$ 
as our fiducial choice.
For the in-situ GC system formed during the slow phase,
its size is expected to be the size of the stellar disk, which
is derived theoretically by, 
e.g. \citet{fallFormationRotationDisc1980},
\citet{fallGalaxyFormationComparisons1983}
and \citet{moFormationGalacticDiscs1998},
and observed to be about $0.02 R_{\rm v}$ 
with a moderate uncertainty
\citep[e.g.][]{kravtsovSizeVirialRadiusRelation2013,
huangRelationsSizesGalaxies2017,somervilleRelationshipGalaxyDark2018}.
We thus set $r_{\rm t}^\text{(in,slow)} = 0.02 R_{\rm v}$ 
as our fiducial choice.
The size of the ex-situ GC system is theoretically uncertain due to the 
complexity of tidal stripping and galaxy mergers. 
The typical value found by 
observations is about $0.1$--$0.2 R_{\rm v}$
\citep[e.g.][]{
karthaSLUGGSSurveyGlobular2014,
alabiSLUGGSSurveyMass2016,
forbesMetallicityGradientsGlobular2018,
dornanInvestigatingGCSMRelation2023},
from which we choose a conservative value 
of $r_{\rm t}^\text{(ex)} = 0.1 R_{\rm v}$, as suggested 
by \citet{dornanInvestigatingGCSMRelation2023}.

Another difference between the three components is their dynamical hotness
\citepalias[see][]{moTwophaseModelGalaxy2024}.
GCs formed during the fast phase or assembled via mergers are expected to follow random orbits, as dissipation is not sufficient to settle them down into a dynamically cold disk.
Thus, we randomly sample the galactocentric distance 
from the distribution of Eq.~\eqref{eq:n-r-gc}.
In-situ GCs formed during the slow phase live in the dynamically cold disk,
and are expected to preserve their orbits over a long time.
We thus assume that their orbits are the same as those used 
in modeling the tidal disruption, which is achieved by taking 
the $u_{\rm r}$ value sampled for Eq.~\eqref{eq:tidal-frequency}
as the percentile value, and inverse the CDF of Eq.~\eqref{eq:n-r-gc} to 
get the quantile value.

Fig.~\ref{fig:spatial-example} shows the spatial distribution of GCs 
in an example of Milky Way-size galaxies at $z=0$. Here, 
we take the survived GCs together with their \textsc{BSampling} weights 
from the model output, and resample them to get a set of 
un-weighted GCs for display (see \S\ref{ssec:impl}).
Each of the un-weighted GCs is shown by a circle, with the color representing  
the metallicity and the size proportional to the half-mass radius of the GC
(see \S\ref{ssec:sf-in-scs}). The in-situ population (highlighted by 
bold edges) has a smaller spatial extent than the 
ex-situ one, owing to the difference in their $r_{\rm t}$.
Ex-situ GCs are on average bluer, reflecting the less enriched ISM of 
their hosts at formation (see also \S\ref{ssec:metal-bimodal} and Fig.~\ref{fig:f_red_vs_mhalo}).
The sizes of in-situ GCs are also bigger, which is a consequence
of the higher amplitude of the size-mass relation for Pop-I GCs
(Fig.~\ref{fig:mass_size}) and the similar mass function of both
Pop-I and Pop-II GCs (Fig.~\ref{fig:mass_func}).

Upper panels of Fig.~\ref{fig:profile} shows the number density profiles of GCs
in central galaxies with different halo masses at $z=0$.
For the profile of all survived GCs (black solid curve), 
the amplitude increases rapidly with halo mass, as a result
of the tight, linear relation between $M_*^{\rm (GCs)}$ and 
$M_{\rm v}$ described in \S\ref{ssec:gcs-mass}. In contrast,
the shape of the profile is independent of halo mass, which is
a consequence of our model to assign positions to GCs 
(see Eq.~\ref{eq:n-r-gc}). Such a universal profile has been 
found in observations, and has been modeled using the
S\'ersic profile 
\citep{usherSLUGGSSurveyWide2013,karthaSLUGGSSurveyGlobular2014,
hudsonCorrelationSizesGlobular2018}, 
the modified Hubble profile \citep{casoScalingRelationsGlobular2019,
debortoliScalingRelationsGlobular2022,casoScalingRelationsGlobular2024},
and the power-law profile \citep{hudsonCorrelationSizesGlobular2018}.
The profile shows some flattening in the inner region, which is
due to the tidal disruption of the in-situ component in the slow 
halo assembly phase (see \S\ref{ssec:dyn-evol}). 
The disruption changes the profile of the GC distribution, reduces its 
amplitude and makes its shape less concentrated, in comparison with 
the profile of all GCs selected at their formation (black dashed curve).
As disrupted GCs can contribute to the diffuse stellar component, 
a part of the metal-poor and alpha-enhanced stellar population
observed in the Milky Way may have originated from the disrupted GCs
\citep{wallersteinAbundancesDwarfsVISurvey1962,
weinbergChemicalCartographyAPOGEE2019,
rojas-arriagadaBimodalMgFe2019,
lianAgechemicalAbundanceStructure2020}.
Separated into Pop-I and Pop-II populations, the profile of
Pop-I (red curve) is more centrally concentrated than that of 
the Pop-II (blue curve), which owes to the extension of the profile 
caused by metal-poor GCs brought in by mergers. 
This prediction is consistent with the observed negative metallicity 
gradient of GCs in most galaxies \citep[e.g.][]{usherSLUGGSSurveyWide2013,
fahrionFornax3DProject2020,
harikaneComprehensiveStudyGalaxies2023}.
The outer profiles ($r\sim 100\Kpc$) for both Pop-I and Pop-II
GCs in massive galaxies (right panel) are very similar,   
indicating a flat metallicity gradient in the outskirt 
of massive galaxies. This is indeed seen in the BCG 
sample of \citet[see their figure 24]{harrisPhotometricSurveyGlobular2023}.

The orange symbols in the upper-center panel of Fig.~\ref{fig:profile} show
the number density profile of GCs in the Milky Way, obtained from the data 
compiled by \citet{rodriguezGreatBallsFIRE2023} using the catalog of 
\citet{harrisCatalogParametersGlobular1996,
harrisMassiveStarClusters2010}.
Our prediction for the profile of Milky Way-size galaxies
is in agreement with the observation. 
A noticeable difference is seen around $r = 5\Kpc$, 
roughly the peak position of the number distribution, $\dd{\rm N_{\rm gc}}/\dd{\rm r}$ 
\citep[see e.g. figure 10 of][]{rodriguezGreatBallsFIRE2023}.
This radial range turns out to be the joint region of 
the in-situ and ex-situ components, where the interaction
between infalling satellites and the central galaxy is expected 
to be strong. The less relaxed orbits of the GCs stripped from 
the satellites, and the perturbed orbits of GCs in the central,
may not be described well by the simple superposition assumed in our 
model. Other uncertainties may come from the scatter in $r_{\rm t}$ 
for a given halo mass, which can change the radial distribution 
of GCs systematically. Such scatter may be caused by variances in
both the assembly history and structure among individual 
halos of a given mass
\citep{liangConnectionGalaxyMorphology2024}, or by  
baryonic processes that can modify the dynamics of the galaxy 
\citep{el-badryBreathingFIREHow2016,jiangDarkmatterHaloSpin2019,
hopkinsWhatCausesFormation2023}.
%As no consensus has been reached for the driven factor of the 
%scatter in galaxy size, we do not try to fine-tune our model to 
%match the radial distribution of GCs for individual systems.
Towards the innermost region, the observed profile for the Milky Way 
GCs is truncated at $r \approx 0.5\Kpc$, while that predicted by our model
continues to rise. As the observed number of Milky Way GCs in the innermost 
bin is very small, Poisson noise dominates the observational result. 
Our prediction, which is obtained by averaging over a large 
sample of galaxies, gives a robust estimate of the inner 
profile. The predicted larger scatter in the inner region, 
as represented by the shaded areas and error bars, 
is caused by sampling error.

The square markers in the lower panels of Fig.~\ref{fig:profile} 
show the surface number density profiles of GCs in dwarf galaxies and BCGs,
obtained by staking the observations of
\citet[including those obtained earlier by \citealt{casoScalingRelationsGlobular2019}]{debortoliScalingRelationsGlobular2022}
and \citet{dornanInvestigatingGCSMRelation2023}, respectively,
in comparison with our model prediction in the same ranges of halo mass.
The error bars here are obtained by randomly sampling the profiles, 
taking into account uncertainties of their fitting parameters.
For BCGs, we follow the Appendix of \citet{dornanInvestigatingGCSMRelation2023}
to correct the number of GCs fainter than the detection limit
by assuming a Gaussian distribution for the magnitudes of GCs.
The inner profile of observed dwarfs is flatter at $r \lesssim 1\Kpc$ than the model.
Part of such discrepancy may come from the uncertainty in the correction 
of incomplete detection, while the rest, as suggested by 
\citet{casoScalingRelationsGlobular2019}, 
can be physically caused by the efficient tidal disruption of GCs in 
inner regions of galaxies \citep[see also][for an interpretation with alternative gravity model]{bilekStudyGravitationalFields2019,bilekGalacticAccelerationScale2024}.
The observed GCs around BCGs have an outer profile $\Sigma \sim r^{-1}$, much shallower
than the model prediction. Such a profile indicates that the integration of the GC number 
over radius is divergent, so that a significant contribution can come from 
GCs in the intracluster space due to projection.
In both cases, it is unknown whether or not the simple assumption of 
a truncated isothermal profile in our model (Eq.~\ref{eq:n-r-gc}) can 
capture the variation with halo mass. A forward modeling with precise 
recipes for GC orbits and disruption
\citep[e.g.][]{vitralPropertiesGlobularClusters2022,ferroneETidalGCsProjectModeling2023} 
within realistic galactic environments
is needed to make accurate model-observation comparison.
This can be achieved by embedding GC-formation and evolution models in 
cosmological/zoom-in simulations \citep[e.g.][]{rodriguezGreatBallsFIRE2023}, 
or by parametric methods with calibrations using simulations and 
observations \citep[e.g.][]{liangConstrainDarkmatterDistribution2024}.

\subsection{Globular clusters in the cosmological hierarchy}

The key aspect of our model is its ability to efficiently populate halos with GCs 
on cosmological scales.
Fig.~\ref{fig:tpcf} shows the two-point correlations of GC populations, 
$\xi_{\rm gc,gc}$, obtained from the GC catalog in the full box of TNG100-1-Dark.
The correlation functions cover a large range of spatial scales, from the scale 
of cosmic web ($\gtrsim 1\Mpc$, as indicated by the size of 
Coma cluster, $R_{\rm 200c, Coma}$), to the scale of a sub-cloud ($\lesssim 10 \pc$, as 
indicated by the typical size of sub-clouds, $R_{\rm sc}$), from 
the regime dominated by dark matter, to the one dominated by baryons. 
Thus, the GC-GC correlation functions encode rich information about structure 
formation at different scales and throughout the cosmic history, 
and can potentially serve as powerful summary statistics to constrain
models of galaxy formation, especially subgrid assumptions in hydrodynamic
simulations \citep[see e.g.][]{liEffectsSubgridModels2020}.

Following the terminology of the halo occupation distribution model
\citep[e.g.][]{jingSpatialCorrelationFunction1998,
berlindHaloOccupationDistribution2002,
guoRedshiftspaceClusteringSDSS2015,
guoModellingGalaxyClustering2016,
yuanAbacusHODHighlyEfficient2022,
qinHIHODHalo2022}, we can partition the GC-GC correlation function 
by characteristic scales in the structure hierarchy: a `two-halo' term that represents 
the cross-correlation of GCs in different halos; a `one-halo' term or `two-galaxy' term
that represents the correlation of GCs between two galaxies within a single halo; 
a `one-galaxy' term that represents the auto-correlation of GCs within one galaxy.

The grey squares in Fig.~\ref{fig:tpcf} show $\xi_{\rm g,g}$, the galaxy-galaxy correlation functions
obtained by \citet{shiMappingRealSpace2016} for galaxies of different luminosities. 
These correlation functions have been corrected for the redshift-space distortion, 
and thus approximate the real-space distribution of galaxies.
Our $\xi_{\rm gc,gc}$ for the whole sample of GCs (black curve)
tightly follows the $\xi_{\rm g,g}$ of galaxies, and shows a clear 
separation between the two-halo and one-halo terms at sub $\Mpc$ scales.
These indicate that the large-scale spatial distribution of GCs
reflects the large-scale dark matter distribution. 
The orange plus symbols in Fig.~\ref{fig:tpcf} show $\xi_{\rm g,gc}$, 
the galaxy-GC cross-correlation function obtained by using the Milky Way GC profile 
shown in Fig.~\ref{fig:profile}. This function extends the spatial coverage from the 
halo scale to the galaxy scale, as indicated by $R_{\rm e,GCS}$ and $R_{\rm 80,galaxy}$.
The $\xi_{\rm gc,gc}$ for all GCs is significantly different from that of 
the Milky Way, $\xi_{\rm g,gc}$, indicating that Milky Way-size galaxies are not 
the dominant hosts of GCs in the Universe. This is in contrast to the total stellar 
population, for which Milky Way-size halos dominate the stellar mass throughout 
almost the entire cosmic history 
\citepalias[see e.g. figure 16 and \S5.4 of][]{moTwophaseModelGalaxy2024}. 
One reason for this is clearly shown in Fig.~\ref{fig:gc_history}
and discussed in \S\ref{ssec:mass-function},
where the cosmic GC formation is found to peak at $z\approx 5$,
much earlier than that of the overall stellar population,
and in Fig.~\ref{fig:m_gc_vs_mhalo} and \S\ref{ssec:gcs-mass},
where the GC occupation frequency is found to be the lowest at 
about Milky Way mass. 
At $r \lesssim 1\Kpc$, only a small number of GCs were found by observations 
in the MW, and they are even more difficult to find in extra-galactic systems due to 
the crowded environment. The recent discovery of GCs by Gaia in the bulge of MW
\citep{palmaAnalysisPhysicalNature2019,minnitiIntriguingGlobularCluster2021}
provides a new opportunity to extend the radial distribution of GCs 
to the core of the Galaxy, and to test the GC-GC correlation function
predicted by our model. Another common approach is to use the projected
correlation function, which is measurable in more distant galaxies. 
Observations targeting at dwarf galaxies, especially the (ultra-)diffuse 
dwarfs \citep[e.g.][]{janssensGlobularClustersStar2022,forbesUltraDiffuseGalaxies2024,
jonesGasrichFieldUltradiffuse2023}, may also help us avoid
the crowded environment, and extend the GC clustering measurement to smaller scales.

The red and blue curves in Fig.~\ref{fig:tpcf} show the GC-GC correlation
function for Pop-I and Pop-II GCs, respectively. 
The correlation function of Pop-II GCs closely follows that of all GCs at all scales, 
indicating that Pop-II GCs dominate the GC population.
At the two-halo scale, the correlation function of both populations
are similar to that of all GCs, indicating that the cosmic 
large-scale environment plays a secondary role in GC formation
once the halo-scale environment is fixed. At and below the halo scale,
the correlation function of Pop-II GCs appears to be in two parts, 
separated at the scale of $1$--$5\Kpc$, coincident with the core radius, 
$R_{\rm core,CGS}$, found by \citet{casoScalingRelationsGlobular2019}, 
\citet{debortoliScalingRelationsGlobular2022} 
and \citet{casoScalingRelationsGlobular2024} when fitting the spatial  
distribution of GCs. This characteristic scale is the transition scale
from the `two-galaxy' term to the `one-galaxy' term, and marks the
change of the dominating GC population from the ex-situ origin to the in-situ origin. 
On the other hand, Pop-I GCs are more frequently produced in massive 
galaxies, as shown in Fig.~\ref{fig:m_gc_vs_mhalo} and 
Fig.~\ref{fig:f_red_vs_mhalo}, and thus are concentrated in the 
central galaxies of dark matter halos. Consequently, the correlation function 
of Pop-I GCs falls below that of Pop-II GCs at the two-galaxy scale, 
rises rapidly, and becomes higher than that of Pop-II GCs at the one-galaxy scale. 
The crossing point between the Pop-I and Pop-II correlation functions is at about 
$R_{\rm 80, galaxy}$, the observed boundaries of galaxies at or above the Milky Way mass 
\citep[see][]{mowlaMassdependentSlopeGalaxy2019}, indicating 
the dominance of Pop-I GCs in galaxies more massive than the Milky Way 
(see Fig.~\ref{fig:f_red_vs_mhalo}).

The information encoded in $\xi_{\rm g,g}$ complements the clustering
analysis based only on galaxies, and can be used to extend the test of structure 
formation to small scales and early times. The physical processes in the early Universe 
involved in the formation of GCs, such as the fragmentation of SGC into sub-clouds, 
the compression of sub-clouds by supersonic shocks, and the contraction of sub-clouds 
due to the slow cooling of metal-poor gas (marked by the green arrows 
in Fig.~\ref{fig:tpcf}; see \S\ref{sec:gc-model} for details), are expected to 
make imprints on these compact stellar systems, and may be retrieved from 
observations in the local Universe.

\subsection{Globular clusters at high redshift}
\label{ssec:high-z}

\begin{figure*} \centering
    \includegraphics[width=1.0\textwidth]{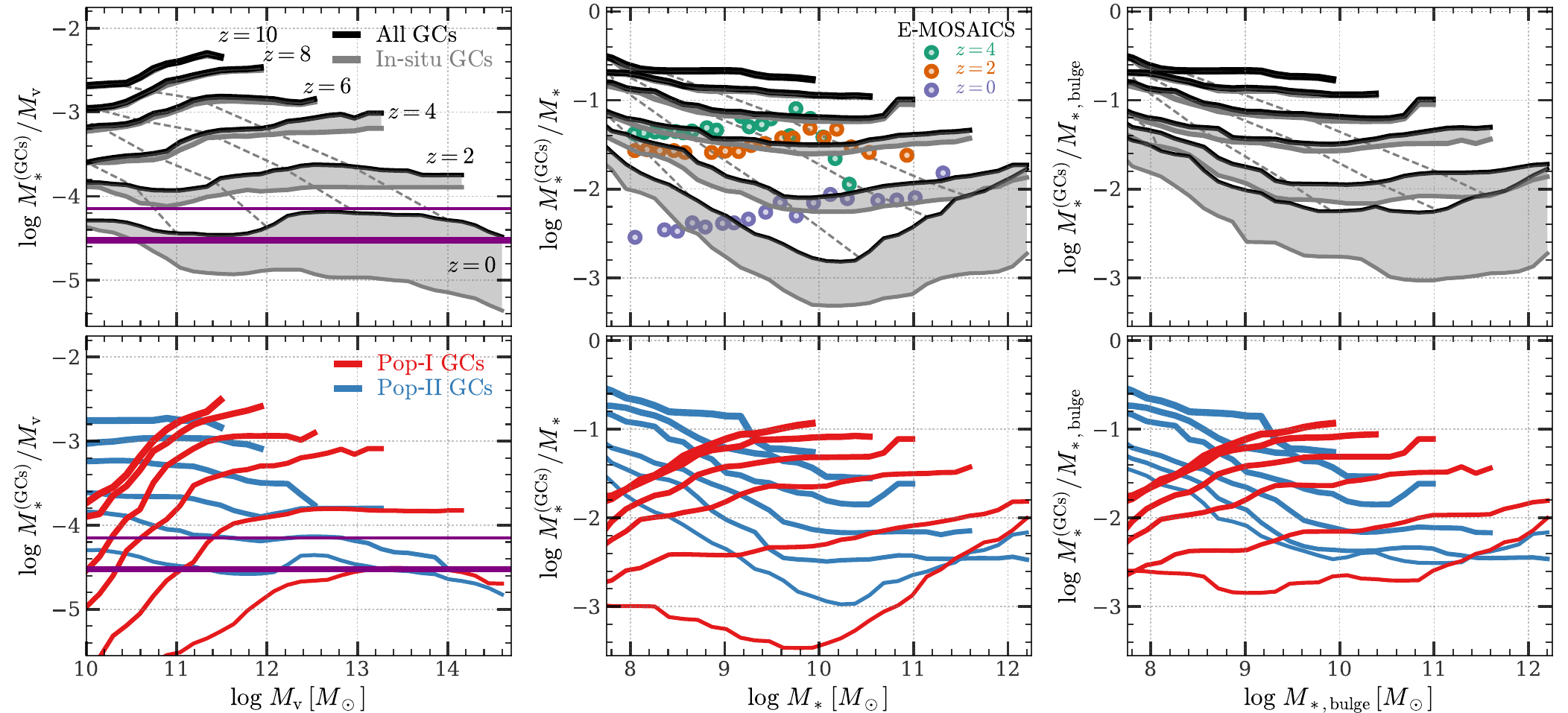}
    \caption{
        GCs mass to host mass relation, as represented by the
        the ratio of total mass of GCs ($M_*^\text{(GCs)}$) and a host mass
        (halo mass $M_{\rm v}$, stellar mass $M_*$, 
        or bulge stellar mass $M_{\rm *,bulge}$) as a function of 
        the host mass.
        Curves of the same color, from top to bottom, and from thick to thin, 
        show the results for central galaxies 
        from $z=10$ to $0$, as indicated in the first panel.
        All galactic GCs (in-situ and ex-situ) are included in the analysis.
        {\bf Upper row} shows the results for all galactic GCs ({\bf black} curve)
        and the in-situ component ({\bf grey solid} curve).
        {\bf Grey dashed} curves show the median evolution paths in the 
        panel for the main branches of halos with given masses at $z=0$, from 
        $M_{{\rm v}, z=0} = 10^{11}\Msun$ to $10^{14}\Msun$.
        {\bf Lower row} shows the results for the Pop-I ({\bf red} curve) 
        and Pop-II GCs ({\bf blue} curve).
        {\bf Purple} lines in the first column indicate the fittings
        of observations, as described 
        in Fig.~\ref{fig:m_gc_vs_mhalo} and \S\ref{ssec:gcs-mass}.
        In the upper center panel, we also show 
        the results obtained by \citet[see their figure 4]{bastianGlobularClusterSystem2020}
        from the E-MOSAICS model at three redshifts.
        See \S\ref{ssec:high-z} for a detailed discussion of this figure.
    }
    \label{fig:high_z_m_gc_vs_mhalo}
\end{figure*}

\begin{figure} \centering
    \includegraphics[width=1.0\columnwidth]{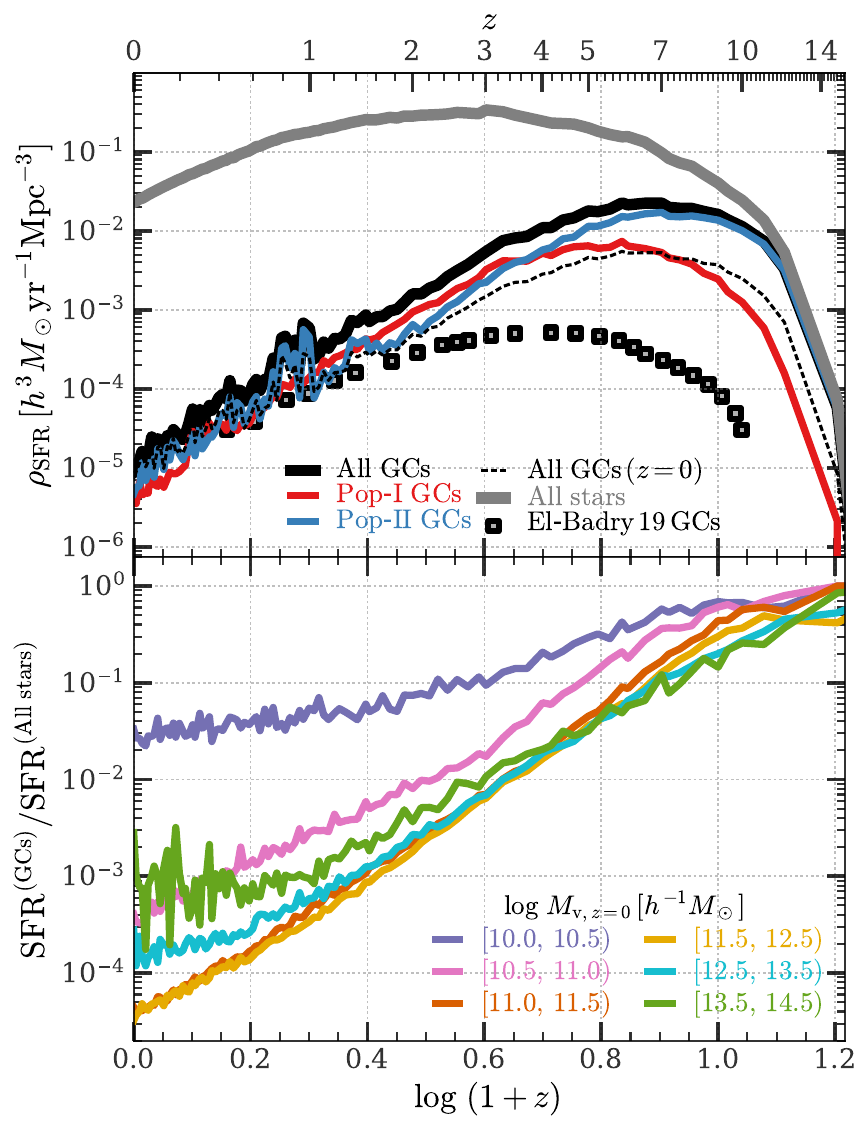}
    \caption{
        Cosmic GC formation history. 
        {\bf Upper panel} shows the cosmic star formation rate density
        $\rho_{\rm SFR}$ obtained by subhalo merger trees rooted in 
        all (central and satellite) subhalos residing in halo with 
        $M_{{\rm v}, z=0} \geqslant 10^{10} \msun$. 
        {\bf Black solid}, {\bf red} and {\bf blue} curves 
        show $\rho_{\rm SFR}$ contributed by all, Pop-I and Pop-II GCs,
        respectively, all selected and evaluated according to their 
        birth-time properties (see \S\ref{ssec:impl} for the selection).
        For comparison, {\bf black dashed} curve is obtained by only counting 
        the retained mass of GCs survived until $z=0$ (see \S\ref{ssec:dyn-evol}
        for the dynamical evolution).
        {\bf Grey} curve accounts for all stars (GC and non-GC).
        {\bf Square} markers show the fiducial results obtained by 
        \citet[see their figure 8]{el-badryFormationHierarchicalAssembly2019} using a
        semi-analytic model.
        {\bf Lower panel} shows the ratio of in-situ star formation rate 
        contributed by GCs and that by all stars 
        in the main branches of central galaxies with different 
        halo masses at $z=0$. This figure suggests that a significant
        fraction of stars in the early Universe is formed in GCs.
        See \S\ref{ssec:high-z} for a detailed discussion of this figure.
    }
    \label{fig:high_z_csfrd}
\end{figure}

\begin{figure} \centering
    \includegraphics[width=1.0\columnwidth]{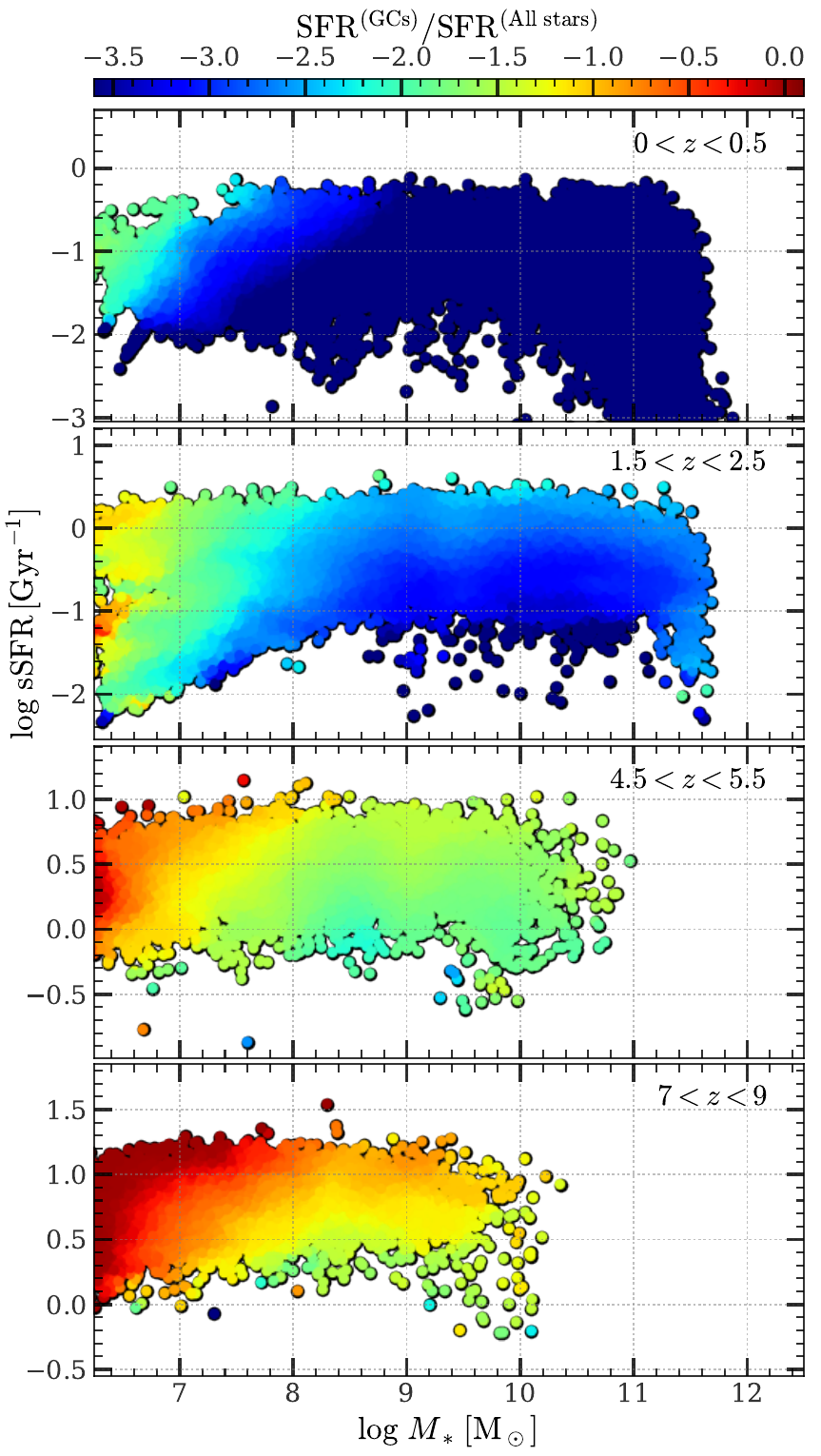}
    \caption{Fraction of in-situ star formation rate contributed by GCs, 
    color-coded according to the color bar,
    as a function of galaxy stellar mass and specific star formation rate (sSFR).
    Four panels include galaxies within different ranges of redshift,
    and each dot represents a galaxy.
    See \S\ref{ssec:high-z} for details.
    }
    \label{fig:CFE}
\end{figure}

The observation of GCs usually requires high spatial resolution so that 
GCs can be separated from other point(-like) sources. This is typically 
feasible only at low redshift, and becomes challenging
at high redshift as the angular sizes involved are small. 
However, under certain conditions, GCs can be observed at high redshift. 
One promising way is to rely on gravitational strong lensing, which can
magnify and stretch background images, making small sources
observable and spatially distinguishable. Such observations
have indeed carried out for a number of young and massive star clusters 
with sizes comparable to those of GCs up to $z\approx 10$ 
\citep[e.g.][]{vanzellaEarlyResultsGLASSJWST2022,
vanzellaJWSTNIRCamProbes2023,
welchRELICSSmallscaleStar2023,
linMetalenrichedNeutralGas2023,
claeyssensStarFormationSmallest2023,
adamoBoundStarClusters2024,
messaPropertiesBrightestYoung2024,
fujimotoPrimordialRotatingDisk2024,
mowlaFireflySparkleEarliest2024,whitakerDiscoveryAncientGlobular2025}. 
The detectability of GCs can probably be enhanced further if the initial mass 
function (IMF) relevant for high-$z$ GCs is top-heavy, as suggested by some theoretical 
studies \citep{larsonEarlyStarFormation1998,raiterPredictedUVProperties2010,
marksEvidenceTopheavyStellar2012,haghiLifetimesStarClusters2020,
chonTransitionInitialMass2021,
shardaWhenDidInitial2022} and hinted by 
some analyses based on observations \citep{zaritskyEvidenceTwoDistinct2014,
zhangStellarPopulationsDominated2018,tangMultiplePopulationsLowmass2021,
dibStellarCollisionsGlobular2022,upadhyayaEvidenceVeryMassive2024}.
Here we make some predictions for the GC population at high redshift,
and provide some guidance for future observations.

\subsubsection{Connection to host galaxies and halos}
\label{sssec:high-z-gcs-mass}

Fig.~\ref{fig:high_z_m_gc_vs_mhalo} shows the GCs mass to host mass 
relation for central galaxies at various redshifts from $z=10$ to $z=0$.
Here, the mass of GCs, $M_*^\text{(GCs)}$, includes
all galactic GCs (in-situ and ex-situ). The host masses shown 
in different columns are halo mass ($M_{\rm v}$), galaxy stellar mass 
($M_*$), and galaxy bulge stellar mass ($M_{\rm *,bulge}$), 
respectively.
The predicted total mass of Pop-II (blue) GCs is nearly a constant fraction
of the host halo mass for halos with $M_{\rm v} \lesssim 10^{10.5} \Msun$
at any given redshift (lower left panel). 
The value of this fraction, however, depends strongly on redshift, 
decreasing from $\sim 10^{-2.8}$ at $z=10$ to about $10^{-4.4}$ at $z=0$.
In contrast, the total mass of Pop-I (red) GCs increases 
with $M_{\rm v}$ in this range of halo mass, although the redshift dependence
is similar to that of Pop-II. 
The redshift dependence of both populations is mainly driven by the evolution of the SGC 
density, as shown in \S\ref{ssec:formation-of-sc},
while the difference between the two populations reflects
the increase of metal enrichment of the host galaxy with time (see Fig.~\ref{fig:mzr}).
At $M_{\rm v} \gtrsim 10^{10.5}\Msun$, the formation of in-situ Pop-II GCs
becomes inefficient due to the enriched ISM, but the ex-situ 
component brought in by mergers drives the total Pop-II GC mass
back to the linear relation with the host halo mass at $z\lesssim 4$.
This is the same conclusion reached in \S\ref{ssec:gcs-mass} from 
the halo mass dependence of different GC populations at $z=0$
shown in Fig.~\ref{fig:m_gc_vs_mhalo}. As the central limit theorem 
also applies to Pop-I GCs, their total mass in a halo also follows the linear 
relation with the halo mass in massive halos at low redshift. 
The behavior of $M_*^\text{(GCs)}$ of all GCs, as shown in the upper left panel, 
is a combination of the two populations, with the amplitude in the  
$M_*^\text{(GCs)}$ - $M_{\rm v}$ relation decreases rapidly with 
decreasing redshift. 

The second column of Fig.~\ref{fig:high_z_m_gc_vs_mhalo} shows
the relation of the total mass of GCs to the stellar mass of the
host galaxy (GC + non-GC). Interestingly, the total mass of all GCs, $M_*^\text{(GCs)}$
shown by black curves, is proportional to $M_*$ at $z \gtrsim 4$. 
At lower $z$, the decrease of $M_*^\text{(GCs)}/M_*$
with decreasing redshift appears to be the fastest at around the 
Milky Way mass, producing a non-linear relation between the two masses.  
A separation of $M_*^\text{(GCs)}$ into contributions from Pop-I (red) 
and Pop-II (blue), plotted in the lower middle panel,   
shows that the linear $M_*^\text{(GCs)}-M_*$ relation at high $z$  
seen in the upper middle panel originates from the balance between the 
two formation channels in halos of different masses
(see Fig.~\ref{fig:gc-criteria}).
A more quantitative analysis can be made using $n_{\rm sgc}\mathcal{M}_2^2$ 
given by Eq.~\eqref{eq:n-sc-post-shock-sn}. As shown there, the high-density tail of 
the sub-cloud distribution depends only weakly on halo mass (via $f_{\rm gas}$
and $f_{\rm str}$) in the SN-driven regime, and so the fraction of stellar mass 
in GCs depends strongly only on the redshift. At lower redshift, however, 
two channels diverge, with the Milky Way mass falling in the 
gap between the two channels (shown by the blank region of Fig.~\ref{fig:gc-criteria}).
For comparison, we show the results of \citet{bastianGlobularClusterSystem2020} 
as symbols in the upper-center panel. Quite different to our results, 
the linear relation between $M_*^\text{(GCs)}$ and $M_*$ in their results 
extends to $z = 0$. They argued that this linear relation is more physical than 
that between $M_*^\text{(GCs)}$ and $M_{\rm v}$. Since the largest difference 
between the predictions is at $z\sim 0$, the two models can be distinguish 
by measuring $M_*^\text{(GCs)}$ in galaxies with $M_*$ between $10^9$ and $10^{11}{\rm M}_\odot$.
The third column of Fig.~\ref{fig:high_z_m_gc_vs_mhalo} is similar 
to the second column, but with $M_*$ replaced by $M_{\rm *,bulge}$, 
the bulge mass of galaxies. Since the bulge of a galaxy typically  
represents the dynamically hot stellar component of the galaxy, 
its mass may be more closely related to $M_*^\text{(GCs)}$, as both 
are affected by early violent collapse of the halo and major mergers 
experienced by the host galaxy. This is indeed seen in the
right column, where the mass ratios at $z \lesssim 2$
become much flatter for galaxies with $M_*>10^{10} {\rm M}_\odot$ 
and for the total, Pop-I and Pop-II populations. 

For reference, the grey dashed curves in the upper row of Fig.~\ref{fig:high_z_m_gc_vs_mhalo}
show the median evolution paths for the main branches of central sub-halos of different 
masses at $z=0$. These are obtained by using the {\sc Diffmah} library 
\citep{hearinDifferentiableModelAssembly2021} to generate main-branch 
halo mass assembly histories, and using the fittings of our 
model predictions to map halo histories to different stellar masses of galaxies. 
The meshes built in this way provide a data cube, so that one can start with 
a given mass $M$, which can be $M_{\rm v}$, $M_*$ or $M_{\rm *,bulge}$, 
%a given mass pair, $(M_*^\text{(GCs)}, M)$ at given redshift, where  
%$M$ stands for $M_{\rm v}$, $M_*$ or $M_{\rm *,bulge}$,
and build the expected evolution history of the halo/stellar/GC mass components of 
the galaxy over the entire cosmic history.

\subsubsection{The cosmic formation history of globular clusters}
\label{sssec:high-z-cosmic-history}

Our results suggest that GCs are preferentially formed at high redshift. 
It is thus interesting to see how GCs contribute to the cosmic
star formation history of the universe.
The top panel of Fig.~\ref{fig:high_z_csfrd} shows $\rho_{\rm SFR}^\text{(GCs)}$, 
defined as the total mass density of stars formed in GCs per unit time. 
The peak of the SFR density from GCs (black curve) is peaked at $z\approx 6$--$7$,
significantly earlier than that of the cosmic star formation rate density of all 
stars (grey curve). This is consistent with the expectation that
the average $n_{\rm sgc}$ (see Eq.~\ref{eq:n-sgc}), 
and thus the formation rate of GCs (see \S\ref{ssec:gc-channels}), 
is higher at higher $z$. Separating  
$\rho_{\rm SFR}^\text{(GCs)}$ into contributions from the Pop-I 
and Pop-II channels, we find that the Pop-II channel dominates
the formation at $z \gtrsim 4$, and becomes equally important
as Pop-I at lower redshift. The black dashed curve shows the cosmic 
SFR density obtained by accounting only for GCs that survive to $z=0$ and 
by using only their retained stellar mass in the evaluation of 
$\rho_{\rm SFR}^\text{(GCs)}$.
This curve is significantly lower than the black solid curve at
$z \gtrsim 1$, consistent with the disruption timescales 
estimated in \S\ref{ssec:dyn-evol}. The results shown here 
is broadly consistent with the age distribution of GCs in $z = 0$ 
galaxies, as shown in Fig.~\ref{fig:gc_history} and discussed in 
\S\ref{ssec:mass-function}.
The squares show the prediction of 
\citet{el-badryFormationHierarchicalAssembly2019} using their semi-analytic 
model with fiducial parameters. The $\rho_{\rm SFR}^\text{(GCs)}$ predicted by their model 
follows the total cosmic star formation history that peaks at a lower 
redshift than that predicted by our model, because of the tight 
relation between GC formation and star formation in host galaxies 
assumed in their model. The difference between our model and theirs is 
partly due to the enhanced formation efficiency of Pop-II GCs in the 
metal-poor ISM in our model (see \S\ref{ssec:gc-channels}), and partly 
due to their assumption that the formation rate of GCs scales with
the average gas surface density of the host galaxy (see their \S2.2).

The lower panel of Fig.~\ref{fig:high_z_csfrd} shows the fraction of 
stars formed in GCs throughout the star formation histories of 
galaxies in halos of different masses at $z=0$. 
Here, the total star formation rate, ${\rm SFR}^\text{(All stars)}$, 
is obtained by the galactic-scale model described in 
\S\ref{ssec:star-smbh-formation}, and the GC star formation rate, ${\rm SFR}^\text{(GCs)}$, 
is obtained by accounting for the formation rate of all GCs (see \S\ref{ssec:impl}). 
At $z \approx 0$, only dwarf halos with $M_{\rm v}\lesssim 10^{10.5}$ can have active 
GC formation that consists of $\approx 2\%$ of total SFR, 
while more massive halos have a smaller fraction ($\lesssim 0.1\%$) 
of stars formed in GCs due to their enriched ISM with decreased 
amounts of turbulence. This is consistent with observations that 
only a small number of YMSCs are found in, e.g. the Milky Way 
\citep[see e.g. table 4 and figure 9 of][]{krumholzStarClustersCosmic2019},
and that there seems to be an excess of GCs in (ultra-)diffuse dwarf galaxies
\citep[e.g.][]{forbesGlobularClustersComa2020,
saifollahiImplicationsGalaxyFormation2022,jonesGasrichFieldUltradiffuse2023}. 

At higher redshift, the fraction of stars formed in GCs can be 
significant. At $z\gtrsim 7$, almost all halos have $\gtrsim 10\%$
of stars formed in GCs, and the fraction can reach to unity 
at $z \gtrsim 10$. Our model suggests that this enhanced GC fraction 
at high redshift is due to the high SGC density (\S\ref{ssec:collapse-gas}), 
the strong turbulence that effectively compresses sub-clouds 
(see \S\ref{ssec:formation-of-sc}), and the low metallicity environment
that prevents sub-clouds from fragmentation before becoming very dense 
(see\S\ref{ssec:cooling}). 

In Fig.~\ref{fig:CFE}, we show the fraction of in-situ star formation rate
contributed by GCs for individual galaxies, to highlight the variation of 
GC-formation efficiency in different galactic environments.
In addition to the dependencies on redshift and stellar mass, 
star-forming environment, as characterized by specific star formation rate
(sSFR, defined as SFR per unit stellar mass), also affects the fraction of
stars formed in GCs. At all redshifts and stellar masses, a higher 
sSFR leads to a higher fraction of stars formed in GCs. This is a direct
outcome of our external feedback model in the destruction of sub-clouds
(\S\ref{ssec:external-feedback}), where low-mass sub-clouds are 
preferentially destroyed by the wind feedback and a higher SFR raises 
the threshold mass for sub-clouds to survive.
A similar situation has been found observationally that the formation 
efficiency of bound star clusters appears to be higher in galaxies with
higher SFR surface density, albeit with ongoing debates on the
potential effects of, e.g. cluster selection criteria and 
sample homogeneity \citep[see \S7 of][for a review]{adamoStarClustersFar2020}.
Note that we have not defined bound star clusters in our model,
but rather adopted a conservative selection for GCs. Meanwhile, 
we have not included the small-scale interactions 
among nearby sub-clouds via different types of feedbacks and tidal forces
\citep[e.g.][]{dengRIGELSimulatingDwarf2024}.
A more complete model thus requires modeling correlations of sub-clouds in 
position and velocity spaces, which may be achieved by calibrations
using zoom-in simulations with sufficient resolution to follow ISM physics.

The large fraction of stars that can form in GCs in a high-redshift galaxy 
has important implications. Indeed, if an observed high-$z$ galaxy 
is not properly resolved and its light is dominated by young GCs, the properties 
of the galaxy may not be interpreted correctly in terms of the picture we have from local galaxies. 
For example, the S\'ersic index,  which is traditionally used as a measure of 
the concentration of light in a galaxy with a smooth profile, does not have
a clear meaning for a galaxy made up of a set of star clusters.  
The dependence of GC formation efficiency on the star-forming environment
further complicates the observations, as intense star formation usually
produces dusty ISM that can obscure the radiation from young GCs.
A basic challenge in future observations of high-$z$ galaxies is, therefore,  
to resolve galaxies down to the scales of GCs so as to identify star clusters 
reliably in a given field. To this end, our model, which predicts 
the stellar mass, star formation rate, and size of individual GCs at different redshifts 
in different halos, can be used to predict observable quantities to explore 
in detail the feasibility of observing high-$z$ GCs in current and future observations.   
We will come back to this in a future paper.
%%%%%%%%%%%%%%%%%%%%%%%%%%%%%%%%%%%%%%%%%%%%%%%%%%%%%%%%%%%%%%%%%%%%%%%%%%%%%%%%

\section{Summary and discussion}
\label{sec:summary}

% Here we summarize our main results and make further discussions.
% `Distinctive' is used to describe a special property belonging to someone, 
% while `distinct' is used to describe populations/groups that are separated in 
% some way. See e.g. https://www.britannica.com/dictionary/eb/qa/the-difference-between-distinct-and-distinctive
In this paper, we have developed a model for the formation of GCs. 
The model is based on the `two-phase' scenario presented in 
\citetalias{moTwophaseModelGalaxy2024}, which predicts that the
fast assembly of a halo in its early phase drives the formation of 
SGC that subsequently fragments into a dynamically hot system of sub-clouds.
Here we show that a fraction of these sub-clouds can reach a supernova-free regime 
via two distinctive channels, and lead to the formation of two distinct
populations of GCs. The main findings and conclusions are summarized as follows.
\begin{enumerate}
    \item The average density of an SGC is in general 
    below the typical density of GC formation at 
    $z \leq 10$ (\S\ref{ssec:collapse-gas}). 
    The fragmentation of SGC into sub-clouds (\S\ref{ssec:formation-of-sc}) 
    and compression of sub-clouds are needed to provide conditions 
    for GC formation. 

    \item There are two channels to make a sub-cloud sufficiently dense to form 
         a GC. The first is
    shock compression in a turbulent medium produced by 
    gravity and/or SN feedback. This can deform  
    the log-normal density distribution of sub-clouds, producing  
    a high-density, power-law tail, $n_{\rm sc} \gtrsim n_{\rm sgc}\mathcal{M}^2$, 
    determined by the Mach number ${\cal M}$ (\S\ref{ssec:formation-of-sc}). 
    The second channel is through radiative cooling 
    in metal-poor sub-clouds ($Z \lesssim 0.02 \Zsun$), 
    where cloud fragmentation and star formation are prohibited 
    before a sub-cloud reaches a high density $n_{\rm sf,1}$
    (\S\ref{ssec:cooling}). Both channels are able to lift the
    density of a sub-cloud to the SN-free regime,
    $n_{\rm sc}' \geqslant n_{\rm snf} \approx 10^{3.5} \perccm$, 
    to avoid disruption by SN feedback before forming a GC.

    \item According to the channel in which sub-clouds are 
    compactified, we classify GCs into two populations (\S\ref{ssec:gc-channels}):
    Pop-I, which forms in metal-rich ($n_{\rm sf,1} < n_{\rm snf}$) 
    sub-clouds with the post-shock density $n_{\rm sc} \geqslant n_{\rm snf}$,
    and Pop-II, which forms in the metal-poor 
    ($n_{\rm sf,1} \geqslant n_{\rm snf}$) sub-clouds with 
    a post-cooling density $n'_{\rm sc} > n_{\rm snf}$. The 
    halo mass and redshift where each class of GCs can actively
    form can be summarized in a GC-formation diagram, as shown
    in Fig.~\ref{fig:gc-criteria}. 
    Both channels prefer the formation of GCs at high $z$, but 
    diverge at low $z$ where 
    GC formation is suppressed in halos of different masses.

    \item After formation, a GC experiences dynamical evolution 
    driven by its tidal environment, so that it loses mass and may eventually be disrupted. 
    The dynamical disruption timescale depends on the tidal force (and thus redshift) 
    and the mass of the cluster itself (\S\ref{ssec:dyn-evol}).
\end{enumerate}

We have implemented the `two-phase' galaxy formation scenario 
and the `two-channel' GC formation into halo merger trees
produced by a N-body simulation. Our main results are summarized as follows.
\begin{enumerate}
    \item The implementation follows a hierarchical design strategy 
    so that it can be applied within a cosmological context. 
    Structures in the Universe are separated, according to their spatial scales,
    into four levels: large-scale structure of the Universe, halos, galaxies,
    and sub-clouds. Structures at a large scale are resolved 
    first, and used as boundary conditions for modeling
    structures at the next smaller scale. The overall context of the model 
    is depicted in Fig.~\ref{fig:tpcf}, where relevant structures 
    and processes are shown, in terms of the spatial correlation
    function of GCs. The detailed steps are listed in \S\ref{ssec:impl}.
    \item With calibration by observations, a number of 
    observational results of GCs in the local Universe
    are reproduced, and more results are predicted and to be tested 
    by future observations. 
    The mass and age distributions of GCs produced by the model
    cover those observed in MW and M31, but do not reproduce those seen in M51, 
    an interacting galaxy.
    The size-mass relation of GCs, with its normalization tuned to 
    match observations, can be explained
    as a result of gravity-feedback balance in the star formation of sub-clouds,
    provided that subsequent dynamical evolution does not drive GCs 
    off the birth-time relation.
    The relation between the total mass in GCs ($M_*^\text{(GCs)}$) and 
    the host halo (galaxy) mass is predicted to be linear (non-linear).
    The bimodal distribution in GC metallicity is produced naturally by 
    the two-channel formation, and the fraction of Pop-I/II GCs as 
    a function of halo mass reflects the enrichment history of the host galaxy.
    The predicted spatial profile of GCs within the host galaxy is 
    shaped by both in-situ and ex-situ, and both fast-phase and slow-phase, 
    formation of the host galaxy. The modeled 
    profiles match those of MW and M31, but show discrepancies with those
    observed for dwarfs and BCGs (\S\ref{sec:results}).
    \item The linear relation between $M_*^\text{(GCs)}$ and the host halo 
    mass is set by the metal enrichment process,
    and is preserved by the central limit theorem (\S\ref{ssec:gcs-mass}).
    \item The drop of the Pop-I (red) GC fraction in halos of high mass
    is inferred to be produced by pristine cold streams at high redshift
    (\S\ref{ssec:metal-bimodal}).
    \item A number of predictions are made for GCs at high redshift. 
    These include the relations between $M_*^\text{(GCs)}$ and the
    host mass (host halo mass, total stellar mass and bulge mass of the 
    host galaxy), the cosmic formation history of GCs in halos of different 
    final masses, and the variation of the GC formation efficiency
    with star-forming environment (\S\ref{ssec:high-z}).
%   the GC luminosity functions in UV band. Individual 
%    GCs are possible to be observed by JWST even without strong lensing, 
%    in terms of their UV luminosity. The detection rate of GCs is expected 
%    to be higher in the presence of top-heavy IMF. 
%    However, the small size make GCs hard to be distinguished from 
%    surrounding stellar associations or other GCs.
%    With strong lensing, individual GCs can be resolved, and the GC 
%    luminosity function can be obtained if a large sample has been 
%    accumulated (\S\ref{ssec:high-z}). 
\end{enumerate}

A powerful aspect of our model is its ability to predict the formation of individual GCs 
based on their host sub-clouds with sampled properties. This mimics the
`post-processing' techniques based on resolved cloud properties in high-resolution 
hydrodynamical simulations \citep{grudicModelFormationStellar2021,
grudicGreatBallsFIRE2023}, yet can be applied to a much larger volume
so as to reproduce the statistical properties of GC populations on cosmological scales.
The keys behind this approach are the use of a hierarchical strategy
to reduce the model complexity, the use of analytical approximations
to derive recipes relevant to each step of GC formation, and 
the use of the balanced sampling technique to efficiently sample sub-clouds. 
The sub-cloud-based modeling features only a small number of free parameters 
so as to avoid fine-tuning. 

A number of assumptions in the model of GC have to be verified
or extended by observations and/or hydrodynamical simulations. 
For example, the independent 
sampling of sub-cloud density and mass (\S\ref{ssec:formation-of-sc}) 
is designed to mimic the mass-independent density of virialized dark matter halos 
given by the spherical collapse model, while dissipative processes in SGCs 
may introduce a density-mass correlation. 
Meanwhile, the sub-cloud mass function can be modified by the processes 
discussed in \S\ref{ssec:formation-of-sc}, and may thus depend on the
environment. These can be explored
by hydrodynamical simulation with a resolution 
that is sufficient for resolving the formation of sub-clouds.
Another example is the role of radiative and wind feedback from 
the internal star formation of a sub-cloud. We argued that such feedback   
produces the flat size-mass relation of GCs (\S\ref{ssec:sf-in-scs}). 
These feedback channels have been simulated in hydrodynamic
simulations \citep[e.g.][]{hopkinsRadiativeStellarFeedback2020,liEffectsSubgridModels2020}, 
and can be tested by measuring the correlation of gas clouds and young star clusters
in observations.
The final example is the spatial distribution of GCs within their host galaxies, which 
is modeled semi-empirically with three-components described in \S\ref{ssec:spatial-dist}
and should be improved by directly tracing individual GCs within realistic
gravitational potential. The comparison of the spatial distribution with 
observations can be made by forward modeling that mimics the 
selection and projection in observations.

The treatment of the dynamical evolution of GCs in this paper is clearly 
simplified (\S\ref{ssec:dyn-evol}). One way to make improvements is to trace 
the environment of GCs by assigning them to particles in hydrodynamic 
simulations \citep[e.g.][]{pfefferEMOSAICSProjectSimulating2018,
chenModelingKinematicsGlobular2022,
rodriguezGreatBallsFIRE2023}. Another way is to calibrate a more 
realistic model using finer N-body simulations \citep[e.g.][]{gielesLifeCycleStar2011,
prietoDynamicalEvolutionGlobular2008,
gielesMasslossRatesStar2023}, or to run cluster simulations 
directly for a sample of GCs \citep{rodriguezGreatBallsFIRE2023}.
All these improvements and numerical techniques should be explored in the 
future. The information obtained from our modeling can be used to motivate 
and guide such explorations. 

%This paper relies on a N-body simulation with a trade-off between
%resolution and volume. To make more robust statistics for 
%massive galaxy clusters, a two-step method can be used in the future.
%First, a N-body simulation with a larger volume can be used to represent 
%halo merger trees that can be extended to the required resolution either 
%using a semi-analytic approach \citep{jiangSatGenSemianalyticalSatellite2021} or 
%a statistical learning approach
%\citep{liAIassistedSuperresolutionCosmological2021,
%chenConditionalAbundanceMatching2023,
%niAIassistedSuperresolutionCosmological2021}.
%Second, a N-body simulation constrained from real galaxy survey
%\citep{wangELUCIDEXPLORINGLOCAL2016,dolagSimulatingLOcalWeb2023,
%boruahBayesianReconstructionDark2022,
%liDifferentiableCosmologicalSimulation2022} 
%can be used to control the cosmic variance and to provide a GC catalog 
%to be directly compared with those in massive galaxy clusters.

% GC mergers.
% nuclear star clusters.
% Mini halo.
% multiple role of SN feedbacks.
% GC and large-scale environment. e.g. spatial distribution in terms of local 
% density of GCs and local density of dark matter and stars.  

% \section{Discussion and Summary}
% 
% 
\section*{Acknowledgements}

% Prepend this funding to get a cover of the expense.
We thank the anonymous referees for the useful
reports that significantly improve this paper.
YC is funded by the China Postdoctoral Science Foundation (grant No. 2022TQ0329).
This work is also supported by the National Natural Science Foundation of China 
(NSFC, Nos. 12192224, 11733004 and 11890693) and CAS Project for Young Scientists 
in Basic Research (grant No. YSBR-062).
YC thanks Kai Wang, Fangzhou Jiang, Tao Wang, Yu Rong, Enci Wang and 
Baitian Tang for their valuable insights and discussions, and thanks 
Yi Mao and Chen Chen for their technical support.
The authors would like to express their gratitude to the Tsinghua Astrophysics 
High-Performance Computing platform at Tsinghua University and the 
Supercomputer Center of the University of Science and Technology of China for 
providing the necessary computational and data storage resources that have 
significantly contributed to the research results presented in this paper.

The computations and presentations in this paper are supported by various software 
tools, including the HPC toolkits 
\softwarenamestyle[Hipp] \citep{chenHIPPHIghPerformancePackage2023}
\footnote{\url{https://github.com/ChenYangyao/hipp}}
and 
\softwarenamestyle[PyHipp]\footnote{\url{https://github.com/ChenYangyao/pyhipp}},
interactive computation environment 
\softwarenamestyle[IPython] \citep{perezIPythonSystemInteractive2007},
numerical libraries \softwarenamestyle[NumPy] \citep{harrisArrayProgrammingNumPy2020}, 
\softwarenamestyle[Astropy] \citep{
robitailleAstropyCommunityPython2013,
astropycollaborationAstropyProjectBuilding2018,
astropycollaborationAstropyProjectSustaining2022}
and \softwarenamestyle[SciPy] \citep{virtanenSciPy10Fundamental2020},
the graphical library 
\softwarenamestyle[Matplotlib] \citep{hunterMatplotlib2DGraphics2007},
the halo mass assembly history calculator 
\softwarenamestyle[Diffmah] \citep{hearinDifferentiableModelAssembly2021}\footnote{\url{https://github.com/ArgonneCPAC/diffmah}},
and the two-point correlation function calculator
\citep{sinhaCorrfuncBlazingFast2019,sinhaCORRFUNCSuiteBlazing2020}\footnote{\url{https://github.com/manodeep/Corrfunc}}.
This research has made extensive use of the arXiv and NASA’s Astrophysics Data System.
Data compilations used in this paper have been made much more accurate and 
efficient by the software \softwarenamestyle[WebPlotDigitizer]. 

Credit for the images of the structures at the top of Fig.~\ref{fig:tpcf}: 
Herschel image of gas sub-clouds around young stars 
by ESA, NASA, JPL-Caltech, AURA, NSF, STScl and Univ. of Toledo;
HST and JWST images of the NGC 628 galaxy by NASA, ESA, CSA, STScI, 
Janice Lee (STScI) and Thomas Williams (Oxford);
Simulated image of halos and cosmic large-scale structures by
\cite{liELUCIDVIIUsing2022}.

% The Acknowledgements section is not numbered. Here you can thank helpful
% colleagues, acknowledge funding agencies, telescopes and facilities used etc.

%%%%%%%%%%%%%%%%%%%%%%%%%%%%%%%%%%%%%%%%%%%%%%%%%%
\section*{Data Availability}

% The inclusion of a Data Availability Statement is a requirement for articles published in MNRAS. Data Availability Statements provide a standardised format for readers to understand the availability of data underlying the research results described in the article. The statement may refer to original data generated in the course of the study or to third-party data analysed in the article. The statement should describe and provide means of access, where possible, by linking to the data or providing the required accession numbers for the relevant databases or DOIs.

The code repository \softwarenamestyle[TwoPhaseGalaxyModel]\footnote{\url{https://github.com/ChenYangyao/two-phase-galaxy-model}}
implements the model described in this series of papers.
All data used in this paper, including data points displayed in figures
and observational results used for calibration and comparison, 
will be distributed along with the repository.
The GC catalog of MW and M31 produced by 
the empirical model of \citet{chenCatalogueModelStar2024}
is available online\footnote{\url{https://github.com/ognedin/gc_model_mw}}.
The star cluster catalogs of LEGUS survey are available online\footnote{\url{https://archive.stsci.edu/prepds/legus/}}.
The data of the IllustrisTNG project can be found at 
their website\footnote{\url{https://www.tng-project.org/}}.



\bibliographystyle{mnras}
\bibliography{references} % if your bibtex file is called example.bib

%\begin{thebibliography}{99}
%\end{thebibliography}

%%%%%%%%%%%%%%%%%%%%%%%%%%%%%%%%%%%%%%%%%%%%%%%%%%

%%%%%%%%%%%%%%%%% APPENDICES %%%%%%%%%%%%%%%%%%%%%

\appendix
\section{Balanced Sampling}\label{sec:bsampling}

\begin{figure} \centering
    \includegraphics[width=0.99\columnwidth]{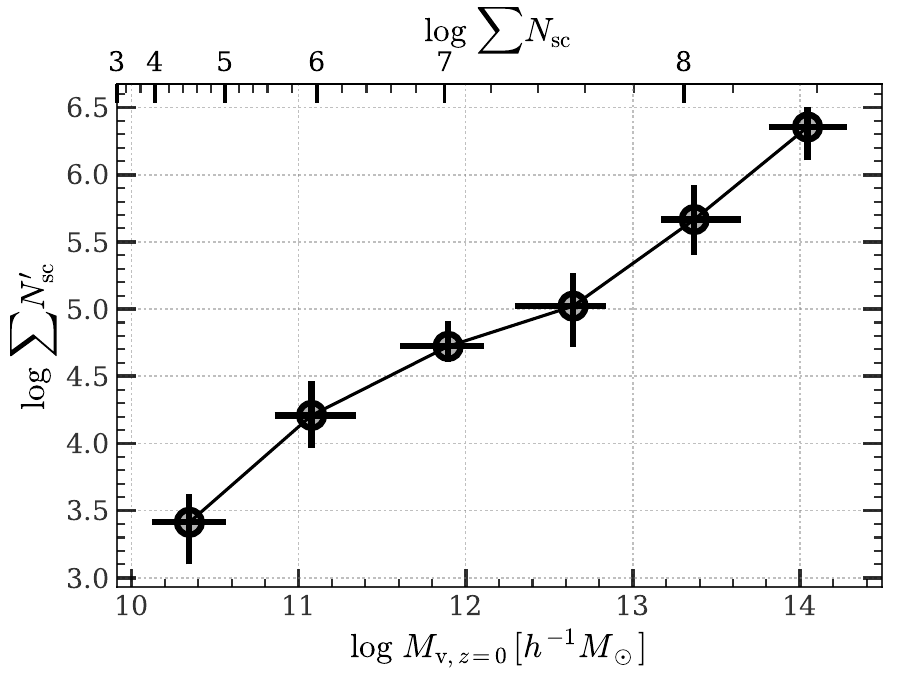}
    \caption{
        The total number of sub-clouds as a function of host 
        halo mass at $z=0$.
        Here, $\sum N_{\rm sc}'$ is the number of B-sampled sub-clouds 
        obtained for each central galaxy 
        at $z=0$ by summing the number of subclouds, $N_{\rm sc}'$, 
        formed in every progenitor in the main and side branches.
        $\sum N_{\rm sc}$ is the real number of sub-clouds.
        {\bf Markers} and {\bf error bars} show the median values and the 1-$\sigma$ 
        ranges, respectively, obtained by a number of trial runs for 
        each bin of halo mass.
        The {\sc BSampling} technique can significantly reduce the 
        computational costs while maintaining the statistical robustness.
        See Appendix~\ref{sec:bsampling} for more details.
    }
    \label{fig:b-sample-scale}
\end{figure}

As mentioned in \S\ref{ssec:impl}, a complete sampling of sub-clouds
on the cosmological scale is expensive in terms of computation
and storage. To reduce the cost while maintaining the statistical 
robustness, here we introduce the balanced sampling
({\sc BSampling}) technique. For more details of the random sampling theory
and its general application, refer to, e.g. \citet{bishopPatternRecognitionMachine2006}.

First, we note that most of the statistical tasks for sub-clouds/GCs 
in this paper can be abstracted as an integration, which can 
be estimated by the standard, unweighted Monte Carlo method as
\begin{equation}
    \mathbb{E}(f) = \int f(\mathbf{x}) p(\mathbf{x}) \dd \mathbf{x} 
        \approx \frac{1}{N_{\rm sc}'} 
        \sum_{i=1}^{N_{\rm sc}'} f(\mathbf{x}_i).
\end{equation}
Here, $\mathbf{x} \sim p(\mathbf{x})$ is a set of sub-cloud/GC properties 
following the target probability density function of $p(\mathbf{x})$, $f(\mathbf{x})$ 
is the target function in interest, 
$\mathbb{E}(f)$ is the expectation value of $f$,
$\{ \mathbf{x}_i \}_{i=1}^{N_{\rm sc}'}$ is a random set of $N_{\rm sc}'$ sub-clouds 
sampled from $p(\mathbf{x})$. 
It can be proved that this estimator of $\mathbb{E}(f)$
has an asymptotic variance of $\sim N_{\rm sc}'^{-1}$, independent
of the dimension of $\mathbf{x}$, and is thus, in principle, an accurate and 
convenient way of solving many statistical problems. The actual speed of 
convergence of this estimator, however, depends on the smoothness of $f$ 
and the uniformity of $p(\mathbf{x})$, and thus can be slow in practice,
especially in our case where the distribution functions
of sub-clouds have a large dynamic range. It is also worth noting that
the formulation of the problem as $\mathbb{E}(f)$ covers not only the 
first-order statistics, but also higher-order ones. For example,
taking $f(\mathbf{x}) = \left[ \mathbf{x} - \mathbb{E}(\mathbf{x}) \right]^2$, 
one can estimate the variance of $\mathbf{x}$.

To address the large dynamic range of $p(\mathbf{x})$, we can propose a more 
balanced distribution, $q(\mathbf{x})$, to replace $p(\mathbf{x})$. Since the 
normalization constants of $p(\mathbf{x})$ and $q(\mathbf{x})$ are usually unknown, 
we express them as $p(\mathbf{x}) = \tilde{p}(\mathbf{x})/Z_p$ and $q(\mathbf{x}) = \tilde{q}(\mathbf{x})/Z_q$,
where $\tilde{p}(\mathbf{x})$ and $\tilde{q}(\mathbf{x})$ are the unnormalized distribution
functions and $Z_p$ and $Z_q$ are their normalization constants (also 
known as partition functions in physics), respectively. With the proposal 
distribution, we can rewrite the Monte Carlo estimator as
\begin{align} \label{eq:e-f-estimator-by-proposal}
    \mathbb{E}(f) 
    & = \int f(\mathbf{x}) p(\mathbf{x}) \dd \mathbf{x} 
    = \int f(\mathbf{x}) \frac{p(\mathbf{x})}{q(\mathbf{x})} q(\mathbf{x}) \dd \mathbf{x}        \nonumber\\
    & = \frac{Z_q}{Z_p} \int f(\mathbf{x}) \frac{\tilde{p}(\mathbf{x})}{\tilde{q}(\mathbf{x})} q(\mathbf{x}) \dd \mathbf{x}  \nonumber \\
    & \approx \frac{Z_q}{Z_p} \frac{1}{N_{\rm sc}'} 
        \sum_{i=1}^{N_{\rm sc}'} f(\mathbf{x}_i) \tilde{r}_i   \,,
\end{align}
where $\tilde{r}_i \equiv \tilde{p}(\mathbf{x}_i) / \tilde{q}(\mathbf{x}_i)$ is 
the probability ratio, and $\mathbf{x}_i$ is now drawn from $q(\mathbf{x})$. The factor
$Z_q / Z_p$ can be estimated in the same way using the same sample, because  
\begin{align} \label{eq:partition-function}
    \frac{Z_p}{Z_q}
    = \frac{1}{Z_q} \int \tilde{p}(\mathbf{x}) \dd \mathbf{x}
    = \int \frac{\tilde{p}(\mathbf{x})}{\tilde{q}(\mathbf{x})} q(\mathbf{x}) \dd \mathbf{x}
    \approx \frac{1}{N_{\rm sc}'} \sum_{i=1}^{N_{\rm sc}'} \tilde{r}_i \,.
\end{align}
Substituting this back into Eq.~\eqref{eq:e-f-estimator-by-proposal}, we 
have 
\begin{equation} \label{eq:e-f-estimator-by-w}
    \mathbb{E}(f) \approx \sum_{i=1}^{N_{\rm sc}'} f(\mathbf{x}_i) \tilde{w}_i \,,
\end{equation}
where $\tilde{w}_i = \tilde{r}_i / \sum_{j=1}^{N_{\rm sc}'} \tilde{r}_j$.
Here we define the weight $w_i \equiv N_{\rm sc} \tilde{w}_i $,
with $N_{\rm sc}$ being the real number of sub-clouds formed in a snapshot 
in the history of a galaxy, as constrained by the total stellar mass, 
$\Delta M_*$, formed in the entire galaxy (see \S\ref{ssec:impl}). 
Thus, $w_i$ can be considered as the number of real 
sub-clouds represented by a B-sampled sub-cloud. 
Note that using $q(\mathbf{x})$ as a replacement can not only improve the speed of 
convergence, but also allow us to avoid drawing samples directly from $p(\mathbf{x})$,
in cases where $p(\mathbf{x})$ is an over-complicated distribution. 

The proper choice of $q(\mathbf{x})$ turns out to be crucial to the sampling 
efficiency. Two requirements for $q(\mathbf{x})$ are that it should be 
nearly uniform in regions of interest, and that it should be easy to draw. 
Thus, we factorize the distribution of $\mathbf{x}$ into a product of
the marginal distributions:
\begin{equation}
    p(\mathbf{x}) = \prod_j p^{(j)}(x_i^{(j)})\,,
\end{equation}
where each $x_i^{(j)}$ is a sub-cloud/GC property.
The properties interested in this paper are
sub-cloud mass ($M_{\rm sc}$; see \S\ref{ssec:formation-of-sc} and Eq.~\ref{eq:m-sc-pdf}), 
density contrast ($s$; see \S\ref{ssec:formation-of-sc} and Eq.~\ref{eq:ln-density-pdf}), 
star formation threshold ($n_{\rm sc}'$; see \S\ref{ssec:cooling} and Eq.~\ref{eq:sampling-n-sc}) 
and the metallicity of the stream inflow ($Z_{\rm sc}$; see \S\ref{ssec:metal-bimodal} and Eq.~\ref{eq:pdf-Z-sc}). 
We choose a power-law form for each of the proposal distributions, 
$q(x_i^{(j)})$, with the power-law index close to $-1$, to ensure the uniformity 
in the logarithmic space. After a number of trials, we find that 
\begin{align}
    q(M_{\rm sc}) & \propto M_{\rm sc}^{-0.75};         \\
    q(n_{\rm sc}') & \propto n_{\rm sc}'^{-0.75};       \\
    q(Z_{\rm sc}) & \propto Z_{\rm sc}^{-0.75};         \\
    q(s) & \propto s^{-1.25}
\end{align}
yields stable results, where the more negative index of $s$ accounts for 
its log-normal piece that dominates the sub-cloud populations in most cases.
The final probability ratio of each sub-cloud, $\tilde{r}_i$, 
is the product of each factor, $\tilde{r}_i^{(j)} = \tilde{p}(x_i^{(j)})/\tilde{q}(x_i^{(j)})$, namely
\begin{equation}
    \tilde{r}_i = \tilde{r}_i^{(M_{\rm sc})} \tilde{r}_i^{(n_{\rm sc}')} \tilde{r}_i^{(s)}
\end{equation}
for the fiducial model, and 
\begin{equation}
    \tilde{r}_i = \tilde{r}_i^{(M_{\rm sc})} \tilde{r}_i^{(n_{\rm sc}')} \tilde{r}_i^{(s)} \tilde{r}_i^{(Z_{\rm sc})}
\end{equation}
for the variant with cold stream inflow (see \S\ref{ssec:metal-bimodal}). 

With the balanced proposal distributions, the number of Monte Carlo samples,
$N_{\rm sc}'$, can be reduced much compared to the real number of sub-clouds,
$N_{\rm sc}$. We thus set
\begin{equation}
    N_{\rm sc}' = \max \left\{ 
        \lfloor \left(\frac{\Delta M_*}{M_{\rm B}}\right)^{\beta_{\rm B}}  \rfloor , 
        N_{\rm sc, min}' 
    \right\} \,.
\end{equation}
Here, we adopt the power-index $\beta_{\rm B} = 1/2$, a value less than 1 to ensure
$N_{\rm sc}' < N_{\rm sc}$ asymptotically, and $\geq 1/2$ to ensure that the
Poisson noise is not dominant.
The lower bound $N_{\rm sc, min}' = 4$ is set to ensure that sub-clouds in 
low-mass galaxies are well represented.
The batch mass $M_{\rm B}$ is empirically set to be $10^{3} \msun$ for halos with 
$M_{{\rm v},z=0} < 10^{12} \msun$, and $2\times 10^{5} \msun$ otherwise, which 
we have verified to be stable. 
Fig.~\ref{fig:b-sample-scale} shows the total number of sub-clouds
B-sampled in the history of a galaxy at $z=0$ as a function of its host 
halo mass and the real number of sub-clouds. $N_{\rm sc}'$ is about 
two orders of magnitude smaller than $N_{\rm sc}$ for massive galaxies,
and thus significantly reduces the computation and storage requirements.

Other target distributions that do not cause imbalance, 
such as that of $u_{\rm r}$ for the effective orbit of dynamical evolution (see \S\ref{ssec:dyn-evol}
and Eq.~\ref{eq:tidal-frequency}), are sampled by the usual Monte Carlo method.
In cases where a real, un-weighted sample of sub-clouds is needed
(see e.g. \S\ref{ssec:spatial-dist} and Fig.~\ref{fig:spatial-example}), 
a resampling of $N_{\rm sc}$ sub-clouds can be made from the 
$N_{\rm sc}'$ B-sampled sub-clouds, $\{ \mathbf{x}_i \}_{i=1}^{N_{\rm sc}'}$, 
using the associated importance weights, $\{w_i\}_{i=1}^{N_{\rm sc}'}$,
as discussed in \S\ref{ssec:impl}.

\section{Calibration of the Model}\label{sec:calibration}

\begin{figure} \centering
    \includegraphics[width=0.99\columnwidth]{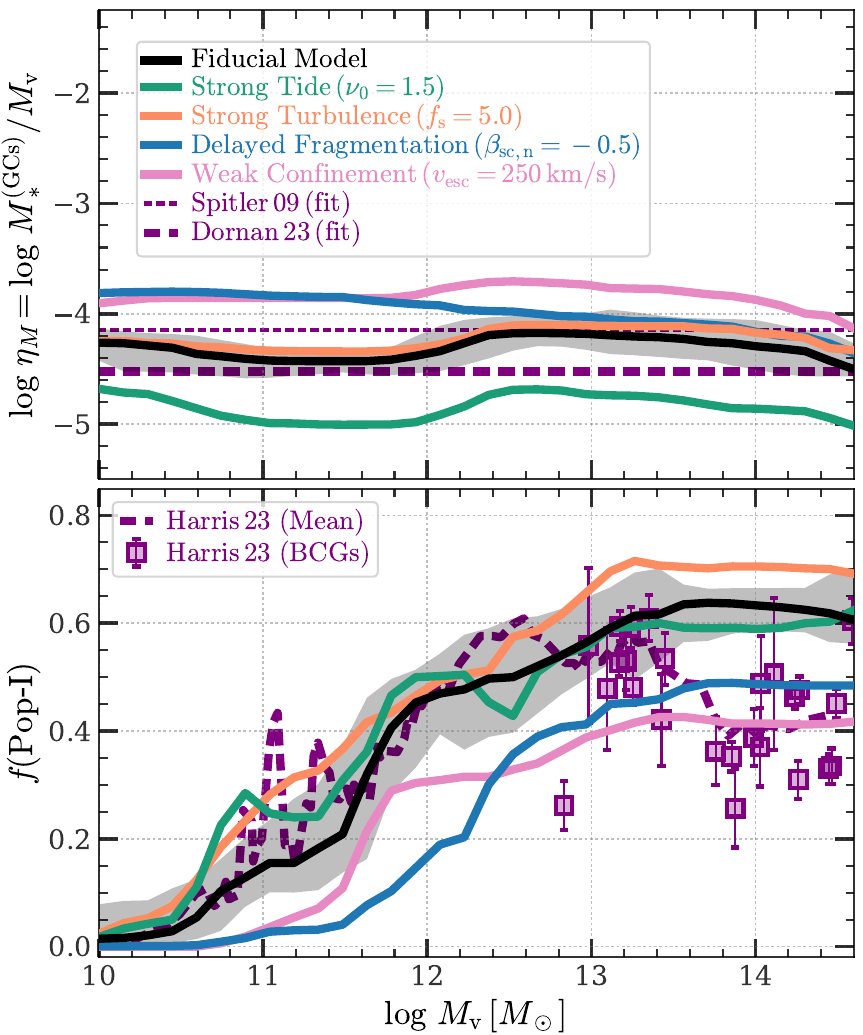}
    \caption{
        Calibration of the model parameters. 
        {\bf Solid curves} show the median relations from different 
        variants of the model.
        The shaded area of the `Fiducial Model' indicates the 1-$\sigma$ range.
        {\bf Purple dashed curves and markers} show the observations, the same 
        as those shown in Fig.~\ref{fig:m_gc_vs_mhalo} and 
        Fig.~\ref{fig:f_red_vs_mhalo}.
        {\bf Upper panel} shows the mass ratio of the GC system to the 
        host halo, as a function of the host halo mass. 
        {\bf Lower panel} shows the fraction of Pop-I (red) GCs as 
        a function of the host halo mass.
        All galactic GCs (in-situ and ex-situ) of central galaxies at 
        $z=0$ are included in the analysis. For details, see
        Appendix~\ref{sec:calibration}.
    }
    \label{fig:calibration}
\end{figure}

The model presented in this paper includes several free parameters: 
$\beta_{\rm sc,n}$, $\nu_0$, $f_{\rm s}$, and those related to metal enrichment 
(see Table~\ref{tab:parameters} for a complete list). We calibrate these 
parameters using two observational results: the $M_{\rm *}^\text{(GCs)}$-$M_{\rm v}$ 
relation and the $f(\text{Pop-I})$-$M_{\rm v}$ relation for galactic 
GC systems at $z=0$.

The parameter $\beta_{\rm sc, n}$ describes the growth of clumpiness 
in a single sub-cloud during its cooling and contraction 
(see \S\ref{ssec:cooling}). A smaller value of $\beta_{\rm sc, n}$ results 
in a more rapid growth of density perturbation, therefore accelerating the 
decrease in the cooling timescale $t_{\rm cool, sc}$
(see Eq.~\ref{eq:t-cool-sc-cii}) and leading to earlier fragmentation 
(see Eq.~\ref{eq:n-sf-threshold}). Conversely, a larger value of 
$\beta_{\rm sc, n}$ delays fragmentation. Thus, $\beta_{\rm sc, n}$ determines 
the fraction of sub-clouds that can reach the SN-free regime via the Pop-II 
channel, and consequently, the total mass of GCs that can 
form in low-mass galaxies.
The black curve in the upper panel of Fig.~\ref{fig:calibration} shows 
the GC-to-halo mass ratio as a function of host halo mass predicted by 
our fiducial model, with $\beta_{\rm sc, n} = -2.5$. For comparison, 
we define a variant, denoted as `Delayed Fragmentation', 
with $\beta_{\rm sc, n} = -0.5$, shown by the blue curve. The delayed 
fragmentation results in higher GC mass in galaxies below the Milky Way 
mass due to enhanced formation of Pop-II GCs. Consequently, the slope of the 
mass ratio curve changes, becoming inconsistent with the observed 
$M_{\rm *}^\text{(GCs)}$-$M_{\rm v}$ proportionality. Based on this effect of 
$\beta_{\rm sc, n}$, we can determine its optimal value and adopt 
it as the fiducial one.

The parameter $\nu_0$ determines the tidal disruption timescale of 
GCs (see \S\ref{ssec:dyn-evol} and Eq.~\ref{eq:tidal-time-scale}). A larger 
value of $\nu_0$ results in a shorter survival time for GCs, leading to 
fewer observable GCs at any given snapshot. The green curves in 
Fig.~\ref{fig:calibration} show the results of a model variant, 
denoted as `Strong Tide', with $\nu_0 = 1.5$. The change in the GC-to-halo mass 
ratio, compared to the fiducial result with $\nu = 0.6$, is quite significant, 
and the fraction of the change is almost independent of the host halo mass. 
The effect of $\nu_0$ on the $f(\text{Pop-I})$-$M_{\rm v}$ relation is moderate. 
The Pop-I fraction in low-mass halos increases because the in-situ Pop-II GCs, 
which are born earlier (see Fig.~\ref{fig:gc_history}), are more likely to be 
disrupted by the strong tidal field. Based on these features of $\nu_0$, we 
can determine its calibrated value mainly using the GC-to-halo mass ratio, 
with the aid of the Pop-I fraction.

The parameter $f_{\rm s}$ describes the strength of supersonic turbulence 
in the creation of dense sub-clouds, which is the only channel for forming 
Pop-I GCs. A larger value of $f_{\rm s}$ leads to more efficient formation 
of Pop-I GCs, resulting in a higher fraction of Pop-I GCs. However, the total 
mass of GCs does not change significantly, as a broader sub-cloud density 
distribution increases both the probability of a sub-cloud reaching the 
SN-free threshold, $n_{\rm snf}$ (see \S\ref{ssec:gc-channels}), and the production 
of more sub-clouds that are just below the threshold. Because the total 
stellar mass has been well constrained by the boundary condition provided by the 
galactic-scale context (see \S\ref{sec:galaxy-model}), the total mass of GCs 
is not sensitive to $f_{\rm s}$. The orange curves in Fig.~\ref{fig:calibration} 
show the results of a model variant, denoted as `Strong Turbulence', 
with $f_{\rm s} = 5.0$. Compared to the fiducial model with $f_{\rm s} = 1.0$, 
the changes in the GC-to-halo mass ratio and Pop-I fraction are as expected. 
Based on these features of $f_{\rm s}$, we can determine its calibrated value mainly 
using the Pop-I fraction, with the aid of the GC-to-halo mass ratio.

The most uncertain part of the model is the process of metal enrichment, 
partly due to the limited constraints from high-redshift observations
\citep[e.g.][]{curtiMassmetallicityFundamentalMetallicity2020,
sandersMOSDEFSurveyEvolution2021,
curtiJADESInsightsLowmass2024,venturiGasphaseMetallicityGradients2024}, and partly 
due to the potential systematic differences in metallicity in various regions 
of the host galaxy \citep{wangDiscoveryStronglyInverted2019,wangEarlyResultsGLASSJWST2022,
venturiGasphaseMetallicityGradients2024}, particularly within the dense sub-clouds. 
Given the lack of constraints, we propose a heuristic model 
in \S\ref{ssec:metal}, motivated by the gas-regulator scenario, 
and adapt it to account for conditions prevalent at high redshifts. The 
resulting MZR of this model is required to reproduce observed trends, 
including the slow redshift evolution at $z \gtrsim 4$
\citep[e.g.][]{nakajimaJWSTCensusMassMetallicity2023}, 
the slope of $\approx 1/3$–$2/3$ at the low-stellar-mass end
\citep[see e.g. figure 3 of][]{curtiMassmetallicityFundamentalMetallicity2020},
and a bending at the high-stellar-mass end
\citep[e.g.][]{tremontiOriginMassMetallicityRelation2004,
mannucciFundamentalRelationMass2010}.
All parameters involved in the metal enrichment process affects GC formation 
via the Pop-II channel (see \S\ref{ssec:gc-channels}), and we adjust them 
within acceptable ranges that meet the aforementioned
requirements to reproduce the GC-to-halo mass ratio and the Pop-I fraction. 
For example, the pink curves in Fig.~\ref{fig:calibration} show the results of a 
model variant, denoted as `Weak Confinement', with an escape velocity of 
$v_{\rm esc} = 250 \kms$, which is larger than the fiducial value of $75 \kms$. 
This variant tends to produce a more metal-poor environment, therefore enhancing 
the formation of Pop-II GCs. Consequently, the changes in the GC-to-halo 
mass ratio and Pop-I fraction are both significant, resulting in a clear 
inconsistency with observations.

An interesting finding by \citet{harrisPhotometricSurveyGlobular2023}, based on 
HST data, is the decline in the Pop-I fraction of BCGs, as indicated by individual 
markers in the lower panel of Fig.~\ref{fig:calibration}. Should this 
observation be confirmed by future studies with higher resolution and larger 
sample sizes, it could pose a significant challenge to most of the current models
of GC formation. Our analysis suggests that this finding is consistent with a 
scenario where high-redshift galaxies in massive halos are fed by pristine cold 
streams, leading to efficient GC formation via the Pop-II channel. 
Because this scenario provides a promising connection between low-redshift relics and 
high-redshift gas accretion processes, we have defined an additional 
suite of variants and discuss their implications in \S\ref{ssec:metal-bimodal}.

% If you want to present additional material which would interrupt the flow of the main paper,
% it can be placed in an Appendix which appears after the list of references.

% End of mnras_template.tex

% Don't change these lines
\bsp	% typesetting comment
\label{lastpage}

\end{document}